\pdfoutput=1

\documentclass[a4paper, 12pt]{report}
\usepackage[a4paper]{geometry}
\usepackage{epsfig,rotating}
\usepackage{amssymb}
\usepackage{mathrsfs}
\usepackage{amsmath}
\usepackage{amsfonts}
\usepackage{multirow}
\usepackage{eurosym}
\usepackage{url}

\newcommand{\bbonu}{\ensuremath{\beta\beta^{0\nu}}}
\newcommand{\bbtnu}{\ensuremath{\beta\beta^{2\nu}}}
\newcommand{\qbb}{\ensuremath{Q_{\beta\beta}}}
\newcommand{\Qbb}{\ensuremath{Q_{\beta\beta}}}
\newcommand{\mbb}{\ensuremath{m_{\beta\beta}}}
\newcommand{\bb}{\ensuremath{\beta\beta}}
\newcommand{\Tonu}{\ensuremath{T^{0\nu}_{1/2}}}

\newcommand{\MO}{\ensuremath{{}^{100}{\rm Mo}}}
\newcommand{\SE}{\ensuremath{{}^{82}{\rm Se}}}
\newcommand{\ND}{\ensuremath{{}^{150}{\rm Nd}}}
\newcommand{\XE}{\ensuremath{{}^{136}\rm Xe}}
\newcommand{\GE}{\ensuremath{{}^{76}\rm Ge}}
\newcommand{\TE}{\ensuremath{{}^{128}\rm Te}}
\newcommand{\TEX}{\ensuremath{{}^{130}\rm Te}}
\newcommand{\TL}{\ensuremath{{}^{208}\rm{Tl}}}
\newcommand{\CA}{\ensuremath{{}^{48}\rm Ca}}

\newcommand{\CD}{\ensuremath{^{116}{\rm Cd}}}

\newcommand{\BI}{\ensuremath{{}^{214}}Bi}

\newcommand{\Monu}{\ensuremath{\Big|M^{0\nu}\Big|}}

\newcommand{\Gonu}{\ensuremath{G^{0\nu}(E_0, Z)}}

\newcommand{\tetaot}{\ensuremath{\theta_{13}}}
\newcommand{\tetatt}{\ensuremath{\theta_{23}}}
\newcommand{\tetaotw}{\ensuremath{\theta_{12}}}

\newcommand{\dsun}{\ensuremath{\Delta_{\rm sun}^2}}
\newcommand{\datm}{\ensuremath{\Delta_{\rm atm}^2}}

\newcommand{\Bi}{\ensuremath{^{214}}Bi}

\newcommand{\Tl}{\ensuremath{^{208}}Tl}

\newcommand{\Pb}{\ensuremath{^{208}}Pb}

\newcommand{\Po}{\ensuremath{^{214}}Po}

\def\email#1{\def\@email{#1}}

\begin{document}
\pdfoutput=1

\begin{titlepage}

\vspace*{1cm}

\centering

{\Large \sf Letter of Intent to the LSC Scientific Committee} \\ \vspace{1cm}
{\LARGE \bf NEXT, a HPGXe TPC for neutrinoless double beta decay searches} \\ \vspace{1cm}

{\normalsize \sf (3 April 2009)} \\ \vspace{2cm}

\begin{minipage}{14cm}
\begin{abstract}
\small
\noindent
We propose a novel detection concept for neutrinoless double-beta decay searches. This concept is based on a Time Projection Chamber (TPC) filled with high-pressure gaseous xenon, and with separated-function capabilities for calorimetry and tracking. Thanks to its excellent energy resolution, together with its powerful background rejection provided by the distinct double-beta decay topological signature, the design discussed in this Letter Of Intent promises to be competitive and possibly out-perform existing proposals for next-generation neutrinoless double-beta decay experiments. We discuss the detection principles, design specifications, physics potential and R\&D plans to construct a detector with 100 kg fiducial mass in the double-beta decay emitting isotope $^{136}$Xe, to be installed in the Canfranc Underground Laboratory.

\end{abstract}
\end{minipage}

\end{titlepage}

\pagebreak

\thispagestyle{empty}
\parindent 0.cm

{\Large \bf The NEXT Collaboration} \\ 

{\sc 
F.~Gra\~nena, 
T.~Lux,
F.~Nova,  
J.~Rico, 
F.~S\'anchez} \\
{\it Institut de F\'isica d'Altes Energies (IFAE), Barcelona, Spain} \vspace{.5cm} \\ 
{\sc 
D.~R.~Nygren} \\
{\it Lawrence Berkeley National Laboratory, Berkeley, USA} \vspace{.5cm} \\
{\sc 
J.~A.~S.~Barata,
F.~I.~G.~M.~Borges,
C.~A.~N.~Conde,
T.~H.~V.~T.~Dias,
\mbox{L.~M.~P.~Fernandes},
E.~D.~C.~Freitas,
J.~A.~M.~Lopes,
\mbox{C.~M.~B.~Monteiro},
\mbox{J.~M.~F.~dos Santos},
F.~P.~Santos,
L.~M.~N.~T\'avora, 
J.~F.~C.~A.~Veloso} \\ 
{\it Universidade de Coimbra, Portugal} \vspace{.5cm} \\
{\sc 
E.~Calvo, 
I.~Gil-Botella, 
P.~Novella, 
C.~Palomares, 
A.~Verdugo} \\
{\it CIEMAT, Madrid, Spain} \vspace{.5cm} \\
{\sc 
I.~Giomataris, 
E.~Ferrer-Ribas} \\
{\it CEA, IRFU, Saclay, France} \vspace{.5cm} \\
{\sc 
J.~A.~Hernando-Morata, 
D.~Mart\'inez, 
X.~Cid}\\
{\it Universidade de Santiago de Compostela, Spain} \vspace{.5cm} \\
{\sc 
M.~Ball, 
S.~C\'arcel, 
A.~Cervera, 
J.~D\'iaz, 
A.~Gil, 
J.~J.~G\'omez-Cadenas\footnote{Spokesperson: gomez@mail.cern.ch}, 
\mbox{J.~Mart\'in-Albo}, 
F.~Monrabal, 
J.~Mu\~noz-Vidal, 
L.~Serra, 
M.~Sorel, 
\mbox{N.~Yahlali}} \\
{\it Instituto de F\'isica Corpuscular (IFIC), CSIC - U.\ de Valencia, Valencia, Spain} \vspace{.5cm} \\
{\sc 
R.~Esteve Bosch, 
C.~W.~Lerche,
J.~D.~Mart\'inez,
F.~J.~Mora,
\mbox{A.~Sebasti\'a},
A.~Tarazona,
J.~F.~Toledo} \\
{\it Instituto ITACA, U.\ Polit\'ecnica de Valencia, Valencia, Spain} \vspace{.5cm} \\
{\sc M.~L\'azaro, 
J.~L.~P\'erez, 
L.~Ripoll} \\ 
{\it U.\ Polit\'ecnica de Valencia, Spain} \vspace{.5cm} \\
{\sc 
J.~M.~Carmona, 
S.~Cebri\'an, 
T.~Dafni, 
J.~Gal\'an, 
H.~G\'omez,
F.~J.~Iguaz, 
\mbox{I.~G.~Irastorza}, 
G.~Luz\'on, 
J.~Morales,
A.~Rodr\'iguez, 
J.~Ruz, 
A.~Tom\'as, 
\mbox{J.~A.~Villar}} \\
{\it Instituto de F\'isica Nuclear y Altas Energ\'ias, U.\ de Zaragoza, Zaragoza, Spain} \\

\tableofcontents
\chapter*{Overview}
\addcontentsline{toc}{chapter}{Overview}
Neutrinos have mass. This experimental evidence, unambiguously established in the last 
decade, has deep implications in physics and cosmology, and reinforces the possibility that 
Ettore Majorana's insight was true: massive neutrinos could be their own antiparticles.
This would account for the see-saw mechanism necessary to describe the lightness of 
neutrino mass relative to other fermions of the same generation without fine-tuning, and 
would provide a possible mechanism for leptogenesis. If neutrinos are Majorana particles, 
neutrinoless double beta (\bbonu) processes can be observed.

In fact, it could be that \bbonu\ processes have {\em already} been observed. A positive 
evidence, put forward by members of the Heidelberg-Moscow (HM) experiment would imply 
an effective neutrino mass of 0.24--0.58 meV. As it happens often in frontier science, the result is 
controversial and other experiments with similar sensitivities do not confirm it. 

One clear lesson repeated several times in neutrino physics is that, in order to confirm or 
unambiguously refute experimental data, one needs to perform experiments much more 
sensitive that the ones being disputed. Neutrinoless double beta decay searches are no 
exception to this rule, and in fact, a new generation of \bbonu\ experiments is being 
planned, constructed or commissioned. Some of the more sensitive (GERDA and 
{\sc Majorana}) use  the same isotope (\GE) and the same experimental technique 
than the HM experiment, and are therefore subject to similar uncertainties. 

When it comes to search for the Holy Grail more than one knight is necessary. Requirements 
for \bbonu\ experiments are often contradictory: ultimate resolution (achievable, for example, 
with Ge crystals); extra handles, such as a topological signature of the two electrons characterizing 
a \bbonu\ events (boasted by NEMO and SuperNEMO, which, however, have inferior resolution 
to other techniques and are not fully active detectors); and scalability to large masses (difficult 
for all of them and, normally, based on replication). 

In this Letter Of Intent (LOI) we formally propose an experiment, based on a high-pressure
gaseous xenon (HPGXe) TPC, which combines all the above desirable features: excellent resolution, 
at least as good as 1\% FWHM at \Qbb; a fully active detector, easy to scale up as a single
volume; and the availability of a topological signature to provide an extra handle to reject 
backgrounds. In addition, the isotope \XE\ is almost 9\% of natural xenon, and enrichment by
centrifugation is a relatively easy and comparatively cheap technology.

This experiment is called \emph{Neutrino Experiment with a Xenon TPC} (NEXT). We present 
here a full research program whose goal is to build and operate at the Canfranc Underground 
Laboratory (LSC) a 100 kg (fiducial mass) detector capable of confirming (with a large signal) 
or unambiguously refuting the HM claim. Its sensitivity can be improved to fully explore the inverse
hierarchy, competing and possibly outperforming in physics potential other existing or proposed 
next-generation experiments. Operation of this detector is essential to provide a deep understanding
of the experimental techniques to suppress backgrounds that would allow the extrapolation to 
larger (1 ton and beyond) detectors.

This document is organized as follows. Chapter 1 introduces the basic concepts of \bb\
phenomenology and briefly reviews the experimental status of the field. In Chapter 2 the physics
principles of the proposed experimental technique are presented. In Chapter 3 we describe the
possible designs of the NEXT detector. The physics potential of NEXT is discussed in Chapter 4.
Finally, in Chapter 5 we explain the  projected research program for the following years.

\chapter{Introduction}
\section{The importance of \bbonu\ searches} \label{INTRO}
Neutrino oscillation experiments have demonstrated that neutrinos have masses and mix 
\cite{Gonzalez-Garcia:2007}. In the Standard Model, neutrinos were introduced as 
massless particles. Therefore, it is necessary to extend the model to accommodate them. 
The mechanism responsible for generating neutrino masses is in turn 
related to the Majorana or Dirac nature of neutrinos. Neutrinos could be Majorana particles, 
that is, identical to their antiparticles. All other Standard Model fermions, being electrically 
charged, are instead Dirac particles, distinguishable from their own antiparticles. Majorana 
neutrinos provide an attractive explanation for the smallness of neutrino masses, the 
so-called \emph{seesaw} mechanism. Besides, Majorana neutrinos violate lepton-number 
conservation. This, together with CP-violation, is a basic ingredient to help uncover the 
reasons why matter dominates over antimatter in our Universe.

The most promising experimental method to reveal the neutrino nature is the search
for neutrinoless double beta decay (\bbonu). Absent any lepton-number-nonconserving 
interactions, such a process is possible if and only if neutrinos are massive, Majorana particles. 
In addition, the measurement of the half-life of the \bbonu\ decay would provide direct information 
on the absolute scale of neutrino masses. 

In addition to neutrinoless double beta decay searches, information on the absolute scale of 
neutrino masses can be obtained with $\beta$ decay experiments and cosmology. Only upper 
bounds on the neutrino mass, of order $\sim$1 eV, currently exist. 

\begin{figure}[tb]
\centering
\includegraphics[width=0.85\textwidth]{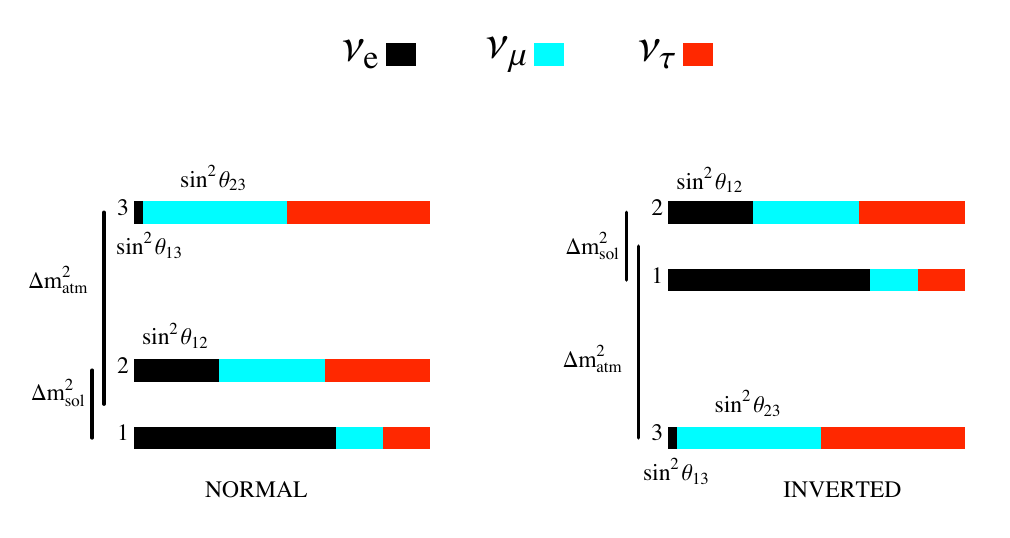} 
\caption{\small The two possible orderings of neutrino masses, as suggested by 
			oscillation experiments. The flavor composition of the mass eigenstates 
			is shown as well \cite{Mena:2005}. \label{fig:mass_hierarchies}}
\end{figure}

Concerning the neutrino mass spectrum, current neutrino oscillation results cannot differentiate 
between the two possible mass orderings, usually referred to as \emph{normal} and 
\emph{inverted} hierarchies (see Figure \ref{fig:mass_hierarchies}). In the former, the gap between 
the two lightest mass eigenstates corresponds to the small mass difference, measured by solar
experiments (\dsun), while in the second case the gap between the two lightest states corresponds 
to the large mass difference, measured by atmospheric experiments (\datm). In the particular case in 
which the neutrino mass differences are very small compared with its absolute scale, we speak of the 
{\it degenerate} spectrum. Oscillation experiments have also measured two mixing angles 
(\tetaotw\ and \tetatt). The third mixing angle (\tetaot) is known to be small. 

\section{Double beta decay rate}
Double beta decay (\bb) is a rare transition between two nuclei with the same mass 
number $A$ that changes the nuclear charge $Z$ by two units. The decay can occur only 
if the initial nucleus is less bound than the final nucleus, and both more than the 
intermediate one.

There are two possible \bb\ decay modes. The standard two-neutrino
double beta decay (\bbtnu),
\begin{equation}
(Z, A) \rightarrow (Z+2, A) + e^-_{1} + e^-_{2} + \overline{\nu}_{e_1} + \overline{\nu}_{e_2},
\end{equation}
was first proposed by Goeppert-Mayer in 1935 \cite{GoeppertMayer:1935qp}, and
has since then been observed for many nuclei, such as \GE, \CA, \MO, \SE\ or \ND.
Typical lifetimes are of the order of 10$^{19}$--10$^{20}$ years.

The neutrinoless mode,
\begin{equation}
(Z, A) \rightarrow (Z+2, A) + e^-_1 + e^-_2,
\end{equation}
is a SM-forbidden process since it violates lepton number conservation. This mode was first 
proposed by Racah in 1937 \cite{Racah:1937qq}, following the fundamental suggestion of 
Majorana \cite{Majorana:1937vz} that same year. To date, no convincing experimental 
evidence for this mode exists.

\begin{figure}[tb]
\begin{center}
\includegraphics[width=0.5\textwidth]{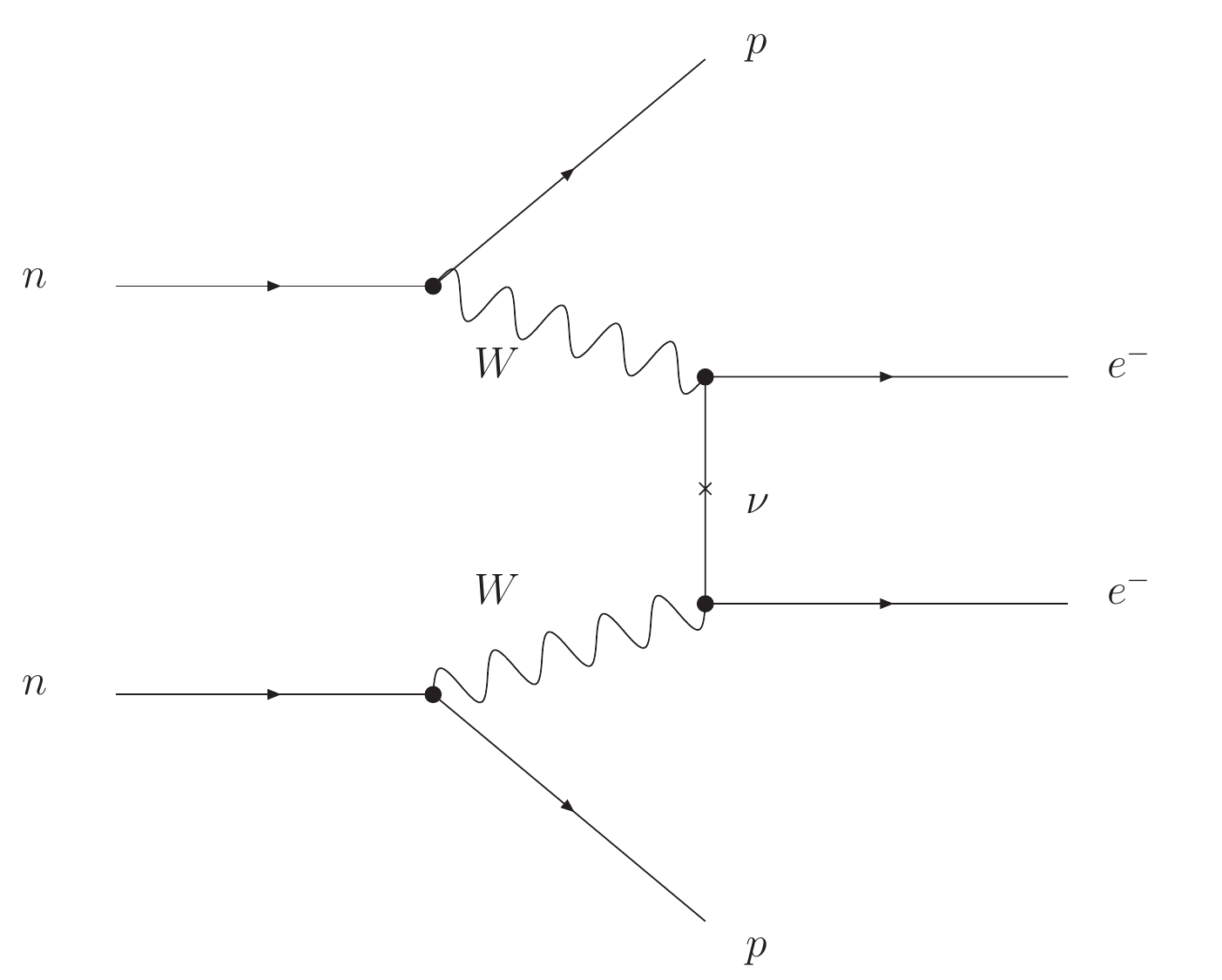} 
\end{center}
\vspace{-1.0cm}
\caption{\small Neutrinoless double beta decay mediated by the exchange of
	light Majorana neutrinos. \label{fig:bb0nu_diagram} }
\end{figure}

Neutrinoless double beta decay can be mediated by many underlying mechanisms
(involving, in general, physics beyond the SM), the simplest one being the exchange 
of light Majorana neutrinos (Figure \ref{fig:bb0nu_diagram}). All of them imply 
nevertheless a Majorana mass term for the neutrino \cite{Schechter:1982}.

The inverse of the lifetime for \bbonu\ processes mediated by the exchange of 
light Majorana neutrinos can be expressed in terms of the phase-space factor \Gonu\ 
and the nuclear matrix element (NME) 
$|M^{0\nu}|$ \cite{Elliott:2002}:
\begin{equation}
\frac{1}{T^{0\nu}} = \mbb^2 \ \Monu \ \Gonu.
\label{eq:Tonu}
\end{equation}

\begin{figure}[t!]
\begin{minipage}[t]{0.47\textwidth}
\begin{center}
\includegraphics*[viewport=104 427 441 749, width=0.95\textwidth]{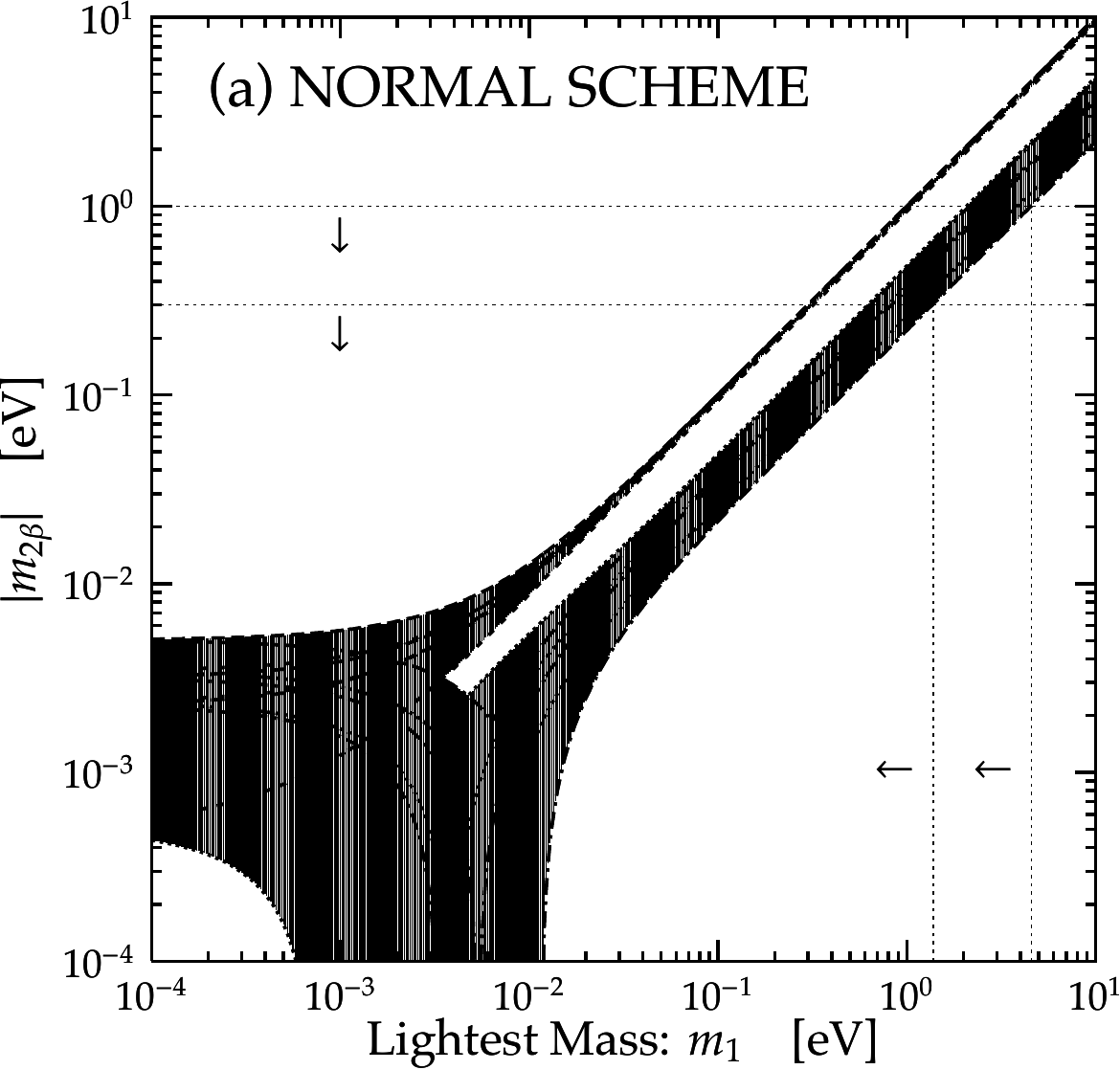}
\end{center}
\end{minipage}
\hfill
\begin{minipage}[t]{0.47\textwidth}
\begin{center}
\includegraphics*[viewport=104 427 441 749, width=0.95\textwidth]{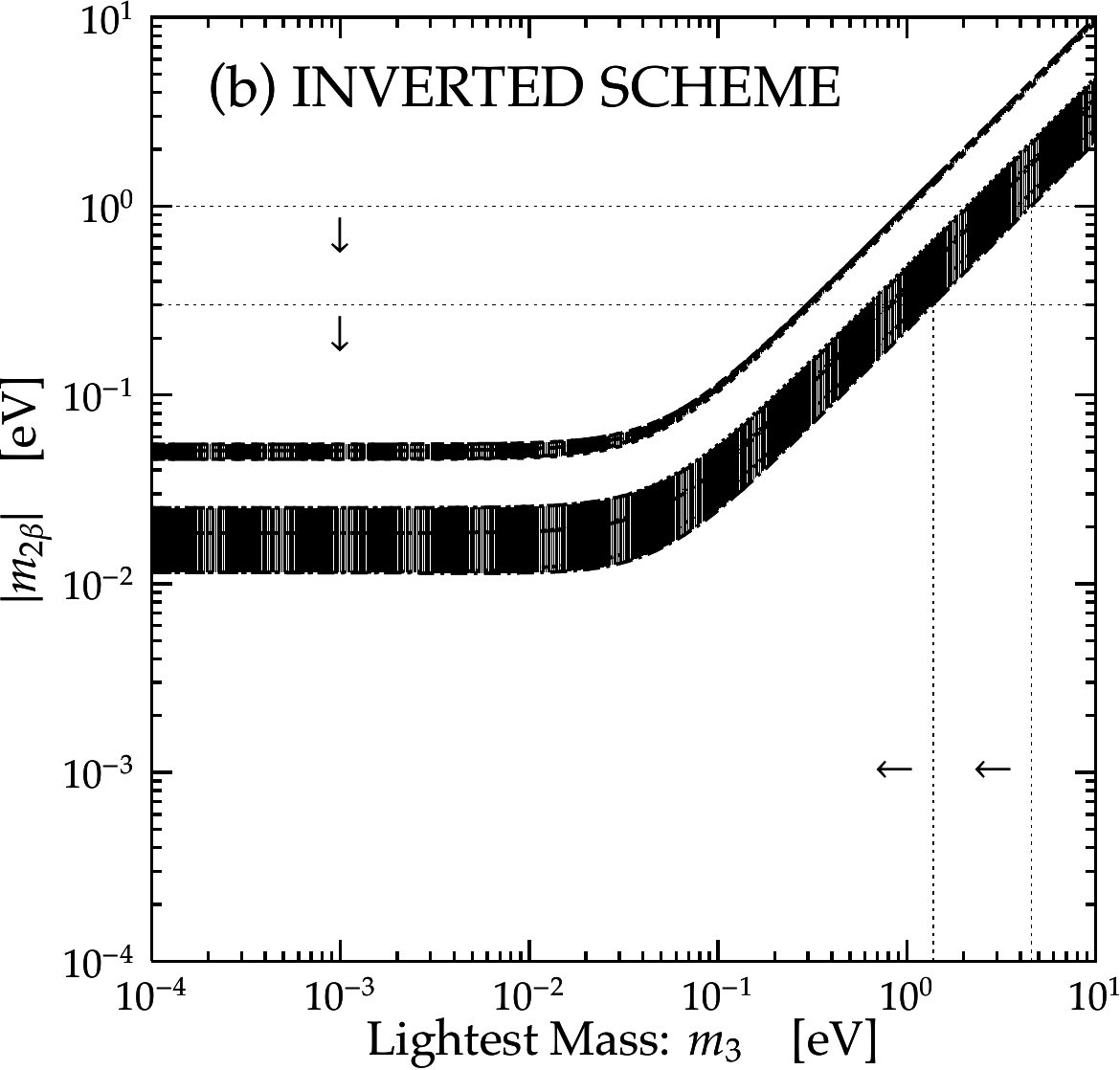}
\end{center}
\end{minipage}
\caption{ \small \label{fig:db}
Absolute value of the effective Majorana neutrino mass in \bbonu\ decay as a function
of the lightest mass: $m_{1}$ in the normal scheme (a) and $m_{3}$ in the inverted scheme (b).
The white areas in the strips need CP violation. The two horizontal dotted lines correspond to 
the 90\% confidence level upper bounds from \bbonu\ searches for two different assumptions on nuclear matrix elements. The two vertical dotted lines show the 
corresponding upper bounds for $m_{1}$ (a) and $m_{3}$ (b), obtained from the intersection of the horizontal lines with the filled areas. From Reference \cite{Giunti:2006}.
}
\end{figure}

The effective neutrino mass, \mbb, is defined as the $m_{11}$ element of the neutrino mass matrix in the flavor basis:
\begin{equation}
\label{eq:mass}
\mbb^2 = \Big| \sum_{i} U^2_{ei}m_{i} \Big|^2 = \Big| \sum_{i} |U_{ei}|^2 e^{\phi_{i}} m_{i} \Big|^2 . 
\end{equation}
The relation of \mbb\ with the mass eigenstates and the oscillation parameters is \cite{Elliott:2002}:
\begin{equation}
\mbb = \Big| \cos^2 \theta_{13} \ (|m_1| \cos^2\theta_{12} + |m_2| e^{2i\phi_1} \sin^2\theta_{12}) + |m_3| e^{2i(\phi_2-\delta)} \sin^2\theta_{13} \Big|.
\end{equation}
Notice that \mbb\ depends on the unknown phases $\delta$, $\phi_1$, $\phi_2$ of the Pontecorvo-Maki-Nakagawa-Sakata (PMNS) matrix. However, the lower and upper limits of \mbb\ are independent of such phases and depend only on the absolute values of the mixing angles. 
Thus, if one observes \bbonu\ processes, the known values of the oscillation parameters can be used to deduce a range of absolute values for neutrino masses, as illustrated in Figure  
\ref{fig:db}. 
Conversely, if a given experiment does not observe the \bbonu\ process, the result can be interpreted in terms of a bound on \mbb.  

\section{Past, present and future \bbonu\ experiments}
\label{sec:bbexp}

\begin{table}[tbp]
\centering
\footnotesize
\begin{tabular}{cccccc}
\hline \hline
Isotope & Experiment, date & Expos.\ (kg$\cdot$y) & Technique & \Tonu\ ($10^{23}$ y) & \mbb\ (eV) \\
\hline
\CA\ & Elegant VI, 2004 & 4.2 & scintillator & 0.14 & $<$(7.2-44.7) \\
\GE\ & HM-Klapdor, 2004 & 71.7 & ionization & $(119^{+299}_{-50})$ & 0.24--0.58\\
\GE\ & IGEX, 2001 & 8.9 & ionization & 157 & $<$(0.33--1.35)\\
\SE\ & NEMO-3, 2007 & 1.8 & tracking & 2.1 & $<$(1.2--3.2)\\
\MO\ & NEMO-3, 2007 & 13.1 & tracking & 5.8 & $<$(0.6--2.7)\\
\CD\ & Solotvina, 2003 & 0.5 & scintillator & 1.7 & $<$1.7\\
\TEX\ & Cuoricino, 2007 & 11.8 & bolometer & 30.0 & $<$(0.41--0.98)\\
\XE\ & DAMA, 2002 & 4.5 & scintillator & 12.0 & $<$(0.8--5.6)\\
\ND\ & Irvine TPC, 1997 & 0.01 & tracking & 0.012 & \\
${\rm ^{160}Gd}$ & Solotvina, 2001 & 1.0 & scintillator & 0.013 & \\ \hline \hline
\end{tabular}
\caption{\small 	A selection of the past and present experiments giving 
			the best result per isotope to date \cite{PDG}. 
			All given \Tonu\ are lower limits, with the exception 
			of the Heidelberg-Moscow experiment where the 99.997\% 
			CL value is given. \label{tab:pastExp}}
\end{table}

Many experiments have searched for the \bbonu\ mode over the last fifty years. 
Lifetimes in the range of $10^{23}-10^{25}$~years have been explored, corresponding 
to \mbox{$\mbb\sim$ 250--1000} meV. See Table \ref{tab:pastExp} for further details.

In particular, the Heidelberg-Moscow experiment \cite{Heidelberg-Moscow} searched 
the \bbonu\ decay of \GE\ using five high-purity Ge semiconductor detectors enriched to 87\% in
\GE. The experiment achieved an exposure of 71.7 kg$\cdot$y and set the most stringent limit on 
the lifetime of the \bbonu\ process. 

A group from this experiment, led by H.~V.~Klapdor-Kleingrothaus, also claims 
\cite{Klapdor03, Klapdor06} a controversial $4\sigma$ evidence for \GE\ \bbonu\ decay 
with a lifetime of about $1.2\times10^{25}$ y, corresponding to a \mbb\ of 240--580 meV 
(best value: 440 meV). This claim (HM claim, in the following) sparked a debate (see, for example, References 
\cite{antiKlapdorA, antiKlapdorB}) in the community because the signal peak is faint 
and the spectrum contains other unexplained peaks. 

In addition, a similar experiment, which took data at Canfranc, called IGEX \cite{IGEX} did 
not see any evidence for a signal, although its exposure (8.9 kg$\cdot$y) was not enough 
to exclude this claim.

Clearly, this controversial evidence must be unambiguously confirmed or refuted. This leads to the goal of the 
next-generation \bbonu\ experiments, namely the exploration of the {\em degenerate} 
hierarchy (\mbb\ $\sim$ 50--200 meV) or, in terms of the lifetime, which is not affected by 
uncertainties on the NME, one would like to explore the region of $10^{26}$~y.
To reach that sensitivity and do reliable measurements, around 100 kg of the decaying 
isotope should be enough. 

In order to fully explore the {\em inverse} hierarchy one has to 
reach a sensitivity of $\mbb \sim 20$~meV, or lifetimes of the order of  $10^{27}$~y. 
Observation of \bbonu\ decay at this mass scale would imply the inverted neutrino mass 
hierarchy, or a normal ordering  within a quasi-degenerate 
spectrum. To establish the correct mass pattern, additional input from the overall 
neutrino physics program could be necessary. The study of this mass range requires
experiments at the ton-scale. 

The 1--5 meV \mbb-range arises from the solar neutrino oscillation results and corresponds 
to the \emph{normal} hierarchy. This goal would require hundreds of tons of the decaying 
isotope.

\begin{table}[b]
\centering
\begin{tabular}{ccccc}
\hline \hline
Experiment	&	Isotope	&	Technique			&	Reference		\\ \hline
CUORE 		&	\TE\		&	TeO$_2$ bolometers      &	\cite{CUORE}	\\
EXO		&  	\XE\		&	LXe TPC       	        & 	\cite{EXO}		\\
GERDA           &  	\GE\		& 	Enr. Ge semicond. det.	&	\cite{GERDA}		\\
{\sc Majorana}	&  	\GE\		& 	Enr. Ge semicond. det.	&	\cite{MAJORANA}		\\
SNO+		&  	\ND\		& 	Nd loaded liq. scint.	&	\cite{SNO+}		\\
SuperNEMO	&  	\SE\		& 	Se foils in tracko-calo	&	\cite{SUPERNEMO}	\\
\hline \hline
\end{tabular}
\caption{\small Some of the new-generation \bbonu\ projects.}
\label{tab:proposals}
\end{table}

Many experiments have been proposed in the last 10 years. Table \ref{tab:proposals} collects 
some of the most active. See \cite{Avignone:2008} for a recent detailed review.

The GERDA \cite{GERDA} and {\sc Majorana} \cite{MAJORANA} collaborations propose new Ge detectors to take
advantage of the Heidelberg-Moscow and IGEX experiences. This is a well-established 
technique that offers an outstanding energy resolution (0.15\% FWHM at \Qbb) and
high efficiency, but limited methods to reject backgrounds (basically, segmentation of 
the detector and pulse shape analysis).

CUORE \cite{CUORE} and its demonstrator, CUORICINO, are arrays of TeO$_{2}$ bolometers. Because
$^{130}$Te has a large natural isotopic abundance ($\sim$34\%), the need for enrichment
is less important. The pros and cons of the technique are similar to those of Ge experiments:
energy resolution is better than 0.5\% at \Qbb, and the efficiency for the signal is $\sim85\%$; background identification is not possible, though the segmentation of the detector allows the 
rejection of multiple-hit events.
 
SNO+ \cite{SNO+} is a proposal to fill the Sudbury Neutrino Observatory (SNO) vessel with liquid scintillator, 
now that its solar neutrino program has finished. A mass of several hundred kg of double beta
decaying material could be added to the experiment by dissolving a neodymium salt in the
scintillator (\ND\ natural abundance is 5.6\%). This would make SNO+ the largest \bbonu\
experiment of the next generation. However, the energy resolution will be modest 
($\sim8\%$ FWHM at 1 MeV) for a pure-calorimetric experiment.

Super-NEMO \cite{SUPERNEMO} is the proposed expansion of the NEMO-3 detector, still in operation at the
Laboratoire Souterrain de Modane.  Its principle of operation is based on the use of thin foils
of the \bb\ emitter surrounded by a tracking chamber and a calorimeter. Track reconstruction
provides a topological signature useful to discriminate signal and background. However, this
design suffers from a poor energy resolution, due to their calorimeter (as reference NEMO-3 plastic scintillator
calorimeter achieved a resolution of about 14\% FWHM at 1 MeV) and from the fact that is not a fully active
detector. The target is a foil located in the middle of the tracking chamber and thus background from Radon isotopes will deposit on it resulting in a serious background that cannot be vetoed (since the signal also emanates from the foil). 

The Enriched Xenon Observatory (EXO, \cite{EXO}) will search for \bbonu\ decay in \XE\ using a liquid xenon
TPC with 200 kg mass. The ultimate goal of the Collaboration is to develop the so-called \emph{barium tagging},
that would allow the detection of the ion product of the \XE\ \bb\ decay, and thus eliminate all
background but the intrinsic \bbtnu.

\section{Xenon TPCs for \bbonu\ searches} \label{XENON-TPCs}
%
A xenon Time Projection Chamber (TPC) is a fully active detector, in which this noble gas acts simultaneously as target and detector. 

Why xenon? Because it is the only one, among the noble gases, that has a \bb\ decaying isotope:
\XE, whose natural abundance is rather high (9 \%). Furthermore it does not have other long-lived
radioactive isotopes. In addition, it can be enriched by centrifugation methods to high concentrations 
of \XE : for example, an enrichment of 80\% is being used by the EXO-200 experiment.  Its \Qbb\ value, 2480 keV, is acceptably high. The \bbtnu\ mode life-time, not yet measured, 
may be as long as $10^{22}-10^{23}$~y. Finally, the \bbonu\ mode life-time is predicted to be almost 
as short as the one of the other commonly-used \bbonu\ isotope, \GE\ \cite{Avignone:2005cs}.

Why Xe TPCs? One of the most attractive features of a TPC is the fact that it can be scaled up to 
large masses. Some of the most successful recent experiments searching for dark matter are liquid 
xenon TPCs (LXe). In particular, the XENON experiment \cite{Aprile:2008rc} has evolved from a small, 
3-kilogram detector, through a very successful 10-kg intermediate apparatus \cite{Xenon10A},  
to a 100 kg device featuring a sensitivity to the spin-independent WIMP-nucleon cross section of $\sim 10^{-44}$ cm$^2$ for WIMP masses above 20 GeV \cite{Aprile:2009yh}. The XENON experiment is already planning its next phase, a ton-scale device, to further improve its sensitivity by three orders of magnitude \cite{Aprile:2009yh}. Other 
experiments planning or building LXe TPCs for Dark Matter searches are LUX and ZEPLIN 
(for a review, see \cite{Baudis:2007dq}). In addition, the WARP\footnote{See \url{http://warp.lngs.infn.it/}} 
and ArDM experiments \cite{Kaufmann:2007zz} use liquid argon instead of xenon. 

The general theme in all the above detectors is the so-called dual-phase TPC. Charged particles interacting in the liquid will produce both primary scintillation ultra-violet (UV) light and ionization electrons. The primary UV light is a narrow continuum, peaked around 172 nm and with a 14 nm FWHM width for xenon \cite{Suzuki:1979km}. For argon, the spectrum is centered around 120 nm. The detection of the fast, primary light using photosensors (typically PMTs sensitive to UV light or coated with a wavelength shifter) provides a precise start-of-event $t_0$ and a crude measurement proportional to the energy deposited in the chamber. In the dual-phase TPC, electrons drift through the liquid to the gas-liquid interface, where a higher electric field facilitates extraction into the thin gas layer above.  

Most of the detectors mentioned above use {\em electroluminescence} (EL) to amplify the ionization signal. When an electron is accelerated in a moderate electric field (of the order of 3-5 kV/cm/bar), it produces {\em secondary} scintillation UV light. The field can be tuned to generate a large number of photons (of the order of 1000) per electron reaching the anode, thus producing a proportional signal. As discussed in Chapter 2, extremely low fluctuations can be reached with EL, which is crucial for optimal energy resolution. The use of the EL technique allows to measure both the scintillation ($S$) and ionization ($I$) signals with the same photosensors. Normally PMTs are used, although other photosensors are possible. In WIMP searches, the ratio $I/S$ is a powerful discriminator between nuclear recoil and electromagnetic interactions for energies greater than about 10 keV.

Also, a liquid TPC, using 200 kg of enriched xenon, has been built to search for \bbonu\ events, by the EXO collaboration \cite{EXO,Akimov:2005mq}. The ionization of liquefied noble gases is accompanied by fluctuations much larger than predicted by Poisson statistics (see Section \ref{ENERGY-RES}). However, the anti-correlation existing between $S$ and $I$ signals is exploited to improve the energy resolution from 4.2\% FWHM (ionization-only) to 3.3\% FWHM (ionization plus scintillation) at \qbb\ \cite{Conti:2003av}. In the EXO200 detector, this is accomplished by simultaneously reading the $S$ signal by means of APDs, and the $I$ signal using two 60$^{\circ}$ crossed wire planes. 

In contrast with the proliferation of the LXe apparatus, only the Gotthard experiment 
\cite{Gotthard93, Gotthard98} built a small (pressurized) gas TPC (GHPXE) for \bbonu\ searches in the early 1990's. The Gotthard apparatus had a fiducial mass of 5 kg, operated a 5 bar and used conventional gain amplification in a wire plane. Its energy resolution was mediocre and the addition of a quencher (5\% of CH$_4$) to stabilize the gas and increase the drift velocity quenched both the ionization and the scintillation light. Because of the quenched scintillation light, the detector did not measure $t_0$. This configuration resulted in large backgrounds due to tracks emanating from the anode and cathode.

Clearly LXe TPCs have some advantages over HPGXe, the most important one being its compactness. The density of liquid xenon is about 3 g/cm$^3$, or 3 tons for a volume of 1m$^3$. In contrast, the density of gas xenon at 10 bar is 0.05 g/cm$^3$, corresponding to only 50 kg of xenon per cubic meter.
A more compact detector has a smaller volume to surface ratio and offers, therefore, a smaller cross section to external backgrounds, such as gammas emanating from the laboratory walls or the detector vessel. It also requires less instrumentation, which in turn minimizes the internal radioactivity (per unit of active xenon mass). Furthermore, the primary scintillation yield is higher in liquid than in gas, and the apparatus itself provides a good measurement of self-shielding (external gammas will tend to interact near the detector wall) at the cost of a non-negligible efficiency. 

Why consider a HPGXe TPC for \bbonu\ searches, then? The answer has to do with the different properties of the liquid and the gas phases. The high density of the liquid---usually a blessing---is a curse when it comes to the observation of the signal topology. The characteristic signature from a \bb\ event is two electrons whose energies add up to \qbb\ (2480 keV). Electrons in this energy range are easily tracked in gas, but deposit all their energy in a blob in the much denser liquid. Therefore, while the gas TPC can resolve blobs within a single track topology, the liquid TPC cannot. Because of this, it is much more difficult for a LXe detector to distinguish between a \bb\ event and a gamma interaction that deposits by photoelectric or Compton effect an energy in the vicinity of \Qbb\footnote{In practice, one can have some level of rejection in the case of Compton interactions if a second, separated energy deposit due to the absorption of the scattered gamma is observed.}. 

\begin{figure}[p]
\begin{center}
\includegraphics[width=0.65\textwidth]{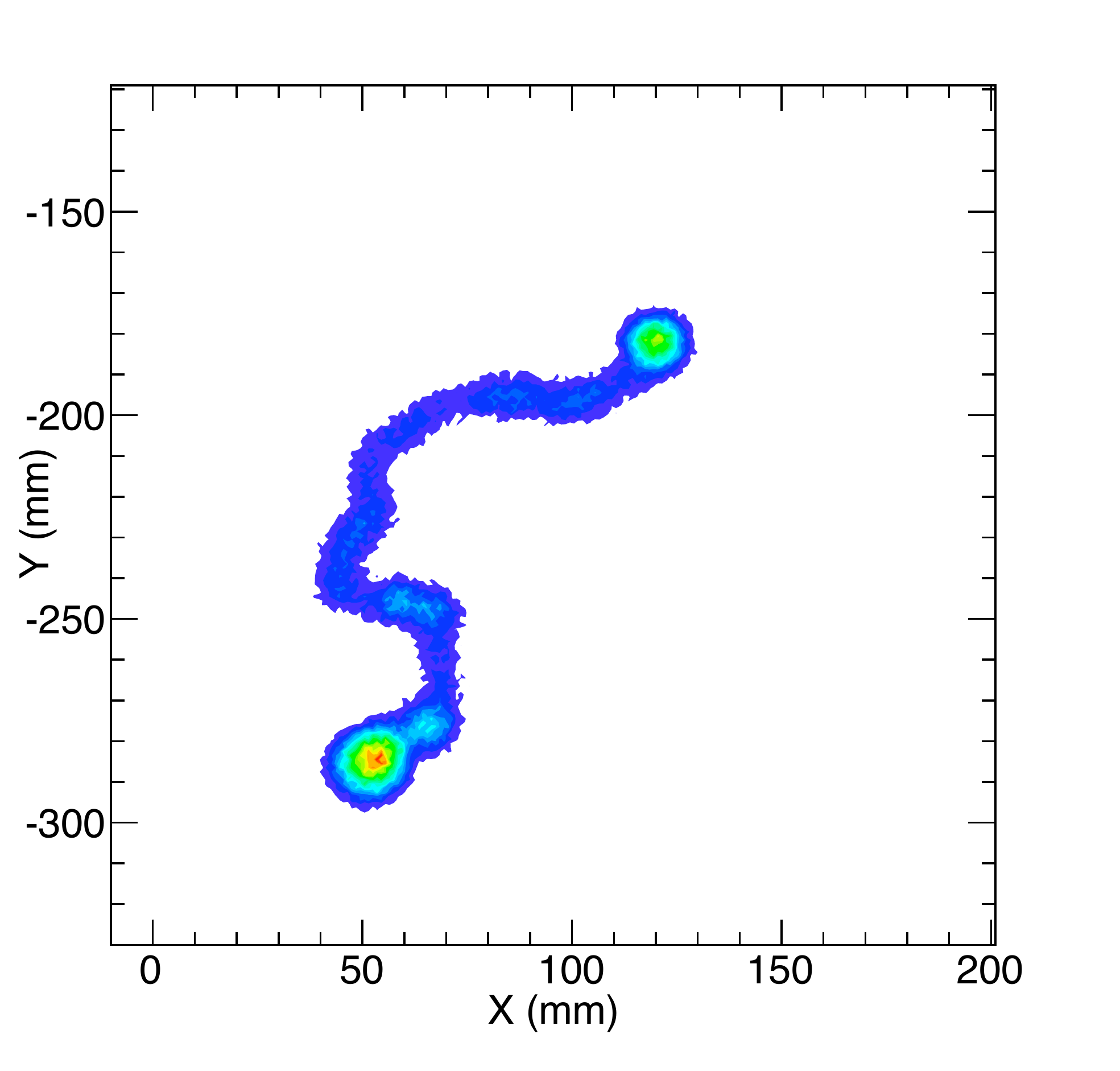} 
\end{center}
\vspace{-1.0cm}
\caption{\small The topological signature in NEXT is a ``spaghetti with two meat balls'', that is,
a track that ends in two ``blobs'' of energy, corresponding to ranging-out electrons. The trajectory of electrons contains no information, being dominated by multiple scattering in the dense gaseous xenon.} 
\label{fig:track} 
\end{figure}

Consider now a \bb\ event produced in a HPGXe. The average energy of the two electrons is about 1250 keV, and at 10 bar each electron travels about 15 cm. The trajectory of the electron in the gas is completely dominated by multiple scattering. The resulting ``topological signature'' is a twisted track that includes both electrons. In terms of ionization, the track behaves like a minimum ionizing particle (MIP), depositing about 70 keV per cm, except at both ends, where each electron deposit 200 keV or more of energy as it ranges out. The picture is that of a ``spaghetti with two meat balls'' (Figure \ref{fig:track}). 


Perhaps even more important than the topological signature, and often not fully recognized as such, is the fact that a HPGXe detector can feature a much better energy resolution than a LXe one, as will be discussed at length in Chapter 2. This turns out to be essential, since backgrounds increase exponentially with the resolution, due to the nearby presence of the 2.6 MeV gamma from \TL\ and the 2.4 MeV gamma line from \BI\ as well as other backgrounds (for a complete discussion, see Chapter 4). 

The fact that a gas xenon TPC offers simultaneously better energy resolution and a distinct kinematical signature indicates that a HPGXe can offer a much better capability for rejecting \bbonu\ backgrounds than LXe.

\chapter{A Gas Xenon TPC for \bbonu\ searches}
\section{Energy resolution and \bbonu\ decay} \label{ENERGY-RES}
\begin{figure}[tbh]
\centering
\includegraphics[width=0.65\textwidth]{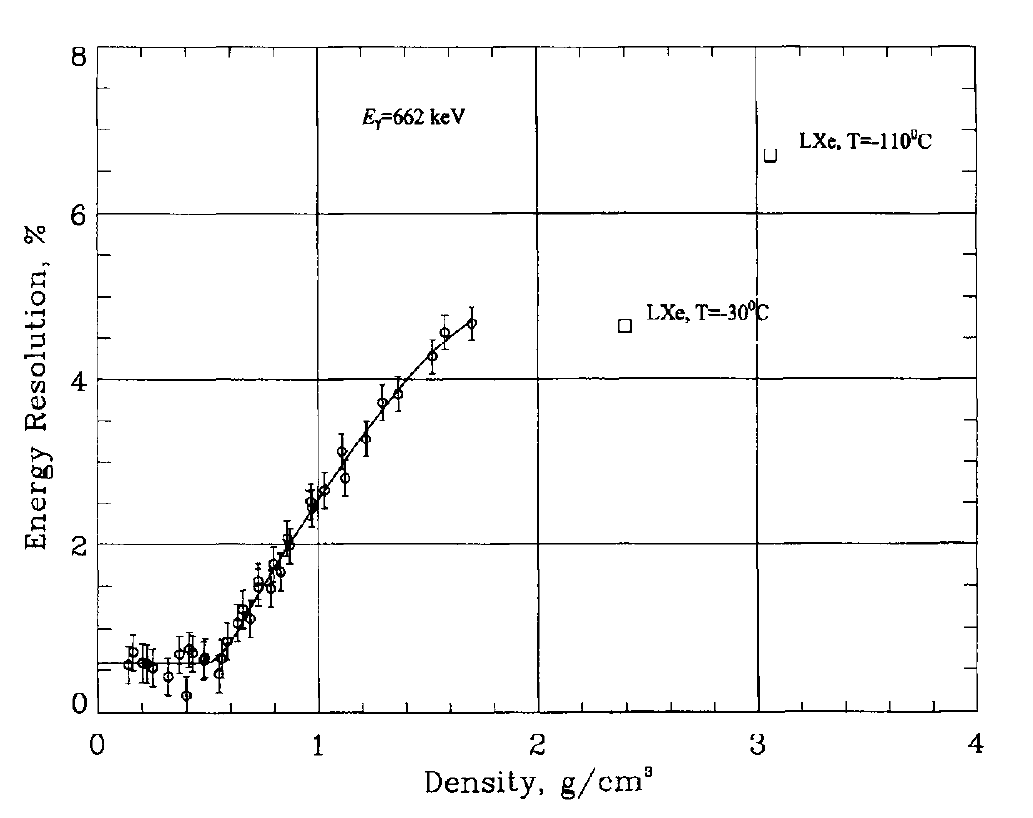} 
\caption{\small The energy resolution, FWHM, is shown for $^{137}$Cs 662 keV 
			gamma rays, as a function of xenon density, for the ionization 
			signal only. Reproduced from \cite{Bolotnikov:97}.}
\label{fig:bolotnikov} 
\end{figure}

As discussed above, excellent energy resolution is a crucial ingredient for a competitive \bbonu\ detector. As will be motivated below, this can be achieved with a HPGXe detector. In addition to the intrinsic properties of xenon, its operational density also matters. Fluctuations due to intrinsic physical processes limit the energy resolution of LXe detectors, disfavoring them compared with HPGXe. This is clearly seen in Figure \ref{fig:bolotnikov}, reproduced from \cite{Bolotnikov:97}. For these data, only the
ionization signal is detected. The resolutions displayed were extracted
from the photo-conversion peak of the 662 keV gamma ray from the $^{137}$Cs
isotope\footnote{As the
photo-peak is asymmetric, the resolution was extracted from the peak upper half only,
and electronic noise was subtracted in quadrature.}.
A striking feature in Figure~\ref{fig:bolotnikov} is an apparent transition at density 
$\rho_t \sim 0.55$ g/cm$^3$. Below this density, the energy resolution is approximately constant:
\begin{equation}
\delta E/E = 6 \times 10^{-3} {\rm ~FWHM}.
\label{eq:intrinsic}
\end{equation}
%

For densities greater than $\rho_t$, energy resolution deteriorates rapidly, approaching a
plateau at LXe density. The most plausible explanation underlying this strange behavior is the appearance, as density increases, of two-phase xenon. A fog, or possibly an
evanescent liquid lacework, coexists with gas. In the liquid phase, a quasi-conduction
band forms in xenon with a higher electron mobility. Regions of high ionization
density in LXe experience nearly complete recombination. The energy in
recombination is returned largely as scintillation light, but some unknown fraction of
energy is converted to heat through excimer-excimer collisions, after which only one
excimer survives (see \cite{Nygren:2007zz} and references therein). 

By supplying an electric field, the recombination fraction can be made
somewhat smaller, but large fluctuations persist. For fast electrons in LXe, the
average fractions of energy spent in ionization and scintillation are roughly equal if an
electric field is present. However, either one of these signals, if taken alone, will show
anomalously large partitioning fluctuations. If both are measured, a strong anti-correlation
will exist for a fixed energy deposition that can be used to improve the energy measurement. This is possible only to a certain extent since, in practice, the primary scintillation signal cannot be measured
with the precision and accuracy needed to recover the intrinsic energy resolution \cite{Aprile-book}. 

In contrast, given the xenon critical density, the intrinsic resolution in the gas phase is very good up to pressures in the vicinity of 50 bar, at room temperature, although practical and technical issues dictate much smaller pressures, in the range of 5 to 20 bar.  

Extrapolating the observed resolution in Figure \ref{fig:bolotnikov} as $E^{-1/2}$ from E = 662 keV to
the \XE\ \qbb\ value (2480 keV), a naive energy resolution is predicted:
\begin{equation}
\delta E/E = 3 \times 10^{-3} {\rm ~FWHM}.
\label{eq:res}
\end{equation}

Based on ionization signals only, the energy resolution in Eq.~\ref{eq:res} reflects an order of
magnitude improvement relative to LXe. For densities less than $\rho_t$, the measured
energy resolution in Figure~\ref{fig:bolotnikov} matches the prediction based on Fano's theory \cite{FANO}. The
Fano factor $F$ reflects a constraint, for a fixed energy deposited, on the fluctuations of
energy partition between excitation and the ionization yield $N_I$. For electrons
depositing a fixed energy $E$, the RMS fluctuations $\sigma_I$~ in the total number of 
free electrons $N_I$ can be expressed as:
\begin{equation}
\sigma_I = (FN_I)^{1/2}.
\label{eq:fano}
\end{equation}

For pure gaseous xenon (GXe), various measurements \cite{Nygren:2007zz} show that:
\begin{equation}
	F_{GXe} = 0.15 \pm 0.02
\label{eq:gas-fano}
\end{equation}

In LXe, however, the anomalously large fluctuations in the partitioning of energy to
ionization produce an anomalous Fano factor:
\begin{equation}
	F_{LXe} \sim 20
\label{eq:liquid-fano}
\end{equation}
Larger than the one corresponding to GXe by about two orders of magnitude.

Of course, realistic detectors do not measure the intrinsic resolution according to the Fano factor and the primary ionization yield, but a convolution of several effects, all of which tend to degrade the intrinsic limit. We can enumerate the different factors affecting the resolution as follows:
\begin{enumerate}
\item Primary ionization yield $N_I$, given by $N_I= E/W_I$, where $W_I$~is the average energy needed to produce an electron-ion pair. The specific conditions of field, density, impurities, etc., can change $N_I$~significantly.
\item Intrinsic fluctuations, represented by the Fano factor.
\item Losses of drifting electrons due to electronegative impurities, volume recombination, grid transparency, etc., represented by a factor $L = 1 - \epsilon$, where $\epsilon$~is the overall electron collection efficiency. 
\item Gain processes such as avalanche multiplication, which multiply the signal by $m$~and introduce fluctuations in the detected signal, represented by a variance $G$.
\item Electronic noise, in electrons RMS at signal processing input, represented by $n$.
\end{enumerate}

In addition there are other important sinks of resolution, such as fluctuations associated to Bremsstrahlung losses, channel equalization, non-linearities, etc. However, an analysis of the previous list is sufficient to understand the main issues to be addressed to approach the intrinsic resolution. Assuming that all the above-mentioned sources are gaussian and uncorrelated, we can combine them in quadrature: 
\begin{equation}
\sigma_n^2 = (F + G + L)N_I + \frac{n^2}{m} .
\label{eq:total-sigma}
\end{equation}

where $\sigma_n$~is the total number of electrons RMS, due to fluctuations in all sources. Then:

\begin{eqnarray}
\delta E/E &= &\frac{2.35 \ \sigma_n}{N_I\epsilon} \\
&=& 2.35 \ \left[\frac{F + G + L + \frac{n^2}{mN_I}}{N_I\epsilon^2}\right]^{1/2},
\label{eq:deltae}
\end{eqnarray}

Let us examine each one of these factors with some detail.

\subsubsection{Primary ionization yield}
Taking $\qbb =2480 {\rm ~keV}$~and the energy  needed to form an electron-ion pair 
\mbox{$W_I=22~{\rm eV}$}, we obtain:
\begin{equation}
N_I = \frac{\qbb}{W_I}= 111\ 200
\end{equation}

This naive number needs to be corrected by the recombination factor of electrons and ions, a quantity that depends on the density (thus the pressure) and the electric field. However, the correction is small. For example, at 20 bar ($\rho = 0.1{~ \rm  g/cm^3}$) and for an electric field
of 1 kV/cm, the corrected $W_I$ factor is 24.8 eV \cite{Nygren:2008}, yielding $N_I \simeq 10^5$.

\subsubsection{Fano factor}
The Fano factor, together with the ionization yield, introduces a lower limit for the energy resolution, and sets the scale for the
other factors in Equation \ref{eq:deltae}. We assume $F= 0.15$~\cite{Nygren:2007zz}.  

\subsubsection{Losses during drift}
Ionization electrons can be lost while drifting due to several reasons, including recombination with positive ions, attachment with electronegative impurities and non-perfect transparency of metallic grids. All these factors can be minimized by using state-of-the-art gas purification methods, strict control of materials to avoid degassing and careful choice of grids. In principle, therefore, the factor $L$~can be made smaller than the Fano factor $F$, although ultimately attachment may limit the maximum drift distance.

\subsubsection{Gain fluctuation factor}
The gain fluctuation factor $G$~is much larger than $F$~in gas proportional counters involving avalanche multiplication of the ionization signal near a wire, with values between 0.6 and 0.8 for typical mixtures and conditions \cite{Kowina:2004} and higher values predicted for higher pressures. However, the use of modern micropattern devices such as Micromegas are likely to offer a smaller gain factor. 

\begin{figure}[tb]
\centering
\includegraphics[width=0.5\textwidth]{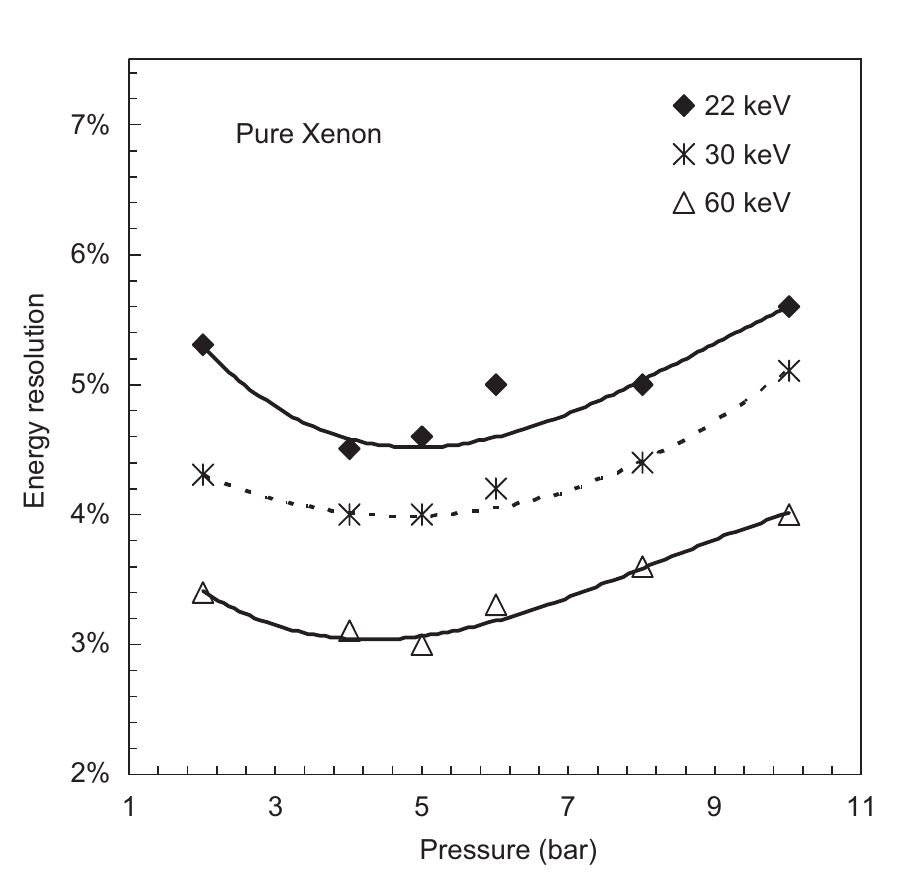} 
\caption{\small Measured resolution $\Delta E/E$, FWHM, from different 
			radioactive sources in a small HPGXe using EL as a function 
			of the pressure (from \cite{Coelho:07}).}
\label{fig:coelho} 
\end{figure}

On the other hand, both theoretical arguments \cite{Nygren:2008} and a variety of experimental measurements suggest that $G$ can be made at least as small as $F$ using {\em electroluminescence} (EL). We will discuss EL in great detail in Section \ref{EL}. However, to gain an insight of the achievable resolution, it is worth to examine the  results obtained by \cite{Coelho:07}, a small HPGXe chamber
equipped with a large area avalanche photodiode (LAAPD) as photosensor. The detector performed measurements for filling pressures from 1 up to 10 bar, for 22, 30 and 60 keV photons. The results, shown in Figure \ref{fig:coelho}, are consistent with resolutions at \qbb\ between 0.42\% and 0.6\%, indicating also a slight dependence with pressure. This dependence with the pressure is not well understood (and not found by \cite{Bolotnikov:97}). One of the early goals of our experimental program
--- see Chapter 5 --- is to reproduce these results with an independent experimental setup.

Expressing the energy resolution as 
\begin{equation}
\delta E/E = 2.35 \cdot [(F+G) W_I/E]^{1/2},
\end{equation}
and substituting, $E = \qbb = 2.48\times 10^6 {~ \rm eV}, W_I = 24.8 {~ \rm eV}, F= G = 0.15$, we obtain:  
\begin{equation}
\delta Q/Q =4.1 \times 10^{-3} {\rm ~FWHM},
\end{equation}
which is close to the best value found in ~\cite{Coelho:07}, and suggests that $G$ can be made as small as the Fano factor $F$, using EL. 

\subsubsection{Electronics noise}
In EL based systems, electronic noise can be made small by designing the system to produce enough light and using low-noise photosensors, such as PMTs. Electronic noise may also be a significant issue in systems based on avalanche gain, due to capacitive microphone formed by the grid-anode structure. 

\subsubsection{Combined expected resolution}

\begin{figure}[tbh]
\centering
\includegraphics[width=0.80\textwidth]{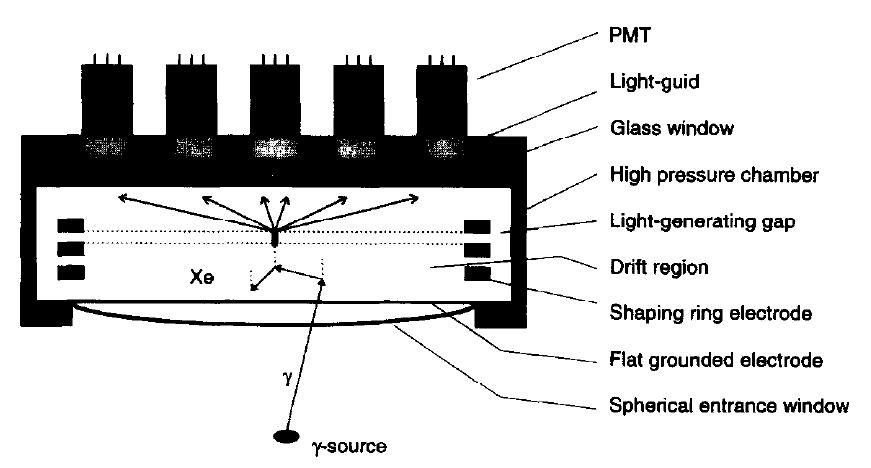} 
\caption{\small A schematic view of the large SDC used by \cite{Bolozdynya:96}.}
\label{fig:sdc} 
\end{figure}

To gain an insight of the combined resolution that can be achieved by a large system, we consider the Scintillation Drift Chamber (SDC), described in ~\cite{Bolozdynya:96} and schematically shown in Figure~\ref{fig:sdc}.

The detector was a
stainless steel pressure vessel which contained a low field
drift region, 4 cm deep, followed by a EL
generating gap, 6 mm deep, defined by two transparent
stainless steel wire electrodes (1 mm spacing, 50 micron
diameter wires). One stainless steel shaping ring was used in
the drift region to provide a more uniform field. The EL light was viewed by a
hexagonal array of nineteen 80 mm diameter glass photomultipliers
coupled to the SDC through glass windows. A
thin layer of wavelength shifter (0.5 mg/cm$^2$), p-terphenyl,
was vacuum-deposited on the internal surfaces of the
glass windows to convert the UV photons emitted both by primary scintillation and by
secondary EL to visible
photons. The quantum efficiency of wave-shifting 172 nm
with p-terphenyl was measured to be 90\%. The distance between light gap and windows was
4 cm. The active area
of the detector (field of view) was 30 cm in diameter.

\begin{figure}[tbh]
\begin{center}
\includegraphics[width=0.99\textwidth]{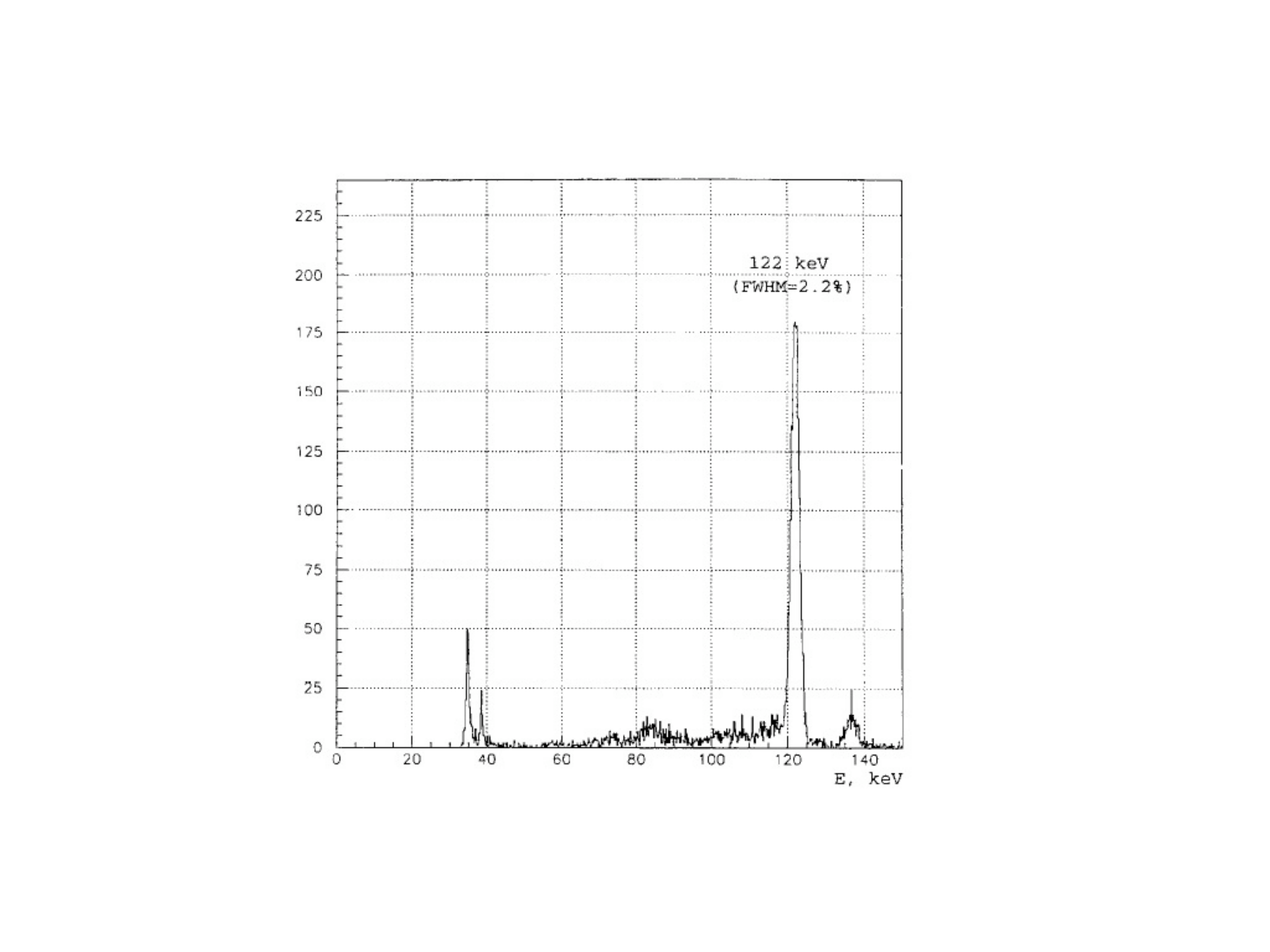} 
\end{center}
\vspace{-2.5cm}
\caption{\small Energy spectrum from $^{57}$Co radioactive sources in a large SDC using EL at 9 bar (from \cite{Bolozdynya:96}). The measured resolution $\Delta E/E$, FWHM, at 122 keV is also given.}
\label{fig:boloz} 
\end{figure}

Figure \ref{fig:boloz}, shows the best resolution measured for a $^{57}$Co radioactive source, corresponding to 2.2\% at 122 keV.  Extrapolation to \qbb\ energies yields a relative resolution of 0.49\% FWHM. This measurement includes the convolution of all experimental effects discussed in this section, included intercalibration of the 19 PMTs, and it corresponds to rough values of 
 $G=0.2, F=0.15, N_I = 10^5, L=0.05, n^2/m =0$~in Equation \ref{eq:deltae}.

\section{Electroluminescence} \label{EL}
\subsection{Brief historical overview}

\begin{figure}[tbhp]
\begin{center}
\begin{tabular}{c}
\includegraphics[width=0.8\textwidth]{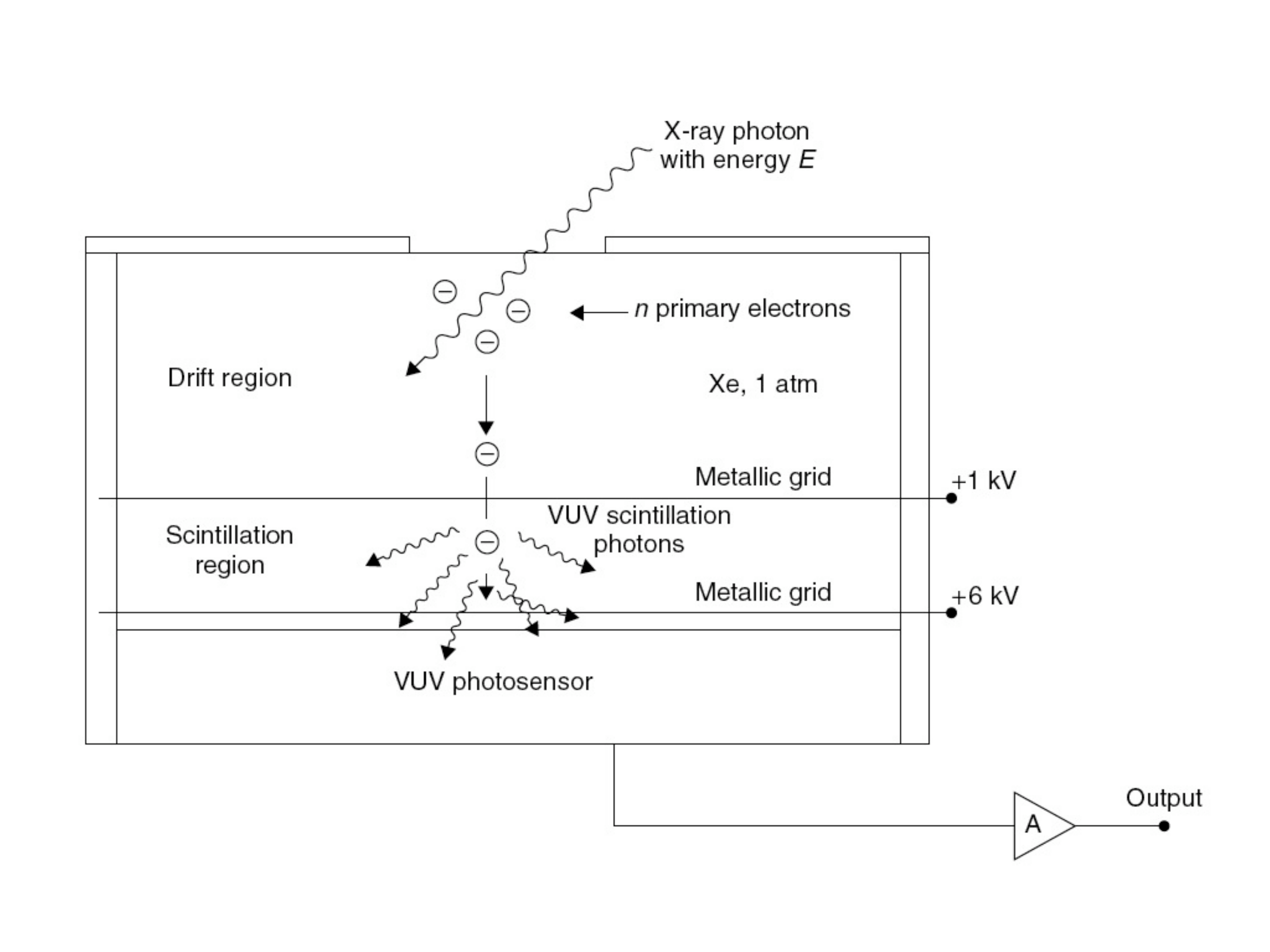} \\
\includegraphics[width=0.8\textwidth]{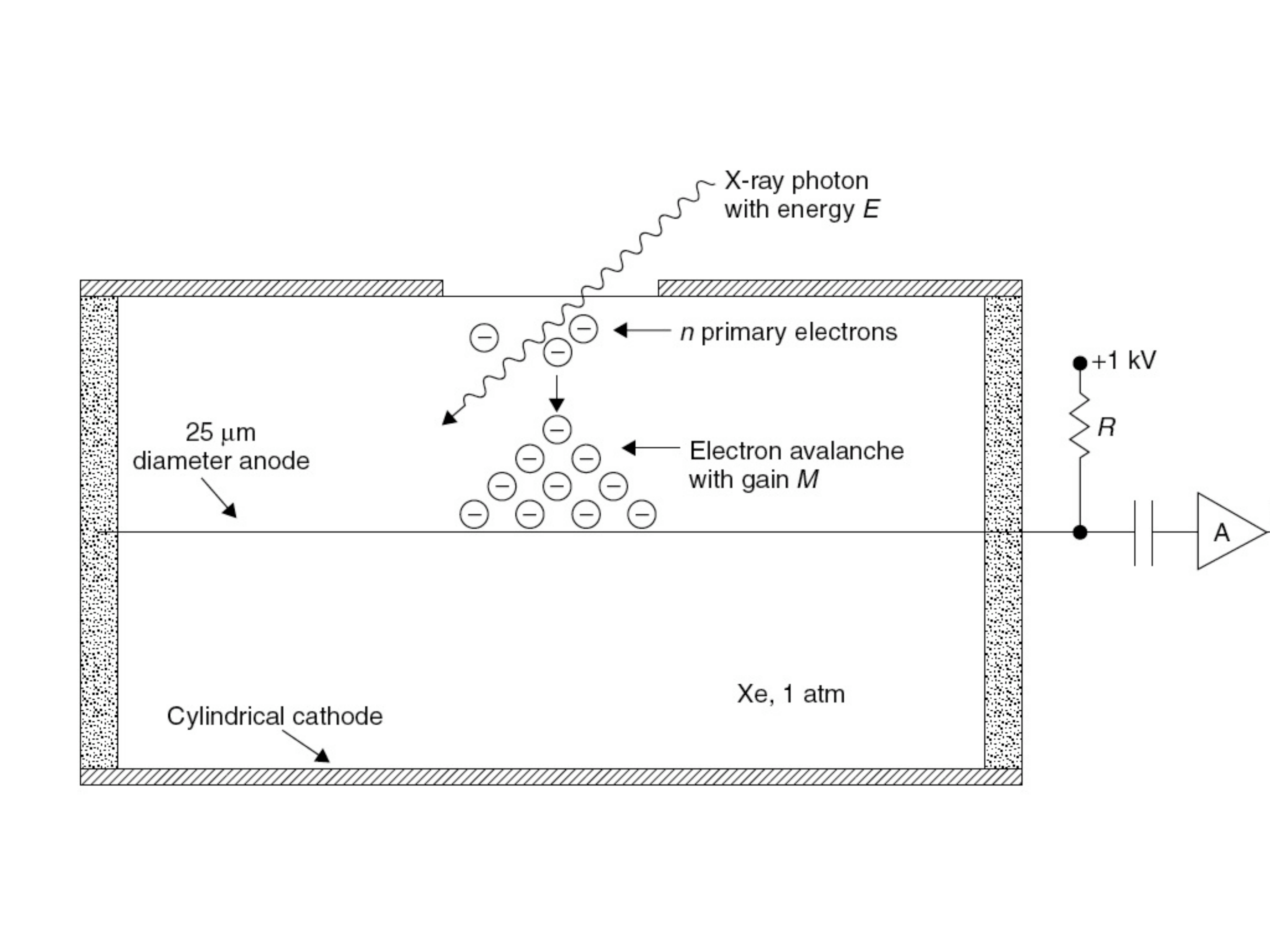} 
\end{tabular}
\end{center}
\vspace{-2.0cm}
\caption{\small Top: principle of a Gas Proportional Scintillation Counter. Bottom: principle of a Gas Proportional Counter with avalanche gain (from \cite{Conde:04}).}
\label{fig:GPSC} 
\end{figure}

With the purpose of
performing high-resolution spectrometry of low-energy X-rays, detector designs exploiting electroluminescence (EL) in noble gases were developed in the last third of the 20th century, especially in Coimbra (Portugal) \cite{Conde:67} and CERN \cite{Charpak:75}. 
 
Figure \ref{fig:GPSC} (top), illustrates the principle of a Gas Proportional Scintillation Chamber (GPSC). An X-ray enters through the chamber window and is absorbed in a region of weak electric field
($>0.8~{\rm kV ~cm^{-1}~ bar^{-1}}$) known as the drift region. The ionization electrons drift under such field to a region of moderately high electric field (around $3-6 ~{\rm kV ~cm^{-1}~ bar^{-1}}$ range), the so-called scintillation
or EL region. In the scintillation region, each electron is accelerated so that it excites, but
does not ionize, the gas atoms/molecules.
The excited atoms decay, emitting UV light (the so-called
secondary scintillation), which is detected by
a photosensor, usually a photomultiplier tube. The intensity of the secondary scintillation light is
two or three orders of magnitude stronger than
that of the primary scintillation. However, since
the secondary scintillation is produced while the
electrons drift, its latency is much longer than that for the primary scintillation, and its rise time is much slower
(a few $\mu$s compared to a few ns). For properly chosen electric field strengths and EL region spatial widths, the number
$n_{ph}$ of secondary scintillation photons produced by
a single primary electron is nearly constant and can
reach values as large as a few thousand photons per electron.

The average total number, $N_t$, of secondary
scintillation photons produced by an X-ray photon
is then $N_t = n_{ph}\cdot N_I$, (recall that $N_I$~is the number of primary 
ionization electrons) so the photosensor signal
amplitude is nearly proportional to E, hence the
name of gas proportional scintillation counter
(GPSC) for this device.

What made the devices extraordinarily attractive was their improved energy resolution compared with conventional Proportional Chambers (PC) --- Figure \ref{fig:GPSC} (bottom). In a PC
the primary electrons are made to drift
towards a strong electric field region, usually
in the vicinity of a small diameter (typically
25$\mu$m) anode wire. In this region,
electrons engage in ionizing collisions that lead to
an avalanche with an average multiplication gain
M of the order of 10$^3$ to 10$^4$. If M is not too
large, space charge effects can be neglected, and
the average number of electrons at the end of the
avalanche, $N_a = M\cdot N_I$, is also proportional to
the energy E of the absorbed X-ray photon (hence
the name proportional (ionization) counter given
to this device).
 However, for PC detectors, there are
fluctuations not only in $N_I$ but also in M; for GPSCs, since
the gain is achieved through a scintillation process
with almost no fluctuations, only fluctuations in
$N_I$ and in the photosensor need to be considered. Thus a better energy
resolution was achieved in the latter case; typical values for 5.9 keV X-rays were 8\%
for GPSC and 14\% for PC. 
 
 The Scintillation Drift Chamber (SDC) was invented in 1978~\cite{Charpak:81}.  An SDC
is a TPC with EL readout instead of charge gain by electron avalanche
multiplication in gas. As examined in the previous section, a large SDC with 19 PMTs ~\cite{Bolozdynya:96} demonstrated excellent energy resolution at high pressure (9 bar), and for high energy X-rays. However, for mainstream particle physics, EL has had application primarily in only one technique: two--phase LXe detectors aimed at direct detection of WIMPs \cite{Aprile-book}. In that
very successful application, the enabling asset of EL is not excellent energy
resolution (limited, as we have seen, by the anomalous Fano factor in liquid), but the capability to detect single electrons.

\subsection{Yield and resolution in EL}

\begin{figure}[tbhp]
\begin{center}
\includegraphics[width=0.99\textwidth]{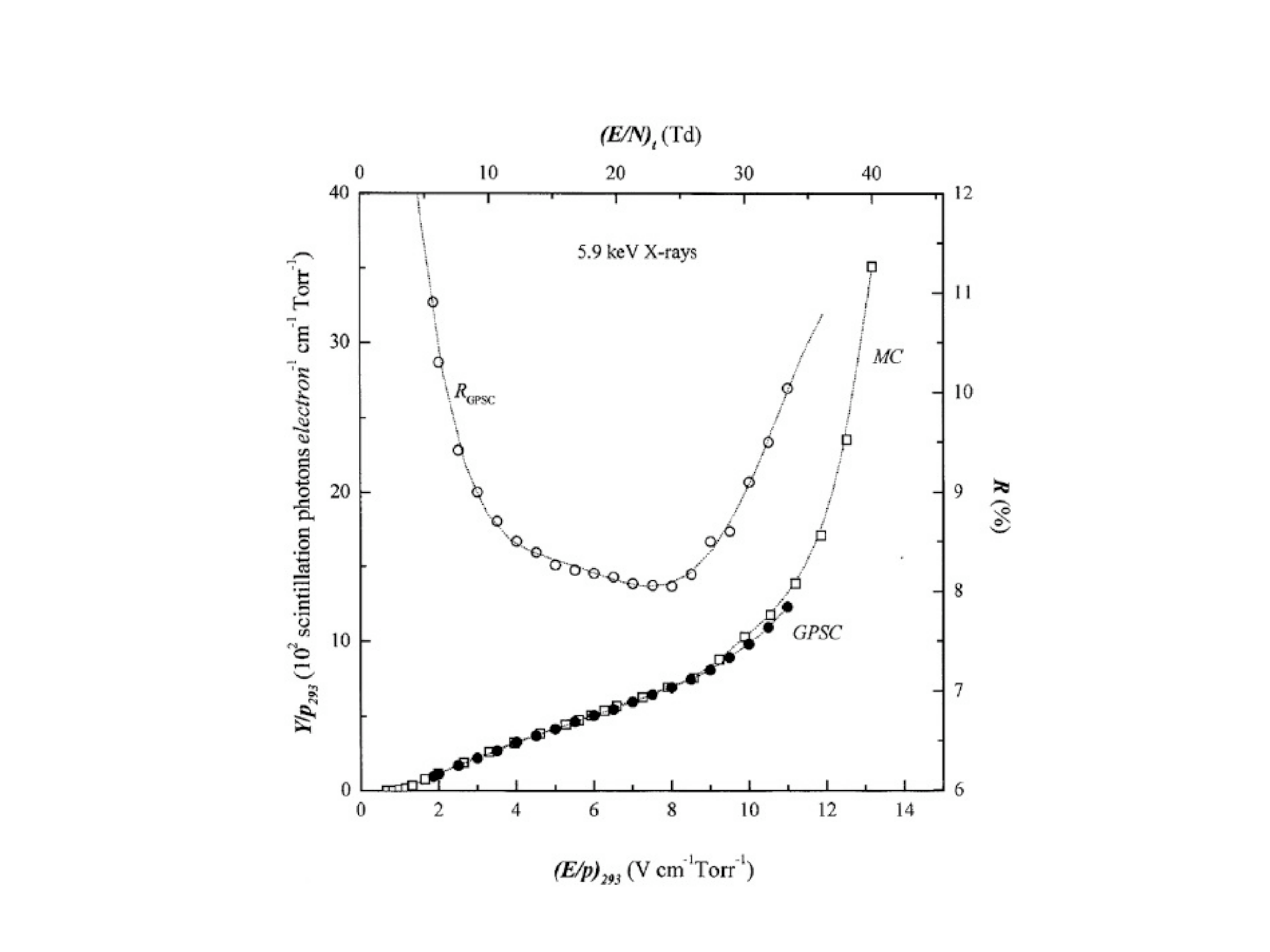} 
\end{center}
\vspace{-1.5cm}
\caption{\small Reduced scintillation yield (open squares--Monte Carlo results, filled
circles--experimental results) and detector energy resolution (open circles) as a
function of reduced electric field in the scintillation region for 5.9 keV x-rays. The
experimental values are normalized to the calculated Monte Carlo values at 
${\rm E/p = 6~V~cm^{-1}~torr^{-1}}$ (${\rm 4.5~kV~cm^{-1}~bar^{-1}}$) (from \cite{Borges:99})}
\label{fig:borges} 
\end{figure}

Electroluminescence occurs in xenon for a fairly wide range of E/p, where E is the electric field responsible for the process, and p is the gas pressure. Both gain and resolution
are shown in Figure \ref{fig:borges}, from~\cite{Borges:99}. 
Above E/p = 6-7 ${\rm kV~cm^{-1}~bar^{-1}}$ (8-9${\rm~V~cm^{-1}~torr^{-1}}$), the effect of charge multiplication is no longer negligible and energy resolution
deteriorates.
As it is evident from Figure \ref{fig:borges}, for $3< E/p < 6{\rm~kV~cm^{-1}~bar^{-1}}$ ($4< E/p < 8{\rm~V~cm^{-1}~torr^{-1}}$), the energy resolution is nearly optimal. In practice, gains of several
hundred have been realized, corresponding to an applied voltage of 5-20 kV
across two closely-separated meshes. Under constant, uniform E/p and total voltage V, the total
EL gain $\eta$~ is the number of photons generated per electron:

\begin{equation}
\eta =V/V_{PH}
\end{equation}
where V$_{PH}$ is the average potential needed to generate one photon at the given E/p
value. 

An empirical formula, describing the EL gain $\eta$~ in xenon gas is given in~\cite{Monteiro:07}:
\begin{equation}
\eta = 140 (E/p  - 0.83)p~ ({\rm UV ~photons/e ~cm^{-1}})\ \Delta x
\label{eq:yield}
\end{equation}
where E/p is given in ${\rm kV~ cm^{-1}~ bar^{-1}}$, p in bar, and the separation $\Delta x$ between the meshes in cm.

The contributions to the gain resolution $G$ must include fluctuations in:
\begin{enumerate}
\item the optical gain $\eta$;
\item $n_{pe}$, the number of {\em detected} photons, or photo-electrons (pe), per incident electron;
\item the gain process in the photo-detector per single photo-electron, whose fluctuation we express by $\sigma_{pd}$.
\end{enumerate}

The variance can be written as \cite{Nygren:2008}:
\begin{equation}
G = 1/\eta + (1+ \sigma_{pd}^2)/n_{pe}
\label{eq:G}
\end{equation}

The first term in (\ref{eq:G}) is much smaller than the second, since $\eta$~is large, while the limited photon detection efficiency results in a smaller number for $n_{pe}$. Assuming $\sigma_{pd}^2= 0.5$ (most PMTs will do better than that) and setting G=0.15 (so that it contributes no more than the Fano factor) one obtains:
\begin{equation}
n_{pe}^{EL} \ge 10
\label{eq:npe}
\end{equation}

At \qbb, $N_I = 10^{5}$, and consequently one needs to detect at least $10^{6}$
photo-electrons to satisfy (\ref{eq:npe}).

\section{Primary signals} \label{PRIM-SIGNALS}
To conclude this chapter we examine in some 
detail the primary signals in a HPGXe. The partition of visible energy in pure xenon is usefully described by $W_S$ and $W_I$, the
average energies spent in the creation of one primary scintillation photon and one
electron-ion pair, respectively, by energetic electrons or photons. The transferred energy is ultimately
converted to scintillation, free ionization, and invisible heat. Section \ref{subsec:PRIM-SIGNALS_primaryscint} deals with the primary scintillation signal, while Sections \ref{subsec:PRIM-SIGNALS_electrontransport} and \ref{subsec:PRIM-SIGNALS_secondaryscint} are relevant for the EL scintillation signals.

\subsection{Primary scintillation and the t$_0$ signal}
\label{subsec:PRIM-SIGNALS_primaryscint}

Detection of the primary scintillation is necessary to obtain the
$t_0$ needed to properly place an event along the TPC drift direction, and therefore in 3-D space. In high-pressure xenon (HPXe) at 10 bars, the primary scintillation decay time constants are 3 ns (independent of pressure) and $\sim$40 ns (slightly dependent on pressure).  Light is also expected to rattle around the detector volume for about $\sim$10 ns.  Overall, a time resolution requirement of 10 ns RMS for the measurement of the scintillation pulse timing is adequate.  The scintillation spectrum lays in the vacuum ultraviolet, peaking at 172 nm. The primary scintillation is weak relative to the secondary scintillation, but arrives within a few tens of ns as opposed to the delayed secondary scintillation. 

Taking $W_S = 88 \pm 10$~eV as the average energy to produce a primary scintillation photon \cite{Nygren:2008}, one obtains:  
\begin{equation}
n_S = (2.8 \pm 0.56) \times 10^4,
\end{equation}
or about 30,000 primary scintillation photons at \qbb. Assume, for simplicity, that every primary scintillation photon reaches a PMT and is transmitted with an efficiency of 10\% (see Chapter \ref{NEXT}, where these assumptions are justified). Thus the number of photo-electrons recorded would be:
\begin{equation}
n_{pe}^{scint} =3\times 10^{4} \times 10^{-1} = 3 \times 10^3.
\end{equation}

These photo-electrons are distributed over a large readout surface, typically 1 m$^2$ (more of this on Chapter \ref{NEXT}), and so, the number of p.e.\ per cm$^2$ is $\sim$0.6. If one would cover completely this surface with PMTs of 1'', ($2.5 \times 2.5$~cm$^2$), each one would record, on average about 4 p.e. In practice one would like to cover only about 10\% of the surface.

\begin{figure}[tbh]
\centering
\includegraphics[width=0.5\textwidth, angle=90]{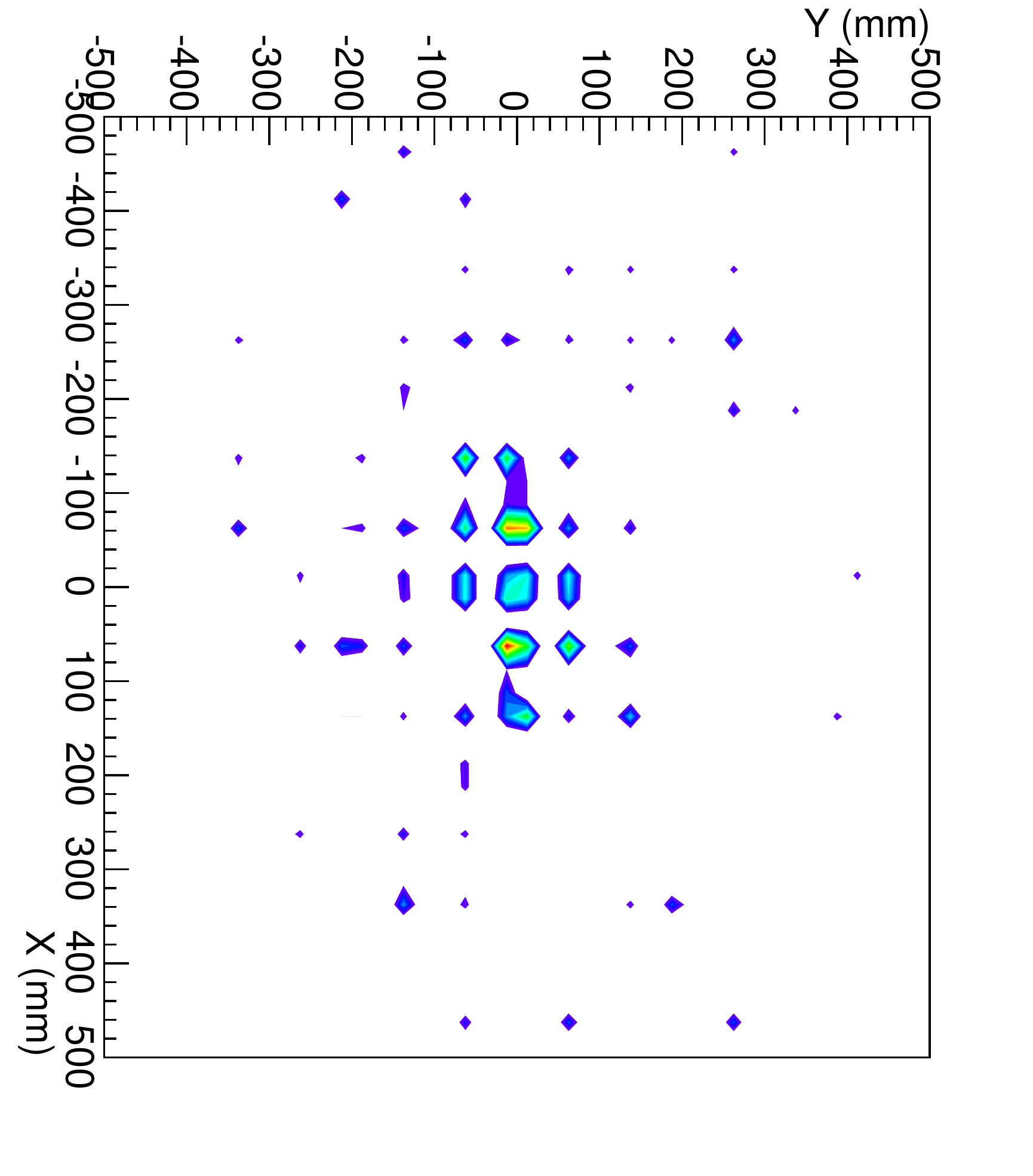} 
\caption{\small Monte Carlo simulation of scintillation light occupancy in the readout plane for a typical \bbonu\ event produced in a cylindrical TPC of 50 cm radius.}
\label{fig:scint} 
\end{figure}

Of course, the photons are not uniformly distributed. As shown in Figure \ref{fig:scint}, most of the scintillation light produced by a typical \bbonu\ event will be recorded in a cluster of PMTs that will have one, or at most a few p.e.,
while many PMTs will have no signal at all.

\subsection{Electron transport and electric field}
\label{subsec:PRIM-SIGNALS_electrontransport}

\begin{figure}[tbhp]
\begin{center}
\begin{tabular}{c}
\includegraphics[width=0.6\textwidth]{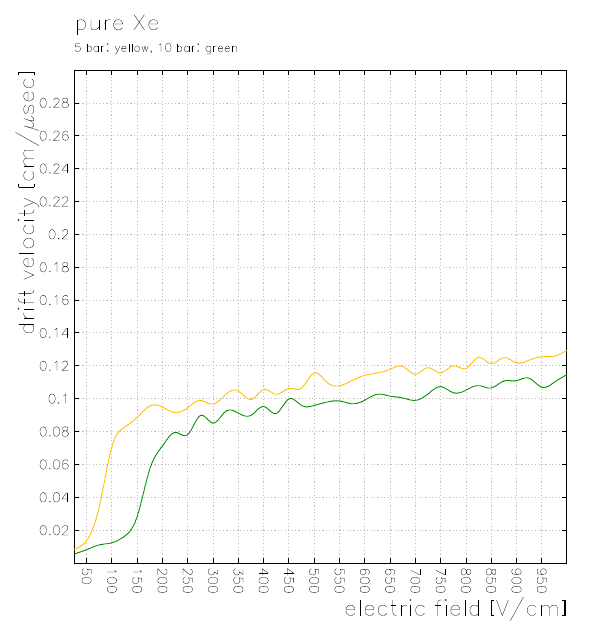} \\
\includegraphics[width=0.59\textwidth]{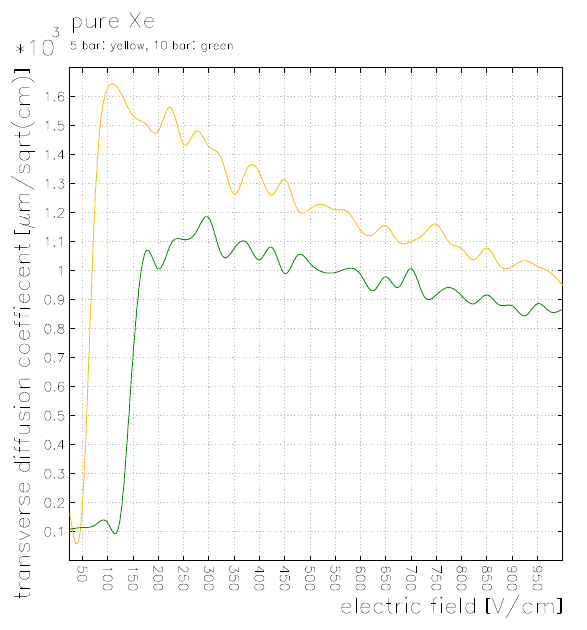}
\end{tabular}
\end{center}
\vspace{-1.0cm}
\caption{\small Top: drift velocity in pure xenon at 5 and 10 bar. Bottom; Transverse diffusion in pure xenon at 5 and 10 bar. Obtained from a Magboltz \cite{Magboltz} simulation.}
\label{fig:xenon-gas} 
\end{figure}

A \bbonu\ event will deposit 2480 keV and produce a twisted track of some 30 cm length, with an average energy deposition of 70 keV per centimeter except at both track ends, where two blobs of about 200 keV appear corresponding to the ranging-out of the electrons. In total, and taking account recombination, about 10$^5$~electrons are produced for a \bbonu\ event.

The ionization will drift towards the anode, with a drift velocity of about 1 mm/$\mu$s for drift electric fields 0.2 $<E<$ 1 kV/cm (see top of Figure \ref{fig:xenon-gas}). Assuming a drift length of 140 cm (our baseline design, see Chapter \ref{NEXT}), this implies a readout time
of around 1.5 ms. The transverse diffusion for a 1 kV/cm drift field, close to what will be our operative value, is 
9 mm/$\sqrt{{\rm m}}$ for 10 bar gas pressure (see bottom of Figure \ref{fig:xenon-gas}).

The value of maximum diffusion  is large, but acceptable. The physical
extent of \bb\ events near the \qbb\ is 200--300 mm at 10 bar, and essential features of 
event topology such as end-point ``blob'' structures are physically larger than this
maximum diffusion. In addition, multiple scattering blurs completely the trajectory of the two electrons, which becomes essentially a ``random walk'' (we will discuss the physical signature of \bb\ events later) with no relevant information except the total energy, the connected topology and the observation of two Bragg peaks or blobs at each end of the trajectory. At the same time, the transverse diffusion sets the scale for the pixelization of the readout plane, suggesting cells of roughly 1 cm size. 

\begin{figure}[tbhp]
\begin{center}
\includegraphics[width=0.6\textwidth]{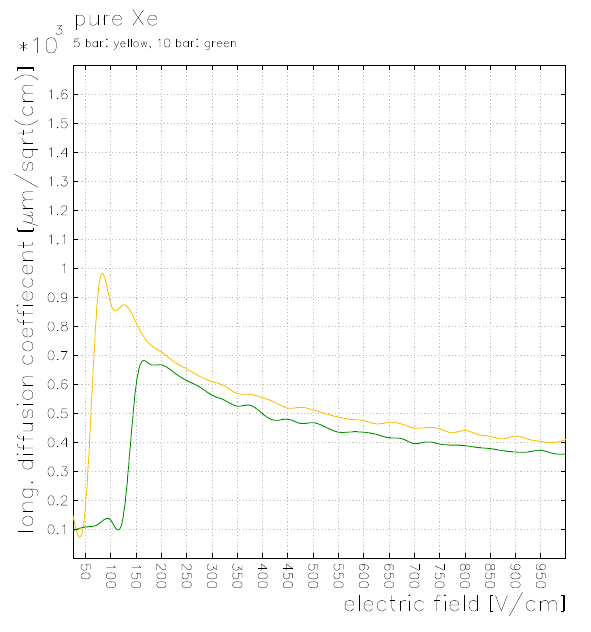} 
\end{center}
\vspace{-1.0cm}
\caption{\small Longitudinal diffusion in pure xenon at 5 and 10 bar. Obtained from a Magboltz \cite{Magboltz} simulation.}
\label{fig:long-dif} 
\end{figure}

Figure \ref{fig:long-dif} shows the longitudinal diffusion coefficient as a function of the 
applied drift field for pure xenon at 5 bar (yellow) and 10 bar (green). At 1 kV/cm one gets 
4 mm/$\sqrt{{\rm m}}$. This implies that, for a cluster of electrons originating far from the 
anode, diffusion during drift causes the signal to be spread over several microseconds.

\subsection{Secondary EL scintillation}
\label{subsec:PRIM-SIGNALS_secondaryscint}

Once the ionization electrons arrive to the anode, secondary, electroluminescent (EL) light can be generated when the electrons traverse a region of moderately high electric field generated between two parallel meshes. As can be seen from Eq.\ \ref{eq:yield}, if one chooses the width of this region to be 5 mm, a gas pressure of 10 bar and E/p $\sim$ 4${\rm~kV ~cm^{-1}~ bar^{-1}}$, (20 kV across the meshes at 10 bar pressure) then each electron yields roughly 2,200 EL photons. The EL signal is expected to spread over about 2 $\mu$s, given that the electron drift velocity is about 2-3 mm/$\mu$s in the region of intense electric field between the two grids. 

The EL light is of the same nature of the primary scintillation, peaked at 172 nm and emitted isotropically. Assuming perfect reflectivity of the TPC walls, roughly half of the photons will reach the readout plane behind the EL mesh, while the other half will reach the opposite readout plane. Assuming, as before, a collection efficiency per PMT of 10\%,  one obtains (per plane):
\begin{equation}
n_{pe}^{EL} = 10^5 \cdot 2.2 \times 10^3 \cdot 5 \times 10^{-1} \cdot 10^{-1} \sim 10^7.
\end{equation}
 
Of course, the light is not distributed in the same way in the readout plane immediately behind the EL grid (the anode) and in the readout plane opposite to the EL grid (which can be the cathode for an asymmetric TPC, or also an anode for a symmetric TPC). In the plane near the EL mesh the light is concentrated following the track path, while the light recorded by the opposite plane is much more spread, with a much fainter memory of the original track topology. 

\begin{figure}[tbhp]
\centering
\includegraphics[width=0.6\textwidth]{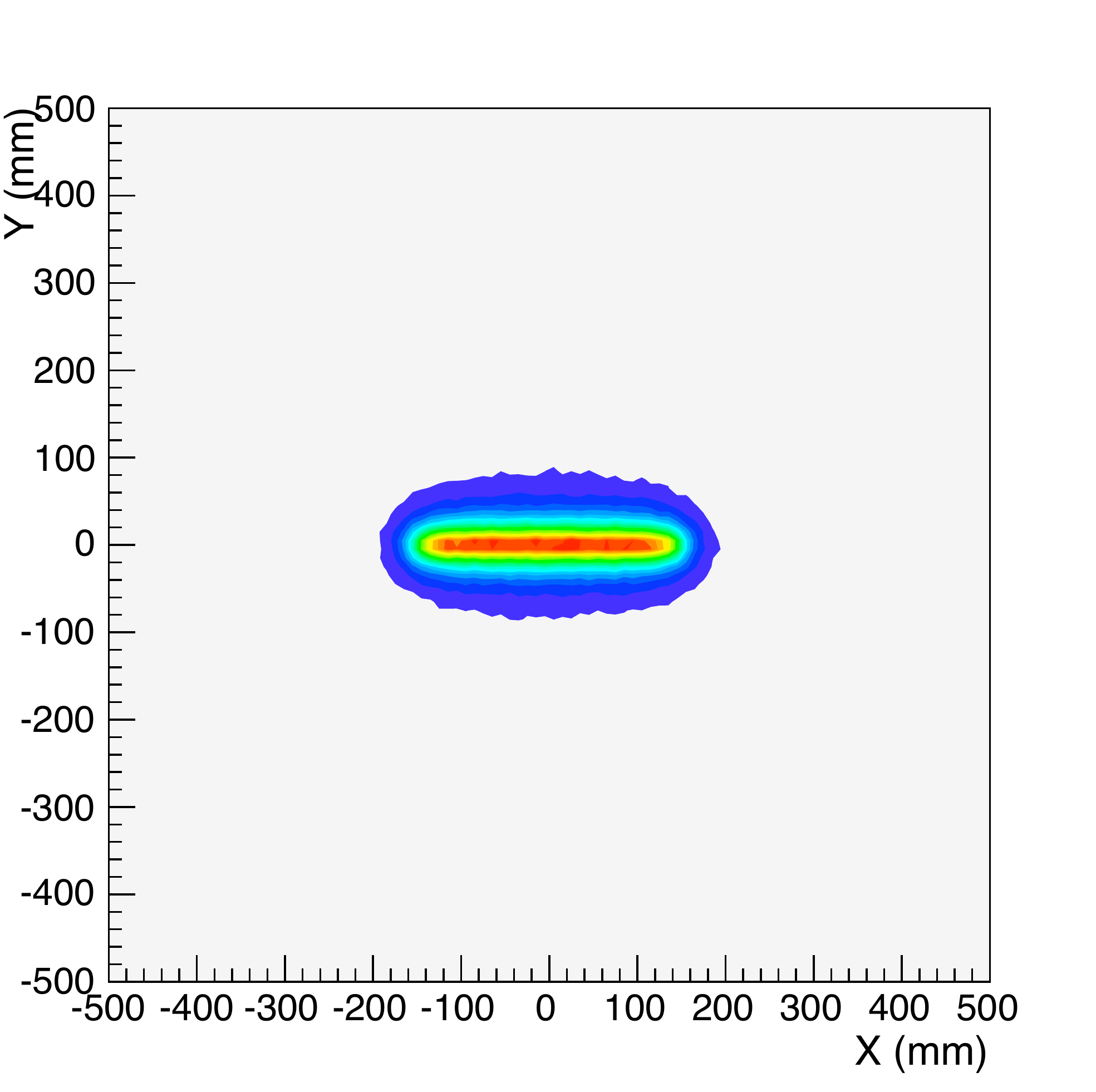} \\ 
\includegraphics[width=0.6\textwidth]{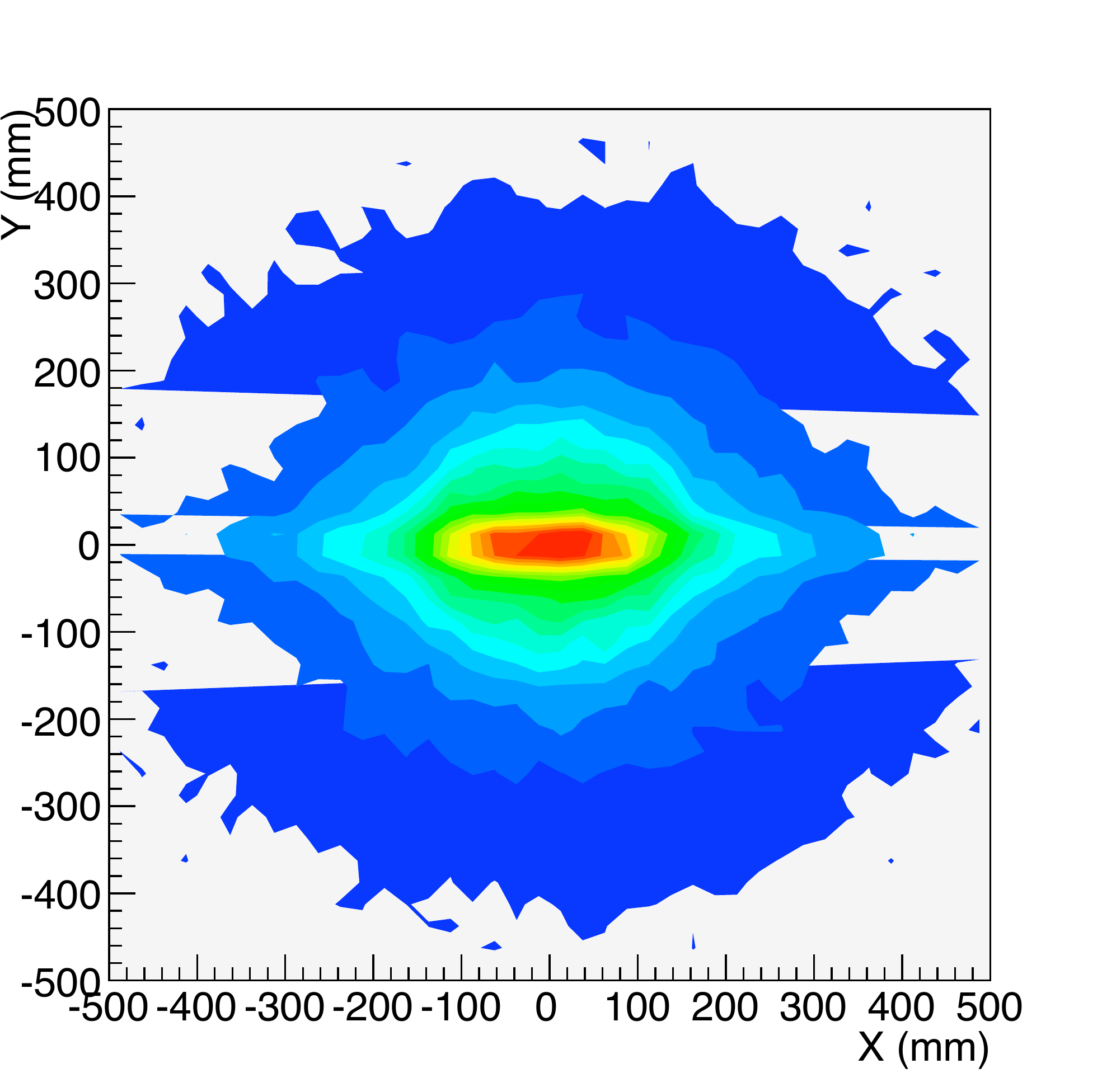}
\caption{\small Top: EL light in the readout plane immediately behind the EL grid. Bottom; EL light in the readout plane opposite to the EL grid. For the purpose of illustration a 30 cm long horizontal track
is generated near the EL mesh. Spatial dimensions are given in mm.}
\label{fig:EL-planes} 
\end{figure}

If one measures the energy in the plane opposite to the EL generating grid, the huge amount of light detected with precision photosensors
guarantees negligible fluctuations, thus a factor $G$ in Eq.\ \ref{eq:G} that is comparable to or less than
the Fano factor. Furthermore, as can be seen in Figure \ref{fig:EL-planes} the measurement has lost
most of the ``memory'' of the boundary contours of the track. One simply collects all available light using
the same PMTs that are needed for collecting primary scintillation light. The fact that EL light signals are far more spread out in time compared to primary scintillation light ones can be used for correlating the two, avoiding to confuse the latter with EL light generated from accidental low-energy backgrounds. Instead, the light spatial
distribution at the plane where EL is produced retains the track topology, and can be recorded to
reconstruct the topological signal. Thus, one plane becomes the ``tracking plane'' and the other the
``energy plane''. This concept, introduced in \cite{Nygren:2008}, is central to the NEXT design and 
we call it the {\em separated function measurement}.

To qualitatively illustrate these concepts, we consider here an idealized asymmetric EL TPC that bears some resemblance to a possible NEXT design presented in the next Chapter. We consider a 70 cm radius cylindrical chamber, with a drift length of 140 cm, filled with pure xenon gas at 10 bar pressure. The chamber inner walls are assumed to be fully reflective, with perfectly diffuse reflected light. The mesh generating the EL region is assumed to be perfectly transparent. Both cathode are anode planes are instrumented with a squared, 1 m$^2$ sensitive area centered on the TPC axis, and no photo-detection efficiency effects are taken into account. A 30 cm long track is simulated near and parallel to the EL mesh, with constant energy deposition along the track. The isotropic EL light reaching the two planes in such a case is expected to be distributed as shown in Fig.\ \ref{fig:EL-planes}, where red (blue) indicates high (low) light levels. As can be seen in the bottom figure, the distribution of light in the plane opposite to EL generation is fairly
uniform, with a mild, cosine-like dependence on the radial distance from the TPC axis of the light production point. On the contrary, the distribution of light in the plane near EL generation (top figure) provides an accurate imaging of the track.

Among other advantages, the separation of tracking and energy functions result in a more robust scheme for PMT calibration.
Since the dependence of the light in the energy plane is mostly due to geometry, the calibration, once made, should be extremely stable, as will be discussed more quantitatively in Section \ref{sec:next_photsensors}. A major advantage then is that no dilemma exists
concerning a track/event boundary for the energy measurement. Assuming that a
reconstructed track lies well within fiducial boundaries (as determined by the tracking plane), including diffusion effects, no
fiducial boundary is needed for the energy plane: every electron contributes
to the total EL signal with equal statistical weight.

\chapter{The NEXT TPC} \label{NEXT}
\section{Physics-driven requirements for NEXT}
\label{sec:next_requirements}
A HPGXe TPC design for \bbonu\ searches must capture true events with high efficiency while rejecting backgrounds to the greatest extent possible. From this, three main challenges emerge:
\begin{enumerate}
\item Determination of the total energy of each candidate event with near-intrinsic resolution, 1\% FWHM, or better, at \qbb. This energy resolution goal can be met using secondary scintillation or electroluminescence (EL).
\item Determination of the complete topology of each event in 3-D, based on energy-sensitive tracking of the \bb\ decay electrons and identification of ``satellite'' deposits of energy. The 3-D localization requires efficient detection of the primary scintillation light to accurately define the start-of-the-event time, $t_0$.
\item Fabrication materials of sufficient radio-purity such that the background rejection capabilities of the HPGXe TPC provide the desired sensitivity. This essential goal is a primary subject of chapter 4.
\end{enumerate}

We argue in the following that these requirements can be met with an EL TPC filled with high-pressure xenon gas, which we adopt as the baseline design for NEXT. Based on performance and practical considerations, we have fixed the density to $\rho \sim 0.05 ~{\rm g/cm^3}$ (50 kg/m$^3$). This corresponds to approximately 10 bars at room temperature.  The TPC design optimization requires not only a detailed definition of TPC chamber layout, but realistic trade-off considerations of readout technologies, configurations, mechanics, gas system, high voltage (HV), costs, safety aspects, and risk, etc. In this chapter, we present a specific design concept we believe is a significant step toward this goal. The further steps we believe will lead us to a practical, cost-effective optimized HPGXe TPC are presented in Chapter \ref{R-D}).


\section{The SOFT TPC: a new concept}
\label{sec:next_soft}

\subsection{Transcending the ``conventional'' TPC }

\begin{figure}[p]
\begin{center}
\includegraphics[width=0.45\textwidth]{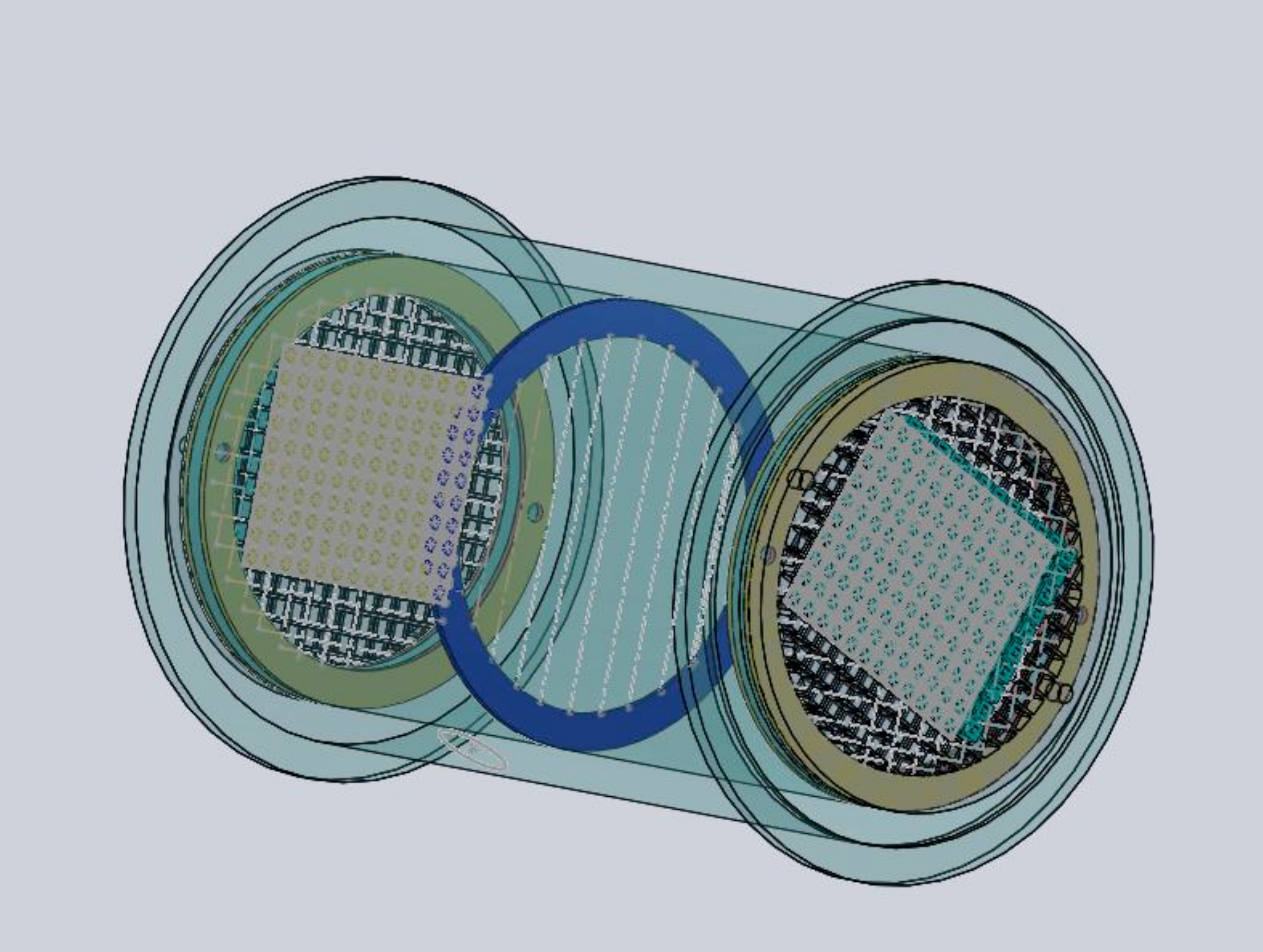} 
\includegraphics[width=0.45\textwidth]{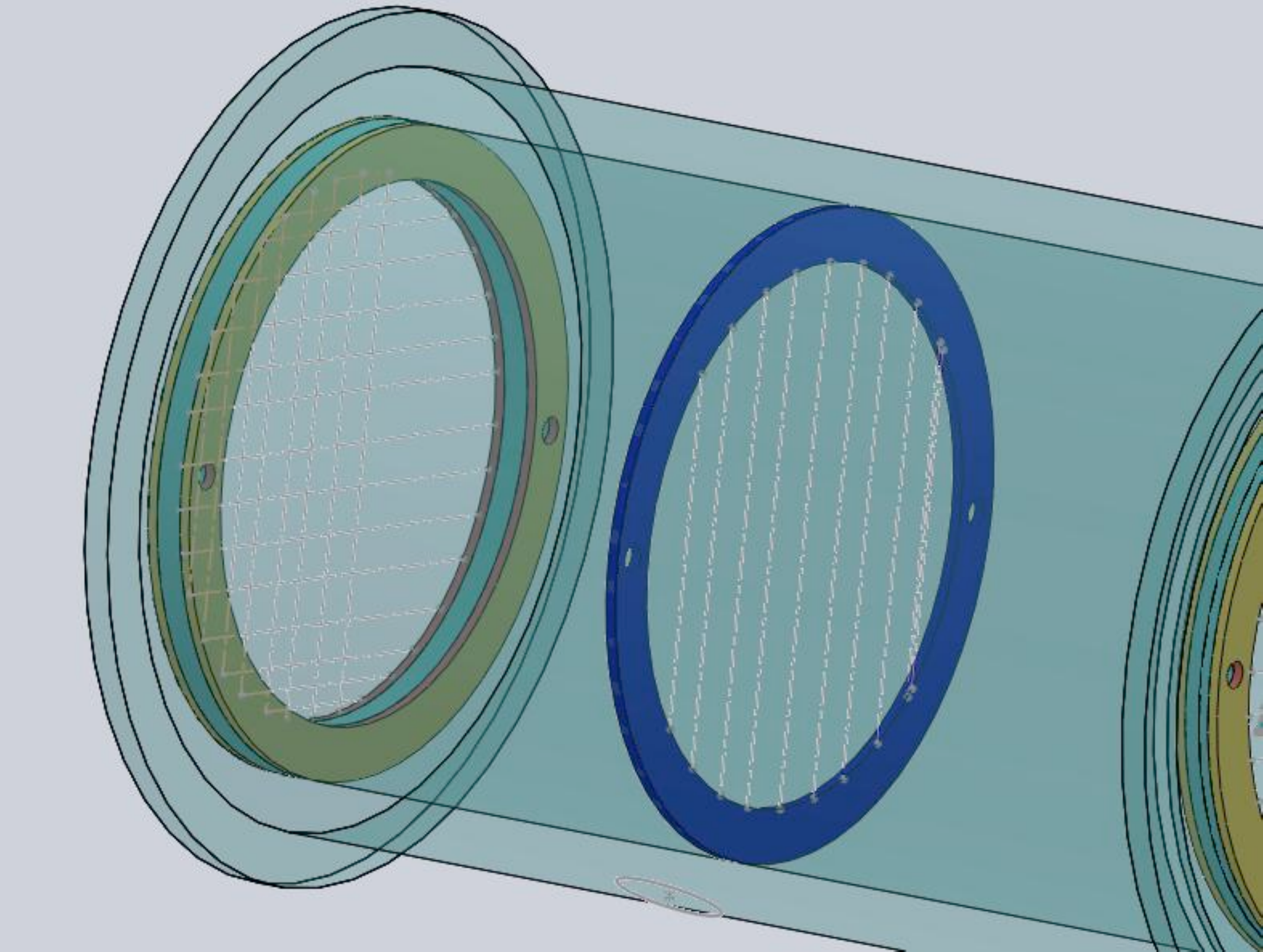}\\[.08cm]
\includegraphics[width=0.43\textwidth]{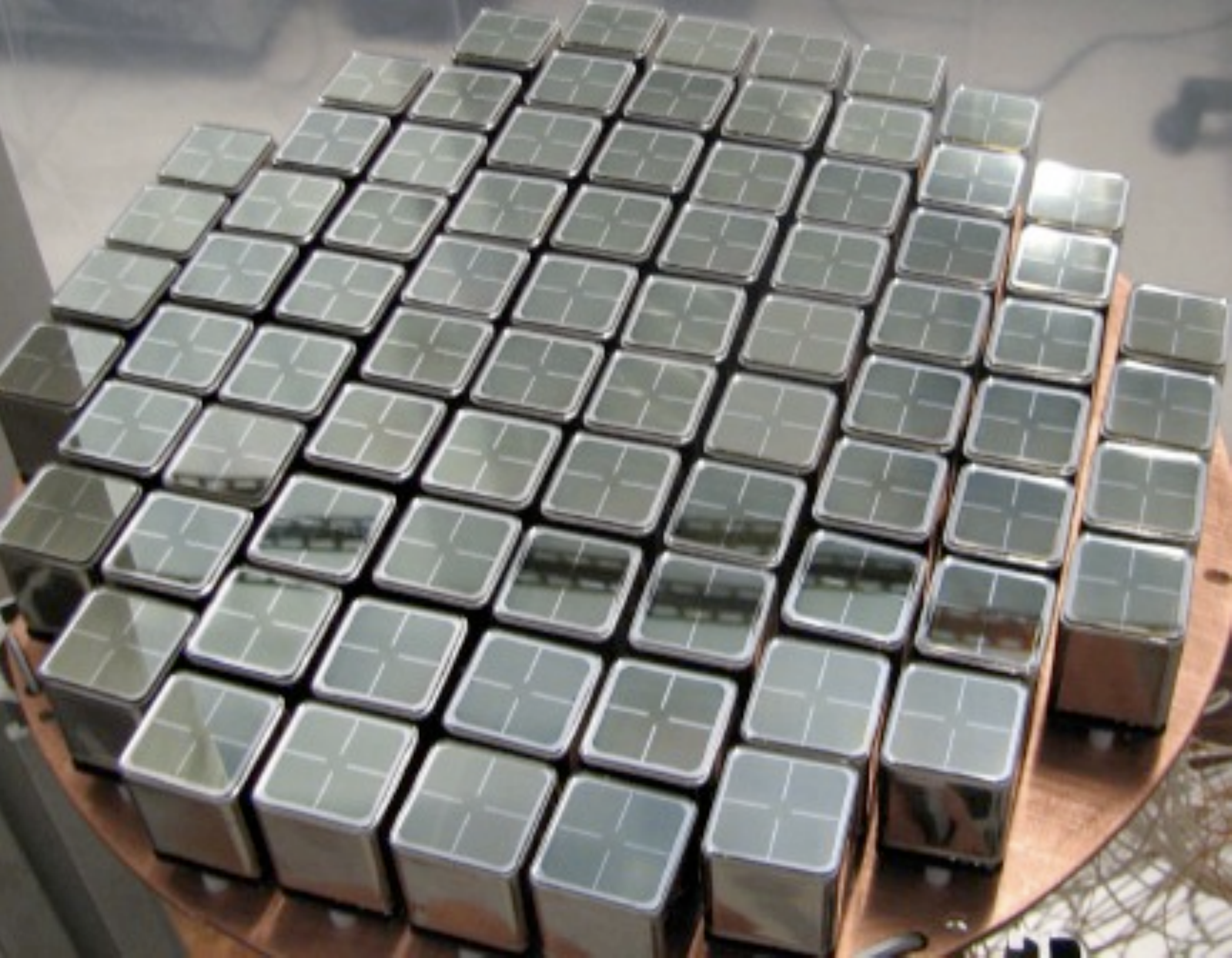} 
\caption{\small Top left: a symmetric, conventional TPC. Top right: Detail of the cathode and the EL grids at the anodes. Bottom: Dense array of R8520 PMTs, a low radioactivity PMT used by the XENON experiment, and a good candidate for NEXT PMTs. \label{fig:stpc-conv}}
\end{center}
\end{figure}


By definition, a ``conventional'' EL TPC \cite{Nygren:2007zz,Nygren:2008} (see Figure \ref{fig:stpc-conv}) is a cylindrically-shaped, ``symmetric'' TPC (STPC), with a central, transparent cathode and a multi-wire anode plane at each end-cap to provide an optical gain through EL of $\sim$300. To match this modest optical gain, dense arrays of PMTs to detect secondary scintillation are needed in each anode plane to measure energy with required precision. As a corollary, the conventional TPC has excellent sensitivity for detection of primary scintillation for $t_0$. However, for the tracking function, the PMTs nearest the secondary scintillation receive large amounts of signal and may occasionally saturate.

A good candidate for NEXT PMT is the Hamamatsu Photonics R8520, used by the XENON experiment. These rather radio-pure PMTs are of relatively small size (2.5 cm) with square-shaped photocathodes.

\begin{figure}[tbhp]
\begin{center}
\begin{tabular}{c}
\includegraphics[width=0.6\textwidth]{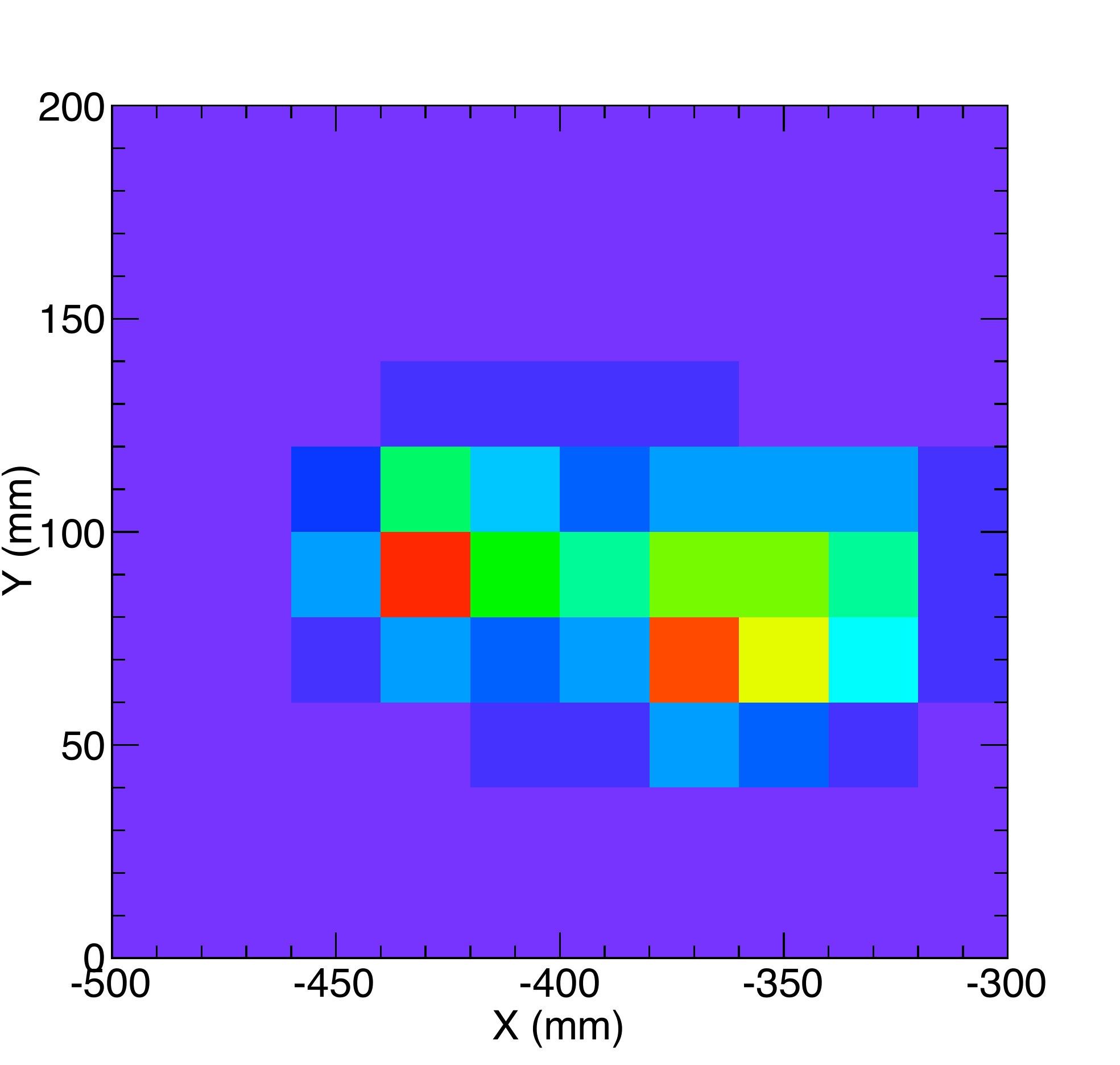} \\ 
\includegraphics[width=0.6\textwidth]{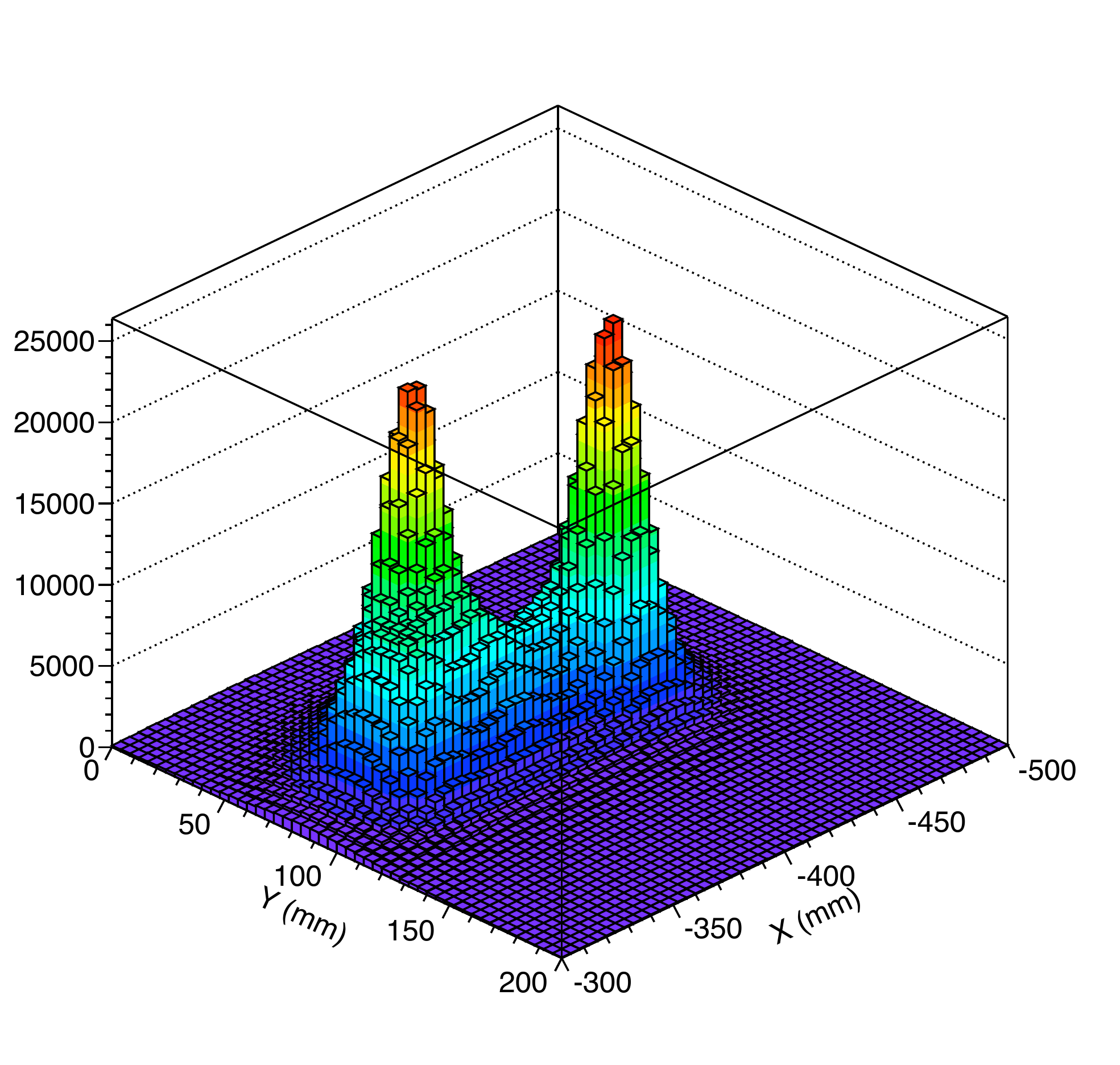}
\end{tabular}
\end{center}
\vspace{-1.0cm}
\caption{\small Top: Monte Carlo simulation of a \bbonu\ event display projected on a PMT plane. Bottom: 3D view for the same simulated \bbonu\ event. The tracks and the energy blobs at track ends are clearly visible.}
\label{fig:EL-tracking} 
\end{figure}

For 10 bar xenon gas pressure, a possible layout for a cylindrically-shaped conventional STPC of about 100 kg active mass has two 140 cm long drift regions, and two 100 cm diameter anode planes. In order to provide full photocathode coverage for the two 0.785 m$^2$ anode surfaces, one would then need about 1250 at each anode, or 2500 PMTs. This is a large number, implying a cost of about 2 million euros and an undesirable radioactivity burden of about 1250 mBq (assuming 0.5 mBq per PMT, as measured by the XENON collaboration \cite{Aprile:2009yh}). Due to complexity in event topology, tracking with pixel sizes of 2.5 cm does not appear optimum. In addition, delivery times for such a large number of PMTs may be problematic.

The use of such dense PMT arrays is clearly not fully satisfactory. Accepting the energy resolution constraint that the number of p.e.\ at \qbb\ is fixed, a better solution is to increase secondary scintillation to match a smaller number of PMTs. The increase in EL can be achieved by returning to the familiar ``parallel mesh'' geometry for secondary scintillation instead of a wire plane anode. Although this approach is constrained by the need to maintain an adequate primary scintillation signal, it appears that most of the \bb\ spectrum can be captured above noise limits (see below). 

At the same time, a smaller pixel size, of the order of the transverse diffusion (about 1 cm), in the ``tracking plane'' is highly desirable. Thus, 2.5 cm PMTs are too big in the tracking plane. This is clearly seen in Figure \ref{fig:EL-tracking}, where the reconstruction of a \bbonu\ event in the tracking plane is shown. In the figure, it is assumed that the TPC surface is covered with highly-reflective material. The sensors near the track receive much more light than the rest, despite the isotropic nature of the EL emission and the diffusely reflective TPC surface. This background light is very small. PMTs outside the track path contain, at most, 1\% of the light relative to what seen by the PMTs closer to the track. The 3D picture shows clearly the energy blobs at both track ends. Nevertheless, detailed tracking information is lost because of the pixel size. Furthermore, with such large PMTs, the ability to detect deposits related to, but disconnected from, the track is poo. 

Finally, the dynamic range also poses a challenge in the conventional TPC. Depending on the event localization, the same PMTs need to record thousands of p.e. (acting as ``tracking cells'' in the plane closest to the event vertex) for certain events, but need to record a single p.e (acting as ``energy cells'' in the plane farthest to the event vertex) for other events. Although the signals are likely to extend over tens of $\mu$s, saturation can occur. All these considerations imply that the conventional STPC is not a pathway for an optimum NEXT concept.

\subsection{Separated-Optimized Function for Tracking (SOFT)}

\begin{figure}[tbh]
\begin{center}
\includegraphics[width=0.6\textwidth]{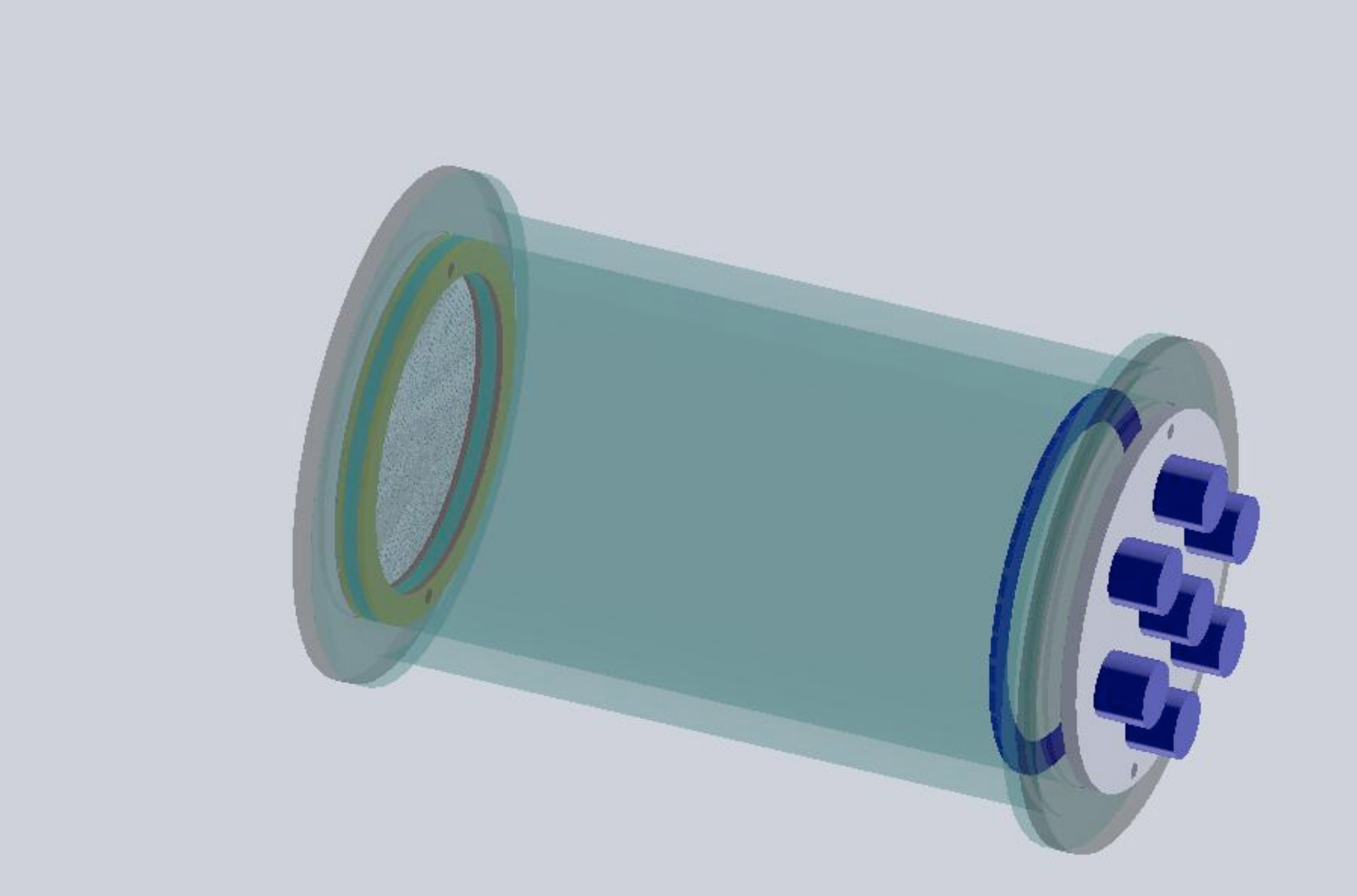} 
\end{center}
\vspace{-0.5cm}
\caption{\small An asymmetric TPC in which the cathode plane performs the energy function and the
anode plane performs the tracking function.}
\label{fig:astpc} 
\end{figure}

The first step in a natural evolution away from the conventional, symmetric TPC would be an asymmetric TPC such as the one sketched in Figure ~\ref{fig:astpc}. In this figure, the drift is only 1.5 m long (rather than about twice as much in the conventional, symmetric TPC discussed above), so it is necessary to compensate with a larger radius to obtain the same active mass within the cylindrical volume. On the other hand, only one surface, the one behind the EL grids, is densely packed with PMTs, while the cathode is sparsely covered with PMTs. In principle, PMTs behind the cathode could be larger in size with respect to the ones behind the anode. In this case the two functions (energy and tracking) are clearly separated, and the dynamic range problem is resolved, since the gain of the PMTs acting as tracking cells can be made lower than the gain of the PMTs acting as energy cells. 

The anode behind the EL grids, acting as tracking plane, would still (for the moment) be instrumented with square 2.5 cm PMTs, for a total of about 2500 PMTs. On the other hand, 10\% photo-cathode coverage in the energy plane, or about 250 2.5 cm PMTs, would be sufficient. In this scenario, the number of p.e. recorded from primary scintillation in the cathode plane is:
\begin{equation}
n_{pe}^{scint} =  n_s\cdot f_{\Omega} \cdot f_{trans} \cdot QE \cdot f_{cathode}
\end{equation}
\begin{equation}
n_{pe}^{scint} = 3\times 10^4 \cdot  5 \times 10^{-2} \cdot 10^{-1} \sim 150
\end{equation}
where we have assumed that of the $(n_s)$ $\sim$ 30,000 primary photons, $(f_{\Omega})$ = 50\%  reach the cathode, $(f_{trans}) = 50\%$ penetrate the quartz window, and that PMTs have a 20\% average quantum efficiency ($QE$) to detect primary scintillation light, and $(f_{cathode}) = 0.1$. With low-noise single-p.e.\ sensitive devices such as PMTs, detecting a 150 p.e.\ signal within 40 ns is easy. Almost all of the \bb\ spectrum is covered by setting the primary charge detection threshold to around 10 p.e, as a chance 10-fold coincidence of PMT noise within the time frame determined by the large secondary scintillation pulse is negligible.

The secondary scintillation light detected at the cathode is:
\begin{equation}
n_{pe}^{EL/cathode} = 10^5 \cdot 2.2 \times 10^3 \cdot 5 \times 10^{-2} \cdot \times10^{-1} \sim 10^6
\end{equation}
where we assume an optical gain of 2200, (using 0.5 cm gap and E/p $\sim$ 4 kV cm$^{-1}$ bar$^{-1}$) and similar detection factors as above. As discussed in Section \ref{EL}, $n_{pe}^{EL/cathode} \geq 1\times 10^6$ is sufficient to ensure that a gain resolution G of $\sim$0.15, of the same order of the Fano factor, can be met. The EL signal is generally spread over several $\mu$s, ranging from about 3 $\mu$s (for an idealized ``point-like’’ event produced at $\sim$1 cm from the EL grids) to many tens of $\mu$s (for long tracks parallel to the drift direction). 

It is reasonable to consider a factor of  2 ``contingency'', both for the detection of the $t_0$ signal (300 p.e. rather than 150) and for the energy measurement ($2 \times 10^6$ rather than $10^6$), by increasing the photo-cathode coverage to 20\%. This would correspond to 500 PMTs in the energy (cathode) plane. The necessity of such a ``contingency'' factor will be determined by our prototypes (see Chapter \ref{R-D}).

Regardless of the above considerations, the number of PMTs needed in the anode plane remains at 2500, with the corresponding problems in radioactivity, complexity, and cost. A realistic solution would likely require even coarser tracking. This undesirable circumstance can, however, be avoided.

The next logical step is to replace PMTs in the tracking plane with a more finely-pixelized readout. This is the {\it SOFT concept} (Separated-Optimized Function for Tracking): a TPC with separately optimized technical solutions for energy and tracking. 

With the above considerations in mind, in the next Section we describe the NEXT baseline design.


\section{The NEXT TPC}
\label{sec:next_tpc}

\subsection{Chamber layout}

The NEXT TPC will be filled with xenon gas at a density of 
$\rho \sim 0.05 ~{\rm g/cm^3}$ (50 kg/m$^3$), corresponding to approximately 10 bar gas pressure at room temperature.  The TPC will have a cylindrical shape. Two configurations are being considered: a) a ``symmetric'' TPC (STPC), with a central, transparent cathode and two anode planes at both end-caps equipped with photosensors; or b) an ``asymmetric'' TPC (ATPC), where the transparent cathode is at one end and a single anode at the other. In both cases, the TPC walls are lined with PTFE (Tetatrex$^{\rm TM}$, or TTX), with near 100\% reflectivity for the 175 nm UV light.

Our baseline detector design contains 100 kg fiducial mass. Table \ref{tab:params} shows the dimensions of the detector for the STPC and the ASTPC. 

\begin{table}[htdp]
\begin{center}
\begin{tabular}{ccc}
\hline \hline
& ASTPC & STPC \\
\hline
L (cm) & 140 & 2$\times$140\\
R (cm) & 70 & 50 \\
Mass (kg) & 108 & 110\\
\hline \hline
\end{tabular}
\caption{\small Detector dimensions for the ASTPC and the STPC.}
\end{center}
\label{tab:params}
\end{table}%

\subsection{The energy function}

In NEXT, the energy measurement will be provided by the detection of EL light via photosensors. An electron-track event near \qbb\, 2480 keV, may traverse 30 cm in HPXe at ten bars, liberating $10^5$ electrons. As previously discussed, in order to meet the energy resolution requirement, the array of photosensors must accurately record at least 10 p.e. per primary ionization electron. At \qbb\, this requires a total of at least 10$^6$ photoelectrons, independent of the number of PMTs. The optical gain must be adjusted to meet this constraint. Due to the isotropic nature of EL, these p.e. are distributed over all photosensors more or less uniformly. Because the light is not distributed with perfect uniformity, small corrections have to be made to the detected signal based on the radial position of the EL during the history of each event. 


In the asymmetric version of the SOFT TPC, the energy function is obtained by instrumenting the chamber with about 250, 2.5 cm square PMTs. This represents  a modest cost (400 k\euro) and relatively small radioactivity budget (200 mBq).

Each PMT must record, on average, $\sim$4000 p.e. to meet the overall goal of $\sim 10^6$ p.e. A typical event may span 100 $\mu$s, generating 40 p.e./$\mu$s, or 1 p.e every $\sim$25 ns.  While extended in time, the instantaneous signal level is thus very low. For a PMT shaping time of $\sim$100 ns (see Section \ref{sec:next_electronics}), this corresponds to about 4 p.e. Such signal characteristics, together with the need for careful intercalibration of the photosensor devices, make PMTs ideal for the energy function.

\begin{figure}[tb]
\begin{center}
\includegraphics[width=0.7\textwidth]{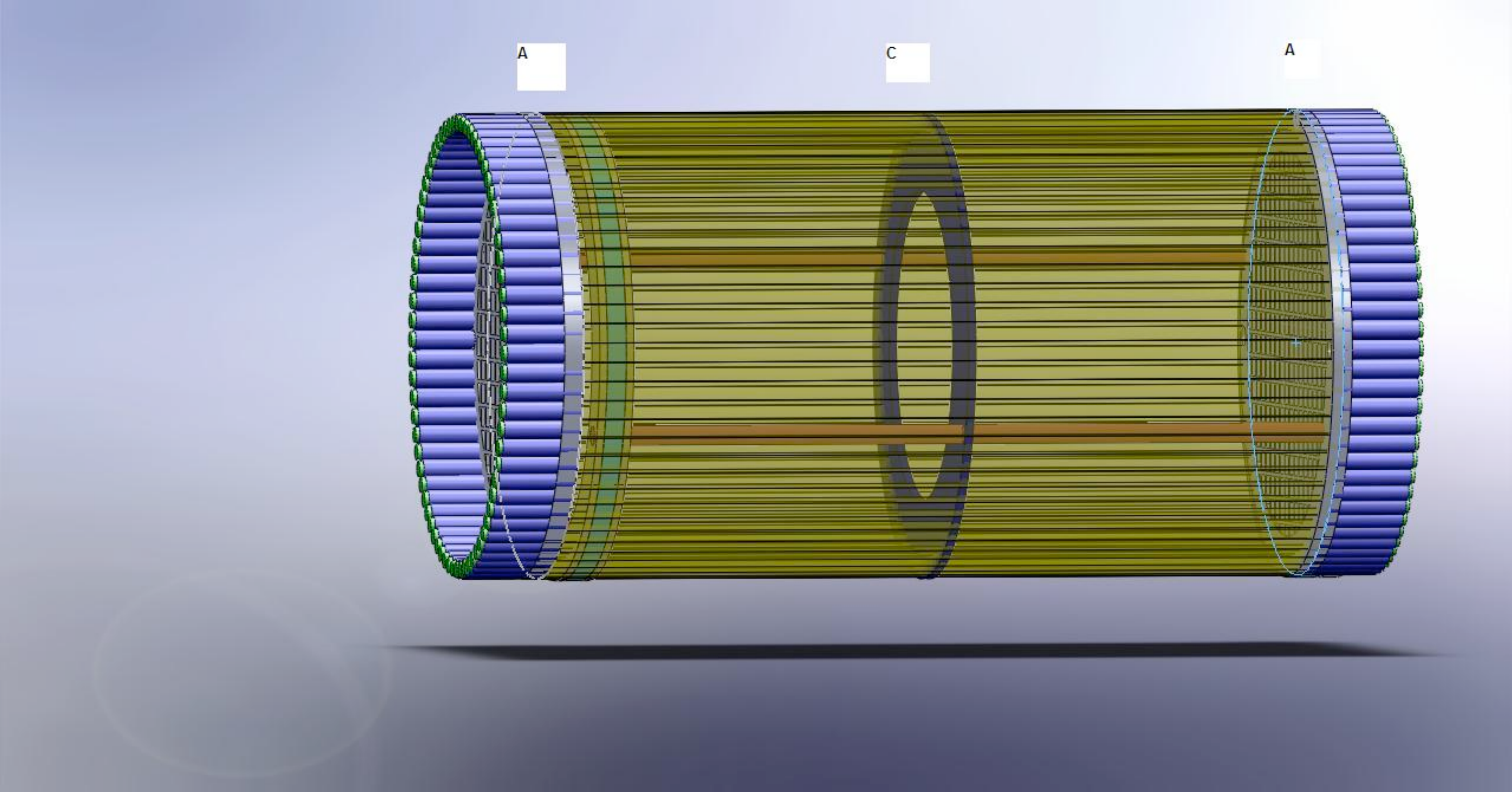} 
\end{center}
\vspace{-0.5cm}
\caption{\small A symmetric TPC with WLS bars for the energy function.}
\label{fig:stpc-soft} 
\end{figure}

The asymmetric TPC is not the only possibility to implement a SOFT TPC. Figure \ref{fig:stpc-soft} shows a concept for our proposed symmetric device. The energy function is provided by a large circumferential array of wavelength-shifting (WLS) scintillator bars to detect both primary scintillation and EL. The perceived advantages include the ability to compare performances of the two supposedly identical readout planes, a more removed location for the PMTs that reduces the impact of radioactivity, and a lower HV requirement for a given mass. While the solid angle fraction of the WLS bars is much larger than the 10\% assumption for the ASTPC, a penalty is paid in detection efficiency since a fraction of the WLS light is not captured (but the loss is not large since the WLS bar is not like fiber optics). On the other hand, the WLS function places the converted light near the maximum QE of modern PMTs, which can now approach 40\% around 400 nm. 

Even though this WLS STPC concept will show a much larger non-uniformity in azimuthal response with radial position of the EL, the integrated signal will not display a substantial dependence on the radial dependence of EL. The large-diameter WLS bars would be cast into Spectrosil$^{\rm TM}$ quartz tubing to eliminate outgassing and maximize UV conversion. In addition to simulations, evaluation of this interesting option will require a significant effort to determine the performance of the WLS bars, for which no experience exists. While it is possible that WLS may also offer a benefit within the ASTPC concept, we prefer not to include further speculation here.
 
\subsection{The tracking function}

The baseline concept for the tracking function is the detection of the EL light via photo-sensors. In this case, the tracking must be provided by an array with 2-d pixelization of 1-1.5 cm and with less radioactivity and less cost per unit sensitive area than PMTs. The signal resolution of the tracking cells does not need to be remotely as good as that for the energy measurement, since diffusion and multiple scattering dominate the spatial information quality.

\begin{figure}[tbhp]
\begin{center}
\begin{tabular}{c}
\includegraphics[width=0.7\textwidth]{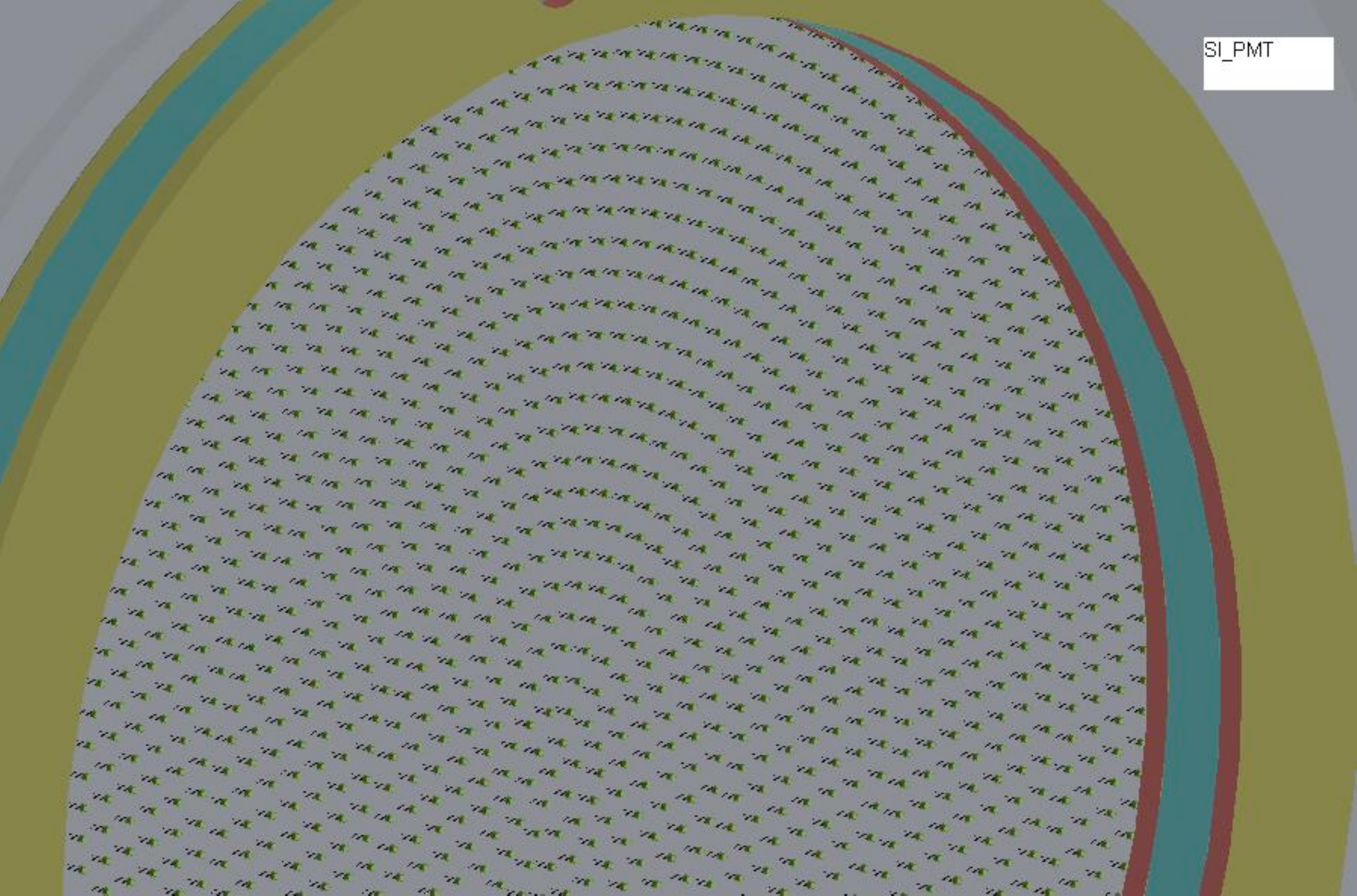} 
\end{tabular}
\end{center}
\vspace{-0.5cm}
\caption{\small The SOFT asymmetric version of NEXT, with silicon photosensors in the anode to perform the tracking function.
The photosensors are positioned into an array at the desired pitch. }
\label{fig:mpcc-tracking} 
\end{figure}


For tracking, we propose to use silicon photosensor devices, with the following advantages with respect to conventional PMT readout: a) smaller pitch, b) smaller cost (as low as 20 \euro, to be compared with some 800 \euro\ per PMT) and c) very low levels of radioactivity. The idea is illustrated in Figure \ref{fig:mpcc-tracking}. The devices are placed inside a matrix formed by a Teflon piece which holds the silicon photosensors and defines the pitch (at about 1-1.5 cm) with hexagonal symmetry. The entrance window is covered with glass or synthetic quartz coated with a WLS such as terphenyl-butadiene (TPB) that shifts light to near the region of maximum silicon photosensor quantum efficiency. The photosensor, together with its holder, is placed in each cell of the matrix and receives the photons impinging directly on its surface. It could turn out that an optimum solution is to directly couple the WLS to the MPPC. In this case, the WLS could be a small disc or rod of inorganic scintillator.

In addition to the above baseline design for tracking, we are also exploring the possibility to perform the tracking function with gain amplification and micro-pattern devices such as Micromegas (MM). This is discussed in Section \ref{sec:mm_in_el}. 

\subsection{$t_0$ measurement}

In the NEXT baseline design, the detection of primary scintillation light is obtained by the same readout plane providing the energy function. This is possible, since the prompt primary scintillation light signal will in fact be well separated in time from the delayed secondary scintillation one. While no fine pixelization is needed, detection of faint light levels by the photo-sensors is needed: we have seen that of the order of 150 p.e. are to be expected at \qbb\ energies, even less (on average) for two-neutrino \bb\ decays. The primary scintillation is always contained within $\sim$50 ns, and is easily distinguished from low-energy backgrounds producing EL, which lasts for several $\mu$s. An affordable number of PMTs can serve this purpose as well as for the energy measurement.


\section{Photosensors for NEXT}
\label{sec:next_photsensors}

The NEXT TPC baseline design includes two different photo-detector technologies: PMTs and silicon photo-sensors. In the following, we discuss in detail requirements and available technologies for both.

\subsection{PMTs}

Photomultiplier tubes will be used for measuring the total energy deposition in the chamber, and to detect the primary scintillation light. As mentioned above, a good candidate for NEXT PMTs are the Hamamatsu Photonics R8520 tubes. These PMTs combine several interesting features:

\begin{itemize}
\item A relative small size (2.5 cm) and square-shaped photocathode, which permits optimal packing for tracking.
\item Good sensitivity to the UV light (20\% quantum efficiency at 170 nm).
\item Low background, 0.5 mBq per PMT plus base on average, from the Th and U chains.
\item Resistant to pressure: rated for 5 bar, for our application they must be reinforced to take up to 10 bar.
\end{itemize}

\subsubsection{Single photoelectron signals}
\begin{figure}[tb]
\begin{center}
\includegraphics[width=0.6\textwidth]{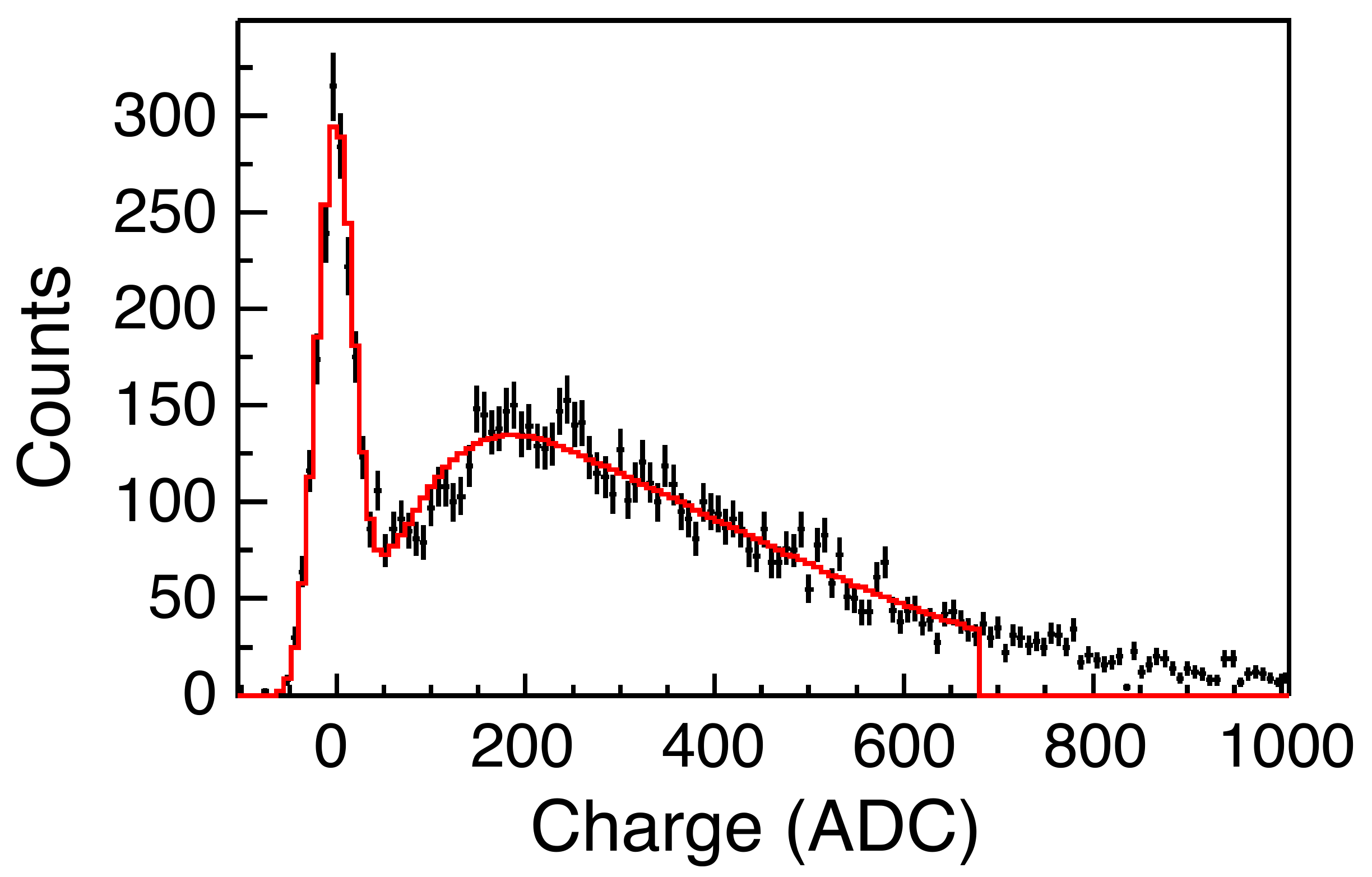} 
\end{center}
\vspace{-0.5cm}
\caption{\small Typical single photoelectron spectrum of a modern PMT.}
\label{fig:spe} 
\end{figure}

PMTs must be operated at a sufficiently high gain $m$~ such that single photoelectron (spe) detection is efficient. This is necessary for both primary scintillation and EL light detection. Figure \ref{fig:spe} shows a typical spe amplitude spectrum of a modern PMT. The broad spe peak can be discriminated against a sharp pedestal peak.  The pedestal peak is suppressed by imposing a threshold at or near the ``valley'' separating typical pedestal and spe ADC values. Threshold effects imply that the true spe detection efficiency for any given PMT is somewhat lower than 100\%, typically in the 80--90\% efficiency range. 

 A typical PMT spe pulse is $\sim$5 ns FWHM, with somewhat faster rise-time than fall-time ($\sim$10 ns wide at the base). We propose to perform PMT signal shaping, stretching the pulse shape to $\sim$100 ns, in order to allow $\sim$50 MHz digitization rates (see Section \ref{sec:next_electronics}). The signal/noise ratio is expected to be high, even for single p.e. signals, and  stretching will not degrade the $\sim$20 ns time span of the primary scintillation signal.  Furthermore, we propose to split the time-stretched PMT signal for separate digitization of the primary scintillation and EL light signals. In the following discussion, we focus on PMT response considerations for the more challenging requirements of accurate, EL-based, energy measurement.

\subsubsection{Operating gain}
For EL light detection, the PMT gain $m$~ must be sufficiently high such that true spe, shaped, pulses produce 4--7 samples above a digital threshold near a low-noise baseline.  Assuming a noise budget of $\sim$50 $\mu$V rms including ADC quantization, one could choose the PMT gain to be around $m=4 \times 10^6$.  This implies a mean spe charge q$_0$ = 0.64 pC and peak current of $\sim$130 $\mu$A, independent of output impedance. For this gain, the average spe pulse into 50 $\Omega$ reaches  $\sim$6 mV peak amplitude.  

A typical PMT will begin to display significant and undesirable nonlinearity above $\sim$1 V. In some PMTs, saturation also produces time-stretching of the signal.  So, roughly speaking, to avoid such nonlinearities at this operating gain, there should be less than 1000/6 $\sim$ 150 p.e. present within the time frame of electron transit time within the PMT, which is about 50 ns. Despite the abundant EL light intensity seen by the PMTs, the EL light waveform extends over $\sim$100 $\mu$s, thus avoiding significant non-linearities. Electronic shaping times longer than the \mbox{$\sim$100 ns} proposed will integrate more signal; such increases will, in turn, impose a greater dynamic range requirement.

\subsubsection{Calibration of gain and quantum efficiency}
PMTs naturally vary in quantum efficiency and in gain. Fortunately, it is not necessary for all PMTs to have absolutely identical quantum efficiency, nor does the primary calibration goal require knowledge of absolute quantum efficiency. It is sufficient to determine the relative quantum efficiencies $\chi$ within the ensemble of PMTs, and the single photoelectron pulse height of each PMT.  The total charge (or energy) Q is:

\begin{equation}
Q=S_0 \sum_{ab} \frac{q_{ab}}{s_a \chi_a(r_b)}
\end{equation}
where 					
$S_0$ is an overall conversion constant, q$_{ab}$ is the signal measured in PMT ``a'' in time slice ``b'', s$_a$ is the average single photoelectron signal for PMT ``a'', and $\chi_a(r)$ is the efficiency for PMT ``a'' to detect a photon originating at distance r (in the anode plane) from the TPC axis.  Because the radius where EL photons originate depends on the particular topology and origin of each event, r is a function of time-slice ``b'' that must be uniquely determined from tracking for each event.  It is important that PMTs display reasonably similar $\chi$ values, since a weak PMT with low $\chi$ contributes a larger variance per detected photon. 

The contribution of errors in $\chi_a(r)$ and s$_a$ to the Q resolution is rather small. With the reasonable assumption that all PMTs in the ensemble contribute similarly: 

\begin{equation}
\delta Q/Q \propto N^{-1/2} [(\delta s/s)^2 + (\delta \chi/\chi)^2)^{1/2}.
\end{equation}

If the maximum allowable error in $\delta Q/Q$ is set to $10^{-3}$, the errors in s$_a$ and $\chi$ must not exceed $\sim$2 \%, for the number of PMTs N $\sim$ 250. Calibrating $\chi$ and s$_a$ for each PMT to better than 2\% accuracy appears feasible by introducing accurately timed very low-intensity blue LED light pulses (wavelength is not important) into the TPC. This should allow calibration of the spe pulses with a negligible contribution of the low-amplitude PMT dynode noise pulses.

\subsection{Silicon photosensors}

For the NEXT baseline design, we propose to use silicon avalanche photosensors as tracking readout devices.

Avalanche photodiodes (APDs) possess a very thin area of very high electric-field strength. When a photo-generated (or thermally generated) electron in the conduction band moves into the avalanche region, the electric-field strength is sufficient to accelerate it to the point at which it can liberate another electron, resulting in detector gain. Typical gains for an APD are in the range of ten to about two hundred. In APDs at room temperature dark current is typically much too high to detect single electron signals.

Geiger-mode operation can increase the gain to a more significant level. Above some breakdown voltage, the APD remains stable only until an electron enters the avalanche region, resulting in the avalanche region breaking down and the APD becoming a conductor. A resistor placed in series with the detector is used to reset the detector. When the junction breaks down, large current flows through the resistor, resulting in a voltage drop across the resistor and the APD. If the voltage drop is sufficient, the APD voltage will drop below the breakdown voltage and will be reset. 

\subsubsection{Geiger-mode APD arrays}

\begin{figure}[tb]
\begin{center}
\begin{tabular}{c}
\includegraphics[width=0.4\textwidth]{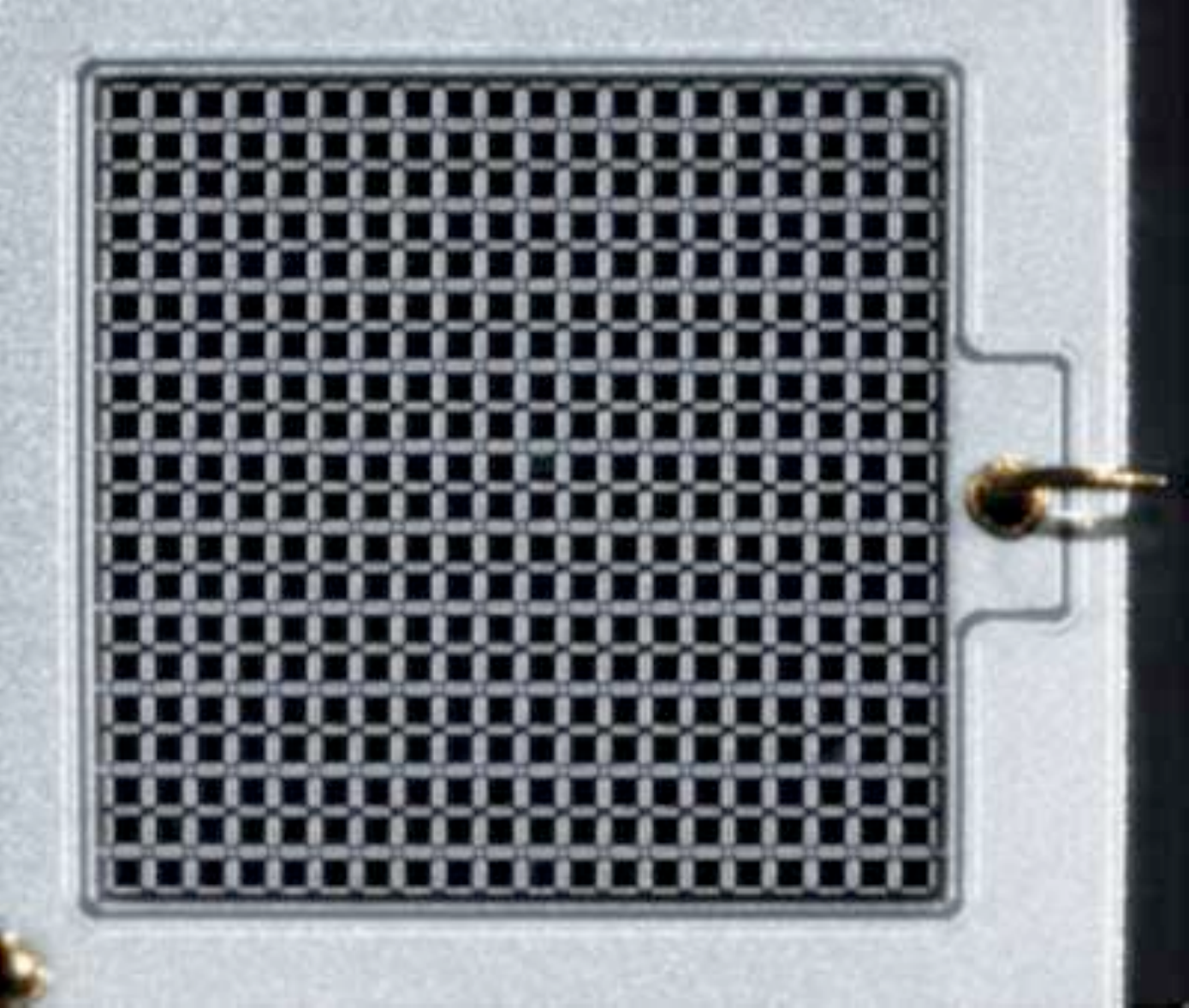} 
\includegraphics[width=0.4\textwidth]{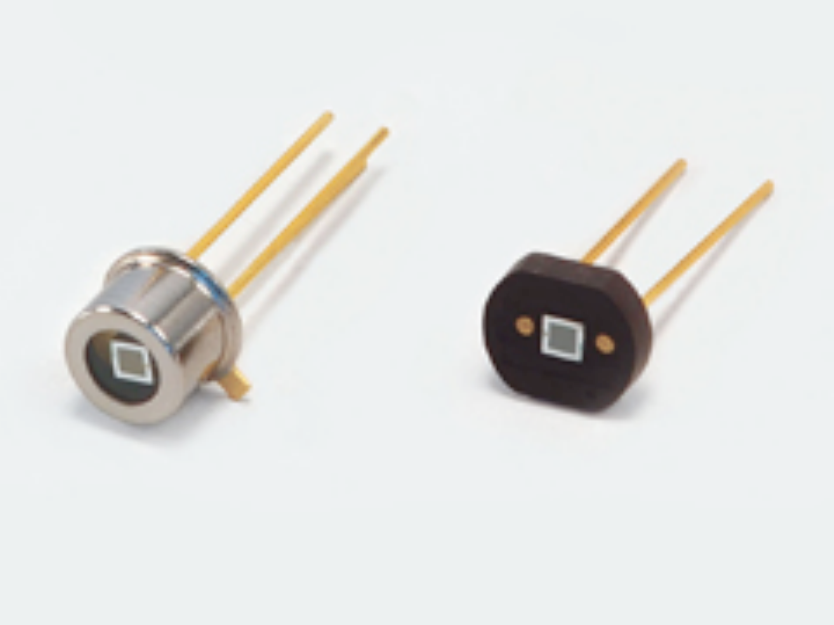}
\end{tabular}
\end{center}
\vspace{-0.75cm}
\caption{\small Left: detail of a multi-pixel photon counter (MPPC), showing the multiple APD pixels composing the MPPC. Right: two MPPC detectors (1 mm$^2$) from Hamamatsu Photonics.}
\label{fig:mpcc} 
\end{figure}

An array of small Geiger-mode APDs connected in parallel can distinguish between a single photon from multiple photons arriving simultaneously. These recently developed arrays can impressively distinguish multi-photon from single-photon light levels. One example of these devices, generically known as silicon photomultipliers, is the Multi-Pixel Photon Counter (MPPC) from Hamamatsu Photonics (Figure \ref{fig:mpcc}).

The sum of the output from each APD pixel in the array forms the MPPC output. When the photon flux is sufficiently low for photons to arrive at a time interval that is longer than the recovery time of a pixel, the MPPC will display a single-pe response. When the instantaneous photon flux is sufficiently high, or the photons arrive in short pulses (that is, pulse widths less than the recovery time), the pixel outputs will add up as multi-pe pulses. In this case, the MPPC is behaving in a quasi-analog manner, providing a measure of the incident number of photons within the integration time of signal processing electronics.

\begin{figure}[tb]
\begin{center}
\includegraphics[width=0.6\textwidth]{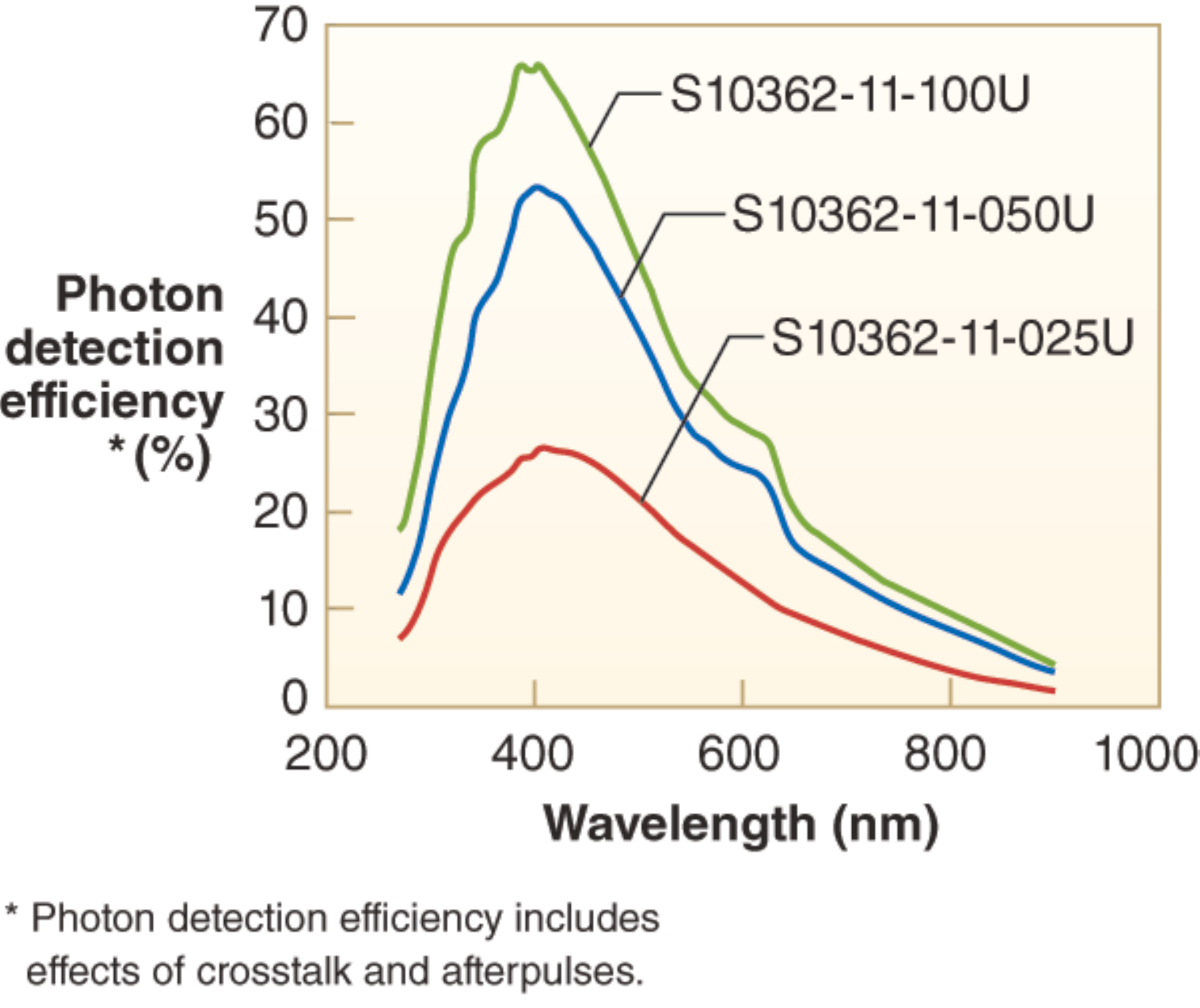}
\end{center}
\vspace{-0.5cm}
\caption{\small The photon detection efficiency (PDE) is a product of the APD quantum efficiency, fill factor, and Geiger avalanche probability. The PDE data are shown for three multi-pixel photon counters with differing numbers of pixels, and include also the effects of cross-talk and afterpulsing. Wavelength-shifting of the 175 nm UV to the \mbox{$\sim$400 nm} band is required for the MPPC.}
\label{fig:pde} 
\end{figure}

One of the biggest advantages of the MPPC is that it is a solid-state device. It is compact, rugged, easy to use (70 V operation), and low cost (about 20 to 50 euros per unit in large quantities). Furthermore the MPPC is capable of high photon detection efficiency (PDE, Figure \ref{fig:pde}).  Additional advantages of the MPPC are high gain (10$^5$-10$^6$) and low multiplication noise (noise added by the multiplication process). Tracking with MPPCs is also free of operational problems associated with gaseous avalanche gain detectors.

A MPPC disadvantage is its sensitivity to temperature, and temperature stabilization or compensation is still required for all applications using MPPCs. The Geiger mode requires a narrow operating voltage range, set for each device (or possibly groups). Another disadvantage is a high-noise rate at the single p.e. level due to thermally generated carriers at room temperature. 

\subsubsection{Tracking with MPPCs}

MPPCs generate nearly uniform single photoelectron (spe) pulses. However, the uniformity of the MPPC noise pulses and the large intensities of tracking signals should allow a digital threshold set high enough to eliminate almost all noise, without degrading the spatial resolution.  A digital threshold at the 4--6 p.e.\ level should lead to an insignificant noise rate.  

To optimize tracking, the optical coupling of each MPPC is arranged to have a limited field of view, approximately perpendicular to the parallel meshes plane.  The MPPC pixel then produces a detectable signal only if the EL light generated in the meshes is within its field of view. The total number of channels in the ASTPC (for a pitch of 1.5 cm) is of the order of $10^4$.

Each primary electron entering the meshes produces EL light for a time interval given by the gap size divided by the drift velocity.  For E/p $\sim$ 4 kV/cm bar, this time interval is about 3 $\mu$s. Typically, 600--1200 electrons contribute to the track imaging at any moment. Tracks less parallel to the TPC axis contribute the higher number of electrons within the EL gap. With so many primary electrons per mm, the statistical contribution to spatial resolution is $\sim$1 mm rms, even for the maximum possible diffusion within the chamber.  

With an optical gain of 2200 and a track population of $\sim$900 electrons within the EL meshes, the total EL luminosity is in the range of 2 $\times 10^6$ photons per $\mu$s. A detection element of 1 cm$^2$ at a distance of 1 cm from the luminous region will subtend a solid angle fraction of $\sim$0.04.  Hence, about 80,000 photons per $\mu$s will impinge on such a 1 cm$^2$ detection area. The shape and extent of the WLS must be matched to the MPPC area of only 1 mm$^2$ to produce an MPPC signal of $\sim$100 p.e.\ per $\mu$s. The required overall WLS conversion/detection factor for UV photons impinging on the 1 cm$^2$ area is 0.00125, a modest goal. 

\begin{figure}[tbhp]
\begin{center}
\includegraphics[width=0.8\textwidth]{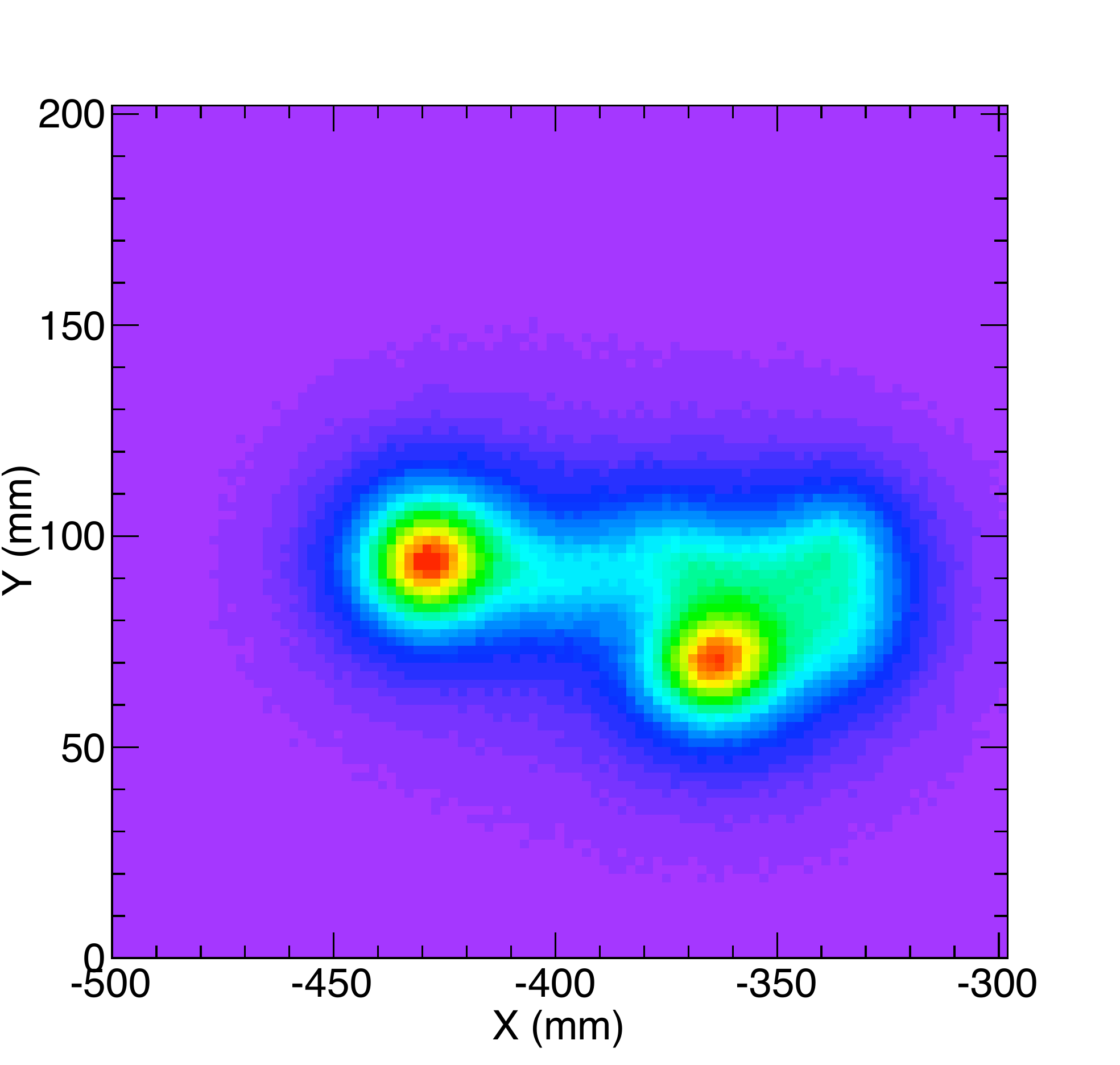} 
\end{center}
\vspace{-0.5cm}
\caption{\small Monte Carlo simulation of the image of \bbonu\ event in a plane of MPPCs.}
\label{fig:eltrk2} 
\end{figure}

Figure \ref{fig:eltrk2} shows a Monte Carlo simulation of an event tracked by MPPCs (compare with Figure \ref{fig:EL-tracking} where the event was tracked with PMTs). The light background collected by cells outside the track is at most 10$^{-4}$ of the total track, and 10$^{-5}$--10$^{-6}$ for most cells. 

Therefore, given that signal levels are ample, MPPCs can be expected to provide more than sufficient spatial and amplitude resolution to distinguish the MIP ``strip'' from the energy blobs at both ends of the track, whose energy deposition is typically 2-3 times higher.

In summary, tracking with MPPCs seem to be a robust and viable option, capable of reducing costs and radioactivity levels compared to conventional PMT readout, and providing optimal granularity. 


\section{Electronics for NEXT}
\label{sec:next_electronics}

In this Section, we further elaborate on basic principles for NEXT electronics with combined PMT plus MPPC optical readout.

The signal processing block diagrams for PMTs and MPPCs will be nearly identical, with differences appearing mainly in analog signal processing, digitization sampling rate and digitization dynamic range. Most of the time, no detector activity will be present, and zero-suppression can greatly reduce data flow.  In both cases, zero suppression is accomplished via a digital threshold well above the noise from the photosensor device and associated electronic circuitry. Time-stamping of the digitized signals will allow time correlations to be determined with a latency convenient for the formation of event triggers.

\subsection{Electronics for the PMT readout} 

\subsubsection{Analog signal processing and sampling rate}

Capture of the intrinsic waveform of the PMT would require $\sim$400 MHz sampling. For NEXT, however, no essential physics information exists in this time domain. A modest 10 - 20 ns RMS timing resolution is expected to be sufficient to correlate primary scintillation signals. A very desirable reduction in electronic system complexity, power dissipation, and data flow can therefore be realized by proper shaping of the PMT signal, ``stretching'' it in time so that it can be sampled at a much lower rate. On the other hand, the PMT pulses cannot be stretched to arbitrarily long shaping times, as the longer shaping time corresponds to longer integration time, and the instantaneous dynamic range requirement for the energy measurement could become severe. 

Although spe pulses can vary considerably in amplitude (see Fig.~\ref{fig:spe}), they are similar in shape. Therefore, provided that the electronics noise is sufficiently low, ordinary waveform sampling of a time-stretched spe pulse can provide timing information that is much more accurate than the sampling time interval. A improvement factor of 10 with respect to sampling clock time interval can be easily accomplished with decent signal/noise. Simple PMT digital signal processing of spe waveforms should therefore be sufficient to meet the NEXT requirement of 10 ns RMS timing resolution, for 100 ns wide time-stretched waveforms.  The natural sampling rate for PMT signals is thus likely to be in the 50 MHz range. This is appropriate for both the energy and $t_0$ measurements.

\subsubsection{Dynamic range considerations}

A 8-bit ADC is sufficient to characterize both spe amplitude spectrum and baseline stability and noise. We foresee a dual-range system, where the analog signal from each PMT is split. The small spe signals from primary scintillation light are digitized with 8-bit resolution, while the much larger amplitudes of the energy signals are measured with a 12 bit ADC, but at lower gain. This allows considerable range overlap to determine the relative gains with good accuracy.  

\subsubsection{Digital signal processing}

The gain of the PMT is adjusted to place the mean amplitude of an spe pulse well above electronic system noise. If the ADC returns a value above this digital threshold, data storage is initiated for that and all subsequent samples that remain continuously above threshold. A noise pulse may cause only a single sample to reach above threshold, whereas a real photoelectron pulse will generally lead to several samples rising above threshold. It is reasonable to require that at least three contiguous samples occur above threshold.  For strings of three or more contiguous samples, a time-stamp is attached to that string of data. A time-stamped string is called a ``Hit'', and constitutes the primary datum. Waveform capture may include some pre- and post-history to facilitate baseline monitoring. 

The digital threshold approach introduces a very useful ``non-linearity'' in the data flow.  Experience shows that an analog sum approach for such a large number of inputs is prone to serious difficulties due to the presence of small but coherent noise sources.  Such coherent noise can arise from power supplies, radio signals, and other electrical or electronic systems nearby.  Such coherent noise may be imperceptible in single channel performance, but, adding coherently in the assembled system, coherent noise may easily produce large and perhaps unmanageable signals in a wide fan-in. Coherent noise is often unseen until the complete system is assembled, when it is more difficult to develop countermeasures. A per-channel digital threshold completely eliminates coherent noise, since the threshold is by construction much higher than per-channel noise. 

To define a t$_0$ signal it is necessary to coordinate the Hits from all PMTs to sense a coincidence in time.  The time resolution needed to recognize the event t$_0$ is modest, about 20 ns, but in practice, it is almost certain that the event trigger is the large energy signal. In this case all data must be buffered for at least one millisecond to recover the t$_0$ information. The Hit data are time-sorted to permit the triggering and event-building functions to occur efficiently. An event trigger keeps all waveforms within the maximum time span of interest.  

\subsection{Electronics for the tracking function}

Because electrons reside in the EL region for 2--3 $\mu$s, a $\sim$2 $\mu$s shaping time and $\sim$1 MHz digitization rate is appropriate for tracking.  A 8-bit ADC seems adequate; this limited resolution and conversion rate minimizes power dissipation, an important consideration given the large number (10$^4$) of tracking channels. 
 
This large number of tracking channels located in the pressurized TPC vessel also poses some challenges for signal transport and processing. Several solutions may be implemented, ranging from extracting all signals out of the TPC via multi-pin feedthroughs, to locating part of the electronics inside the TPC, protected by shielding. In this latter case, the signals will be digitized, serialized, and transmitted via optical fibers for further processing outside the vessel.


\section{Micromegas as tracking plane readout in a EL device} \label{sec:mm_in_el}
\label{sec:next_micromegas}

\begin{figure}[bthp]
\begin{center}
\begin{tabular}{c}
\includegraphics[width=0.7\textwidth]{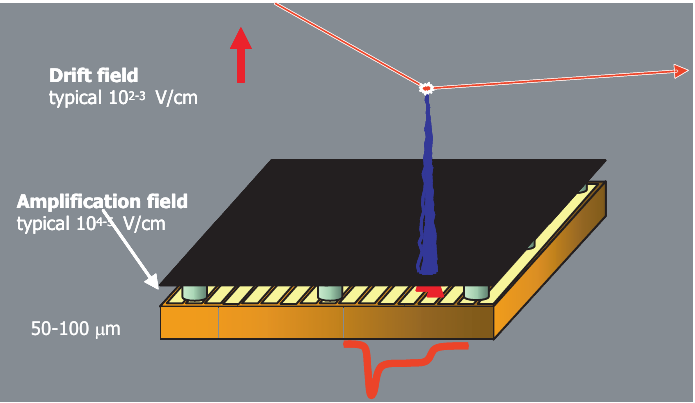} \\
\includegraphics[width=0.7\textwidth]{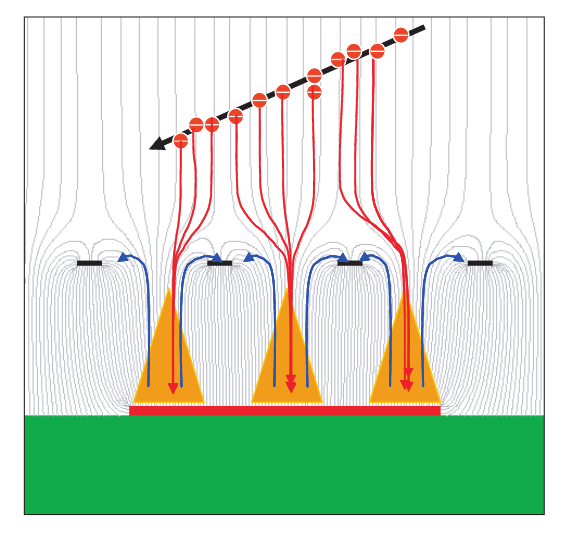} 
\end{tabular}
\end{center}
 \caption{ \small Top: Simplified scheme of a Micromegas structure, and of the detection mechanism: electron drift towards the micromesh and amplification in the amplification  gap. Bottom:
Scheme of the typical electrostatic structure of a Micromegas plane, indicating the drift
trajectories of the primary electrons entering the amplification gap (red lines) and the ions
created in the avalanche (blue lines).}
\label{MMscheme}
\end{figure}

The Micromegas (micromesh gas structure) concept, conceived and
developed at CEA/Saclay \cite{Giomataris:1995}, consists of a
micromesh, suspended over a (pixelised) anode plane, with isolator
pillars defining a high electric field gap of 50--100 microns of
height. This concept is schematically shown in Fig.
\ref{MMscheme}. In the amplification gap, the electron avalanche
occurs in much the same way as in parallel plate chambers, only in
a much thinner and localized spot. Micromegas are being used in
many high energy experiments in which they have proven outstanding
performance in spatial, temporal and energy resolutions, as well
as robustness, homogeneity and stability of operation.

The avalanche process inside a Micromegas structure has a
particular feature, first pointed out in \cite{Giomataris:1998},
which consist of reduced--or even cancelled--dependency on
external geometrical (amplification gap) or environmental (gas
pressure and temperature) factors. This effect favors the
stability of operation over long times as well as the homogeneity
of the gain over large surfaces, making the Micromegas one of the
micropattern concepts with better prospects for scaling up. In
fact, large surfaces of Micromegas readout planes have been
already built by experiments such as COMPASS (1.5 m$^2$) and
T2K (9 m$^2$). 

As amply developed in Chapter 2, electroluminescence is considered the best technology to achieve
ultimate energy resolution in NEXT, using, at the same time, the photosensors needed for the measurement
of $t_0$ (PMTs). Regarding the tracking function, our baseline scenario is based in small, high-gain photosensors
such as the SiPMTs, that also make use of the EL light.

However, we would like to explore also the intriguing possibility of performing the tracking function with Micromegas keeping the energy function with photosensors. Micromegas appears as a competitive possibility in terms of resolution, pixelization, and its capability for covering large surfaces at a reduced cost. Last but not least, the last generation of Micromegas readouts (microbulk), made out of double-clad
 kapton thin planes, are composed of potentially very radiopure
 materials (copper and kapton). 
 
The operation of Micromegas in NEXT requires addressing a set of questions that 
are being studied actively by the collaboration. The
most important ones are:

\begin{itemize}
    \item Operation at sufficiently high gains in pure xenon at high pressure and
    long-term stability. Preliminary tests on a
    small setup at Saclay seems to point that gains close to
    10$^3$ are easily achieved at 4.5 bar. 
      
     \item The possible use of a quencher for xenon, without suppressing
    the production and transmission of scintillation light, both
    primary and secondary (via electroluminescence). A very small amount of CF$_{4}$
    seems to be a good option, with the important additional benefit that
    would reduce the charge diffusion. Tests are foreseen to assess this
    possibility in NEXT.
    
    \item Measurements of actual samples are in progress
    in the Canfranc Laboratory with very encouraging preliminary
    results.
    
    \item Operation of Micromegas planes in conjunction with an
    electroluminescent grid. Study of possible incompatibilities,
    like for example the one potentially caused by the light
    produced in the Micromegas avalanche.
\end{itemize}


\section{Summary}
\label{sec:next_summary}

The NEXT detector is conceived as a high-pressure, radio-pure gas detector apparatus, 
with a fiducial mass of 100 kg of \XE, corresponding to a volume of about 2 m$^3$.
The baseline design is a electroluminescent TPC with separate-optimized function for tracking (SOFT), with PMTs performing the energy 
function and $t_0$ measurement, and Multi-Pixel Photon Counters (MPPCs) performing the tracking function. 

PMTs appear as a mandatory solution to provide a robust t$_0$ measurement from the weak primary scintillation signal, and accurate energy measurement from the intense EL signal. Central to both applications are the PMT capability as single-photoelectron counters. Calibration relies on the relative quantum efficiency of the PMTs that can be measured from the relative height of single photo-electron peaks. Our baseline technology choice for the NEXT PMTs are either a variant of Hamamatsu R8520 (reinforced to take up to 10 bar pressure), or a variant of R7367A (which takes up to 20 bar, but needs to be made radiopure). We foresee about 250 PMTs, at an estimated cost of 200 k\euro\ and about 125 mBq contribution to the radioactivity budget.

MPPCs are our baseline design as tracking sensors. We foresee about 10,000 channels, at a cost of about 20 \euro\ per channel (200 k\euro\ in total). The small (1 mm$^2$), high-density (1600 microcells), S10362-11-025C detector from Hamamatsu Photonics appears to be a good option for NEXT. The radioactivity budget is estimated to be small, likely dominated by the detector holder, and in any case below the level introduced 
by the PMTs.

The possibility of tracking with Micromegas is also being explored as a backup option.

Two fundamental milestones need to be taken before building the NEXT-100 detector. The first one 
is to prove that the SOFT concept provides the desired energy resolution and topological signature. 
The second milestone is to prove that a SOFT TPC with the chosen technological solutions can be built as a
radio-pure detector. These milestones, together with choosing and validating a limited number of technology options, 
set our R\&D program, that we discuss in Chapter 5.

\chapter{Physics potential of NEXT} \label{PYSICS}

\section{Experimental goals}
The search for neutrinoless double beta decay relies on finding a faint signal at the 
transition end-point of the \bbtnu\ energy spectrum. Due to the finite energy resolution 
of detectors, \bbonu\ events, which should pile at \Qbb, spread over a larger region. Any 
background event falling into this region limits dramatically the sensitivity to the effective 
neutrino mass. In a background-limited experiment the sensitivity improves only as 
$(Mt)^{-1/4}$ (where $M$ is the total mass of $\beta\beta$ source and $t$ is the experiment 
run time) instead of the $(Mt)^{-1/2}$ expected in the background-free case. Good energy 
resolution is therefore essential.

Unfortunately, resolution is not enough per se: a continuous spectrum arising from $\alpha$, 
$\beta$ and $\gamma$ radiation from the natural decay chains can overwhelm the signal 
peak, given the enormously long decay times explored. Consequently, extra handles 
to reject backgrounds are required.

In the traditional experimental approach, \bbonu\ searches 
have been carried out with devices such as germanium diodes \cite{Heidelberg-Moscow, IGEX}, 
which rely mainly on an impressive energy resolution (much better than 1\% FWHM at 1 MeV). 
Other experiments \cite{NEMO3A} were designed to exploit the topological signature of a \bb\ 
event to reject backgrounds, but their energy resolution was mediocre (for instance, the energy 
resolution of NEMO-3 is $\sim $14\% FWHM at 1 MeV). 

We believe NEXT offers the best compromise between these two requirements. First, resolution is 
expected to be at least 1\% FWHM at \qbb\ (\mbox{$\sim$2480 KeV}). Second, the topological
signature available on the detector allows the identification and rejection of backgrounds.


Next-generation \bbonu\ experiments aim at exploring the so-called degenerate hierarchy, 
corresponding to effective neutrino mass values down to 50 meV. If no signal is found, the 
inverse hierarchy extending from 20 to 50 meV will be accessible only to experiments 
which can simultaneously achieve large fiducial mass and negligible backgrounds.

In order to confirm or unambiguously refute the signal claimed by the group of the 
Heidelberg-Moscow experiment led by Klapdor-Kleingrothaus, one needs to be sensitive 
to effective neutrino masses in the range of  50--100 meV  or, in terms of the period, to $10^{26}$ years.
In particular, to exclude a signal at 240 meV (lower limit of the HM claim, see Sec.~\ref{sec:bbexp}) with a
confidence level of $\sim$90\%, a sensitivity of $\mbb \sim 110$ meV needs to be achieved. Conversely,
if a detector with such a sensitivity founds a signal at 240 meV, it would be able to measure it with 
$\sim$3$\sigma$.   

As previously said, in case the HM claim is refuted and no signal is found in the degenerate region, a new generation of experiments will explore the inverse hierarchy by means of extrapolating the known technologies to larger detectors. Within this scenario, the goal of NEXT is to provide a deep understanding of the experimental background-suppression techniques which allow the extrapolation to such huge detectors.

According to the above discussion, the goal of the NEXT-100 experiment is to build and operate a 100 kg TPC capable of exploring down to $\mbb \sim 100$ meV, hence confirming or refuting the HM claim. Such a detector will be large enough to prove the feasibility of scaling the technology up to a 1-ton detector. 

In the previous chapters we have defined the baseline for the NEXT TPC. This chapter is devoted to discuss its physics potential as well as to discuss the implications on the radiopurity of the materials. 

\section{Backgrounds in NEXT}
\subsection{Signal and background signatures}
The importance of a background source in NEXT depends on the energy resolution and also on 
the track identification capabilities of the detector. Only those events with energy around or above 
\qbb, and able to mimic a signal track become a background.

Double beta decay events have a distinctive topological signature in HPGXe: a ionization track, 
of about 30 cm length at 10 bar, tortuous because of multiple scattering, and with larger depositions 
or \emph{blobs} in both ends (see Figure \ref{fig:track}). 

Secondary ionization along the track produced by $\delta$-rays complicates the
signal recognition. In addition, in more than a 30\% of the events the electrons produce also 
bremsstrahlung radiation. Those typically low-energy gammas --- below 100 keV in average,
although higher energies are possible too --- may convert within the TPC active volume and 
be detected. Table \ref{tab:brems} shows the probability of radiation together with the energy of
these photons and their mean path in xenon at 10 bar. 

\begin{table}[tb]
\begin{center}
\begin{tabular}{cccc}
\hline \hline
E$_e$ (keV) & Probability (\%) & E$_{\gamma}$ (keV) & Mean path (cm) \\
\hline
500 &  4 & 12 & 0.14 \\
800 &  6 & 27 &  0.8 \\
1240 & 8 & 58 & 1.8 \\
1680 &10 & 95 & 5.5 \\
2000 & 11 &133 & 11\\
2480 &14 &198 & 33\\
\hline \hline
\end{tabular}
\end{center}
\caption{\small Radiation probability for electrons in xenon at 10 bar. Average energy of the emitted photons and their mean free path in the HPXe is also shown.}
\label{tab:brems}
\end{table}

Energetic single-electrons can imitate the two-blob pattern due to fluctuations in energy deposition,
$\delta$-ray secondary tracks or bremsstrahlung radiation. Beta electrons and high-energy gammas
--- which interact in the HPGXe producing electrons --- are therefore the main concern. 

\begin{figure}[ptb]
\centering
\includegraphics[width=15cm]{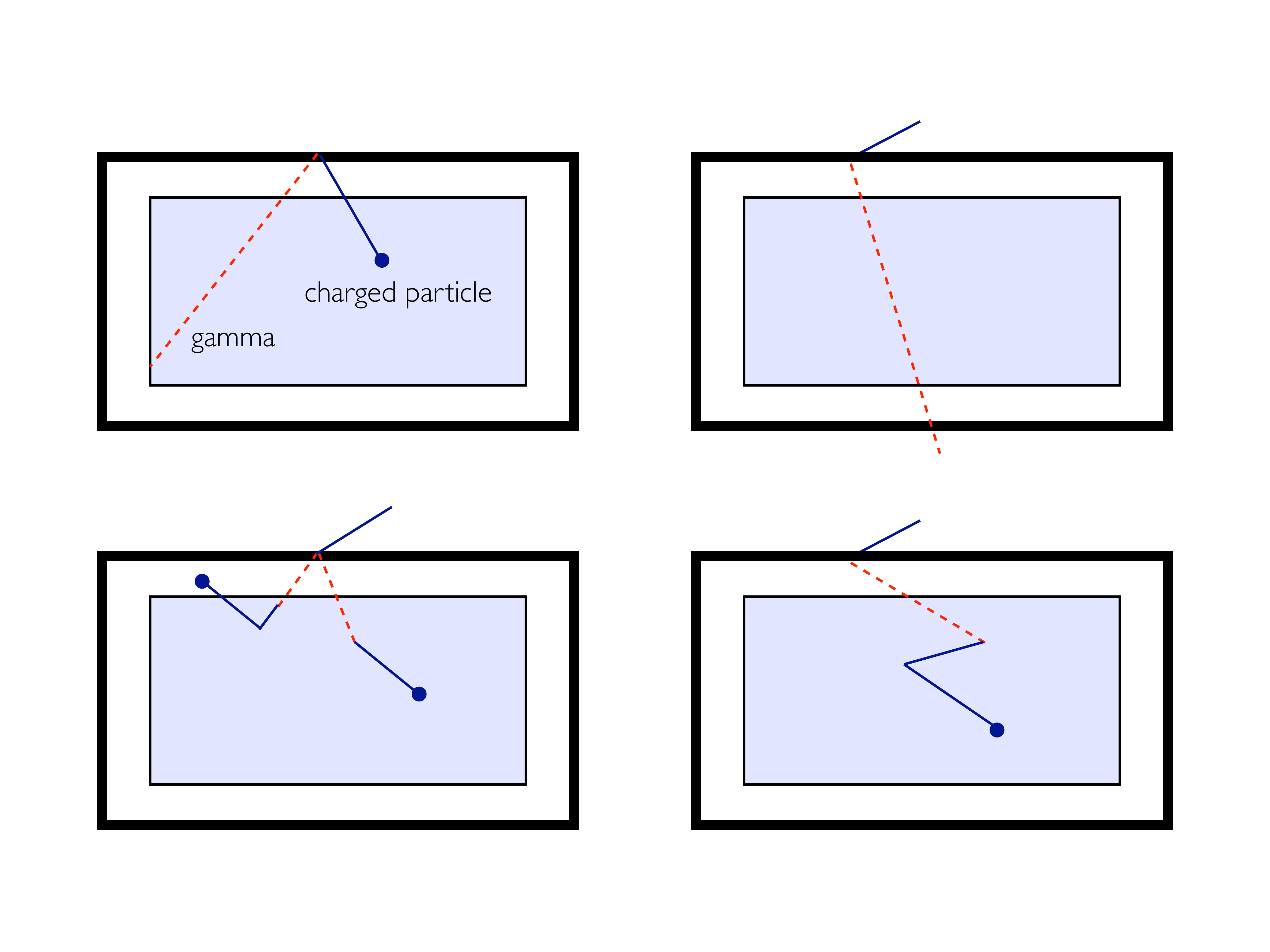}
\caption{\small Charged particle backgrounds entering the detector active volume
		can be rejected with complete 3D-reconstruction (top left). The mean
		free path of xenon for the high-energy gammas emitted in \Bi\ and \Tl\
		decays is $>3$ m, and thus many of them cross the detector without interacting
		(top right). Also, \Bi\ and \Tl\ decay products include low-energy gammas
		which interact in the vetoed region close to the chamber walls (bottom left).
		Only those background events with tracks fully-contained within the fiducial volume
		may mimic the signal (bottom right).
\label{fig:geom_rejection}}
\end{figure}

Background events with charged particles entering the active volume can be rejected defining a 
fiducial volume few centimeters away from the chamber walls (Figure \ref{fig:geom_rejection}). The
reliability of such a veto depends on the accuracy of the $t_{0}$ measurement. 

\begin{figure}[tbp]
\centering
\includegraphics[width=0.7\textwidth]{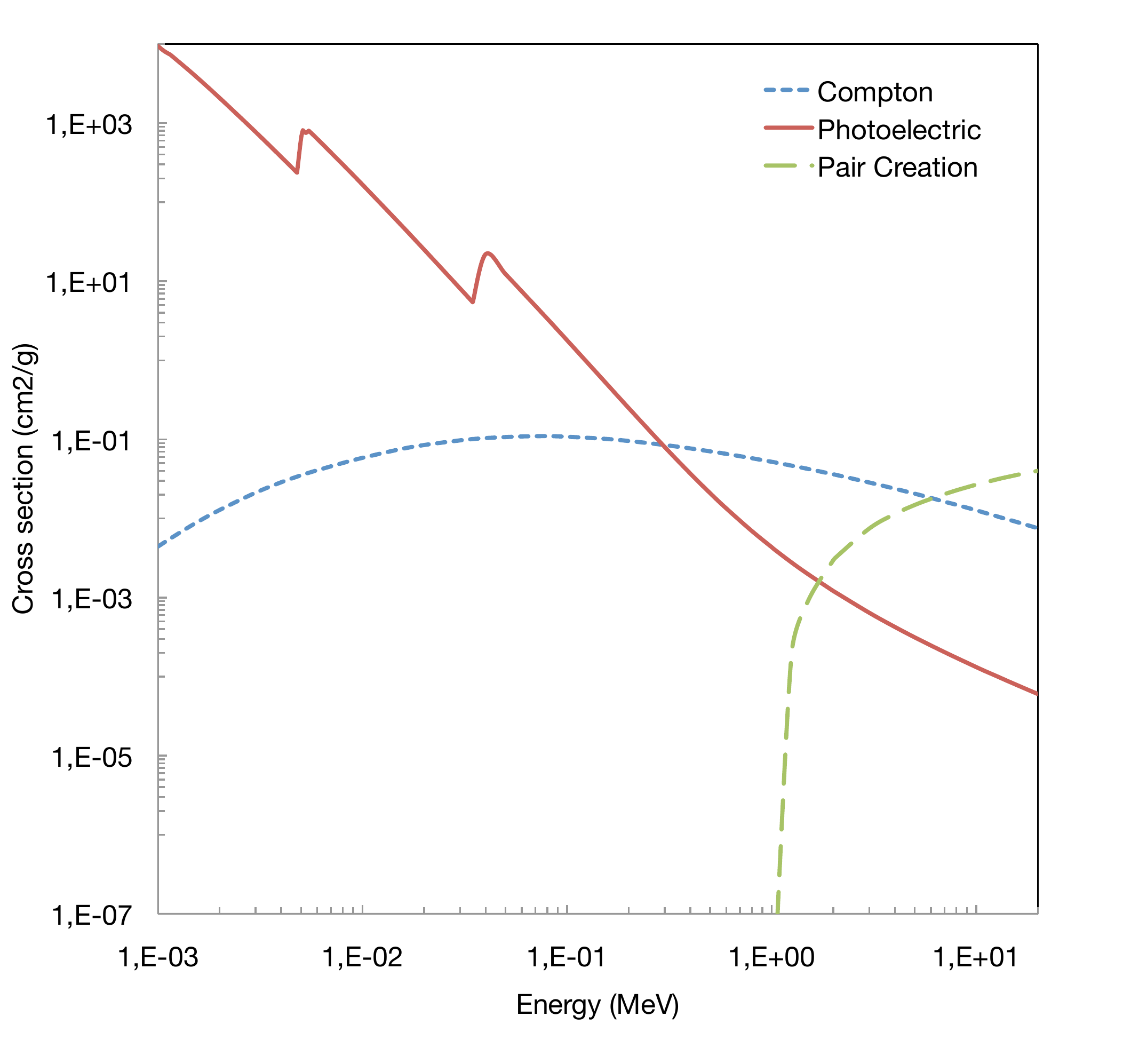}
\caption{\small Cross sections of the main interactions of photons (x-rays and $\gamma$-rays) in xenon.
 		\label{fig:xsec}}
\end{figure}

High-energy photons create electrons far from the detector walls through Compton interactions, 
pair-creation and photoelectric absorption (Figure \ref{fig:xsec}).

\subsection{Sources and types of backgrounds}
The \bbonu\ peak of \XE\ is located in the energy region of the naturally-occurring radioactive 
processes. Although the half-life of the parents of the natural decay chains is comparable 
to the age of the universe, it is very short compared to the desired half-life sensitivity of the new 
\bbonu\ experiments ($10^{26}$ years). For that reason, even a small quantity of these nuclides 
creates significant event rates.

In our case, the dangerous isotopes are \TL\ and \BI, from the thorium and uranium series, respectively.
They emit beta radiation accompanied by alpha and gamma particles due to subsequent decays and nuclear de-excitations 
of their daughter nuclei. 

The daughter of \TL, \Pb, emits a de-excitation photon of 2614 keV with a 100\% intensity.
The Compton edge of this gamma is at 2382 keV, well below \qbb. However, the scattered gamma
can interact and produce other electron tracks close enough to the initial Compton-electron
so they are reconstructed as a single object falling in the energy Region of Interest (ROI). Pair-creation events are not able 
to produce single-track events in the ROI. Photoelectric electrons 
are produced above our ROI but can loose energy via bremsstrahlung and populate the window, 
in case the emitted photons escape out of the detector.

After the decay of \BI, \Po\ emits a number of de-excitation gammas with energies above 2.3 MeV.
The gamma line at 2447 keV (intensity: 1.57\%) is very close to \qbb. The photoelectric peak
may infiltrate into the ROI for resolutions worse than 1.5--2\%. The gamma lines above \qbb\
have low intensity (below 0.1\%), but their Compton spectra can produce background tracks 
in the ROI. Pair-creation events in the window are unlikely (see Section \ref{sec:rejpower}).

\begin{figure}[tbp]
\begin{center}
\includegraphics[width=0.9\textwidth]{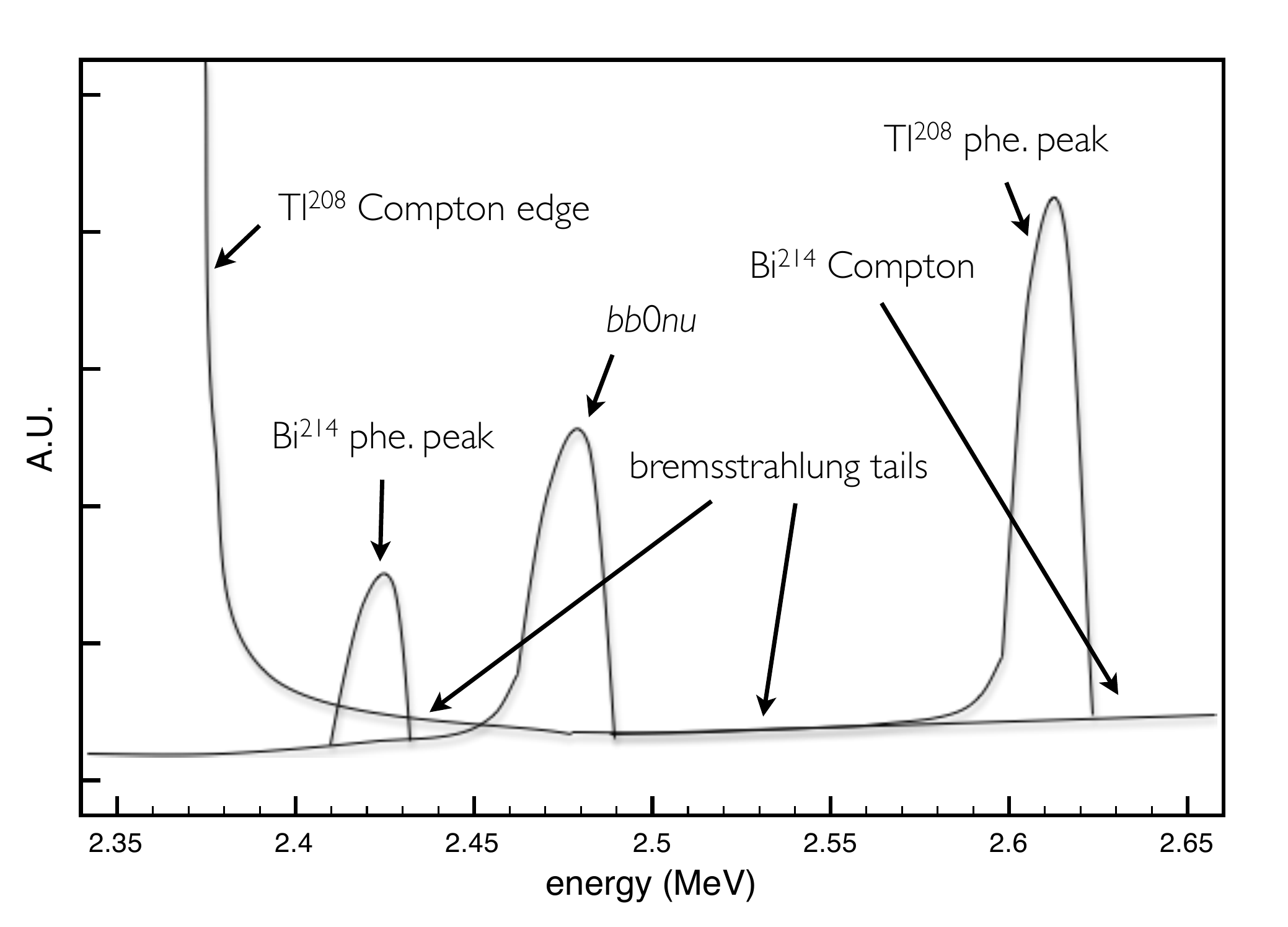} 
\end{center}
\caption{\small The landscape near the end-point of \XE , as a function of deposited energy. 
The normalization of the different peaks is arbitrary. The purpose of the plot is to show how 
the \bbonu\ signal lies between the dominant \BI\ and \TL\ backgrounds.}
\label{fig:landscape2} 
\end{figure}

A sketch of the ROI considering the above description can be seen in Figure \ref{fig:landscape2}.

All materials contain \TL\ and \BI\ impurities in a given amount. Careful selection 
of radiopure materials and purification is mandatory for all double beta decay experiments. 
Indeed, new-generation detectors are being fabricated from amazingly pure components, some with 
activities as low as 10 $\mu$Bq/kg or less.

In addition, another relevant source of background comes from radon gas: either $^{222}$Rn (half-life of 3.8\,d) from the 
$^{238}$U chain or $^{220}$Rn (half-life of 55\,s) from the $^{232}$Th chain. As a gas, 
it diffuses into the air and can enter the detector. Radon contamination 
can be translated into $^{214}$Bi, for the $^{222}$Rn, or into $^{208}$Tl, 
for the $^{220}$Rn, their only decay products affecting the experiment. 
In case of equilibrium, the activity of the radon and that of their products is the same. 

In both cases, the radon suffers from an alpha decay into polonium, producing a negative ion which 
is drifted towards the anode by the electric field of the TPC. Then, $^{214}$Bi and $^{208}$Tl contaminations can be assumed to be deposited on the anode surface. In such a way, 
1 Bq/m$^3$ of radon will mean $\ell$\,Bq/m$^2$ for its daughter, $\ell$ being the length 
of the chamber (assuming an anode covering the full base). Radon may be eliminated 
from the TPC gas mixture by recirculation through appropriate filters. Also, some underground laboratories have installed charcoal Rn scrubbers into the airstream. 

Cosmic particles can also affect our experiment by producing high energy photons or activating materials. This is the reason why double beta decay experiments are conducted deep underground. At these depths, muons are the only surviving cosmic ray particles, but 
their interactions with the rock produce neutrons and electromagnetic showers.
Muon veto detectors can easily eliminate this background contribution.

Before being stored underground, materials can be activated by energetic cosmic neutrons,
which have sizable penetrating power. Once being underground,  the capture of low-energy
neutrons can also produce radioactive isotopes in the detector.

In summary, five types of background are foreseen:
\begin{enumerate}
\item Radioactive contamination of detector materials: vessel, readout plane, etc. Careful selection
of radiopure materials is necessary to suppress this background as much as possible.
\item Radioactive contamination of laboratory walls. In this case, only photons are able to reach 
the detector. This background can be attenuated by shielding.

\item Radioactive contamination of the shielding itself. 

\item  High energy photons due to muon interactions.The muon flux itself is strongly suppressed 
by operating in an underground laboratory.
\item Neutron activation.

\end{enumerate}

Hereafter the first type will be referred to as {\it internal} background, meanwhile the other four, as {\it external}. In this chapter, the impact of all these background sources will be discussed.

\section{Simulation}
\label{sec:mc}
The NEXT collaboration is developing a Geant4-based \cite{GEANT4}, detailed simulation of the 
detector. The simulation is capable of propagating the particles through the physical volume. 
All relevant physical processes, such as multiple scattering, ionization, Compton scattering, 
pair creation, bremsstrahlung, etc., are considered. Particles with very short track-length
(less than a few mm) are not propagated but added as local energy depositions in order to 
speed up the simulation. In addition, the simulation can drift the ionization produced by 
charged particles towards the anode(s), generate and propagate the scintillation UV light 
and EL light, and provide the basic response of both photosensors (PMTs, SiPMTs) and 
Micromegas readout. 

The Monte-Carlo event generator GENBB/DECAY4 \cite{DECAY4} is used to produce the initial 
samples of signal and background. This code can simulate several \bb\ decay modes for 
all 69 known nuclei candidates for such processes, and a variety of potential backgrounds. 
It provides the initial energy, time, and angular distributions of particles emitted in those processes.

The analysis presented in the following sections relies on data samples generated with this simulation.

\section{Event selection} \label{sec:sel}
As shown in previous chapters, the topological signature of the processes taking place inside 
the TPC provides an extra handle to discriminate between signal and background events. A 
minimal set of selection cuts allows to reject background events with high efficiency.



Our Monte-Carlo can simulate the response of both Micromegas and a pixelized photosensor plane. 
In both cases, particles giving a signal on the TPC are reconstructed as a collection of ``hits'', or
``connected pixels'' in three dimensions (3D), two of them available from the pixel/MM plane, and 
the third one from the time information. 

After pixelization, a minimal set of selection cuts can be applied, based on the fact that a $\beta\beta$
event tends to have a single track with two intense deposits of energy (blobs) at both extremes due to
electron absorption. Therefore, the first step is to define the event in terms of segments and blobs.
Different algorithms are being developed in order to identify tracks (or \emph{wires}) and blobs with the
best performance allowed by the experimental parameters. These mathematical procedures take into
account the length and the charge densities to estimate the ends of a track, and treat as different objects those ones separated by one empty pixel. 

Once an event is described as a group of wires and blobs (what we call a reconstructed event), it is
selected as a signal candidate or rejected as a background one by means of the selection cuts. 
A signal candidate is an event whose reconstructed energy is within a window of a few standard
deviations around \qbb\ (the so-called ROI), and whose topology fulfills the following requirements:

\begin{enumerate}
\item	{\em Fully contained in the fiducial volume}: there is no energy depositions closer than 1 cm to the vessel. Thereby, a veto region is defined around vessel surfaces. Figure \ref{fig:geom_rejection} shows the capabilities of this veto.


\item	 {\em Only one reconstructed track}: events with more than one object (tracks or disconnected energy depositions) are rejected, since they may be produced by coincidental background interactions.  


\item	{\em \bbonu\ signature}:  the unique track ends in two blobs of high energy, as expected for a \bb\ event. This cut relies on pattern recognition algorithms currently being under study.

\end{enumerate}
%


\section{Performance of the NEXT detector}


We describe along this section the dependence of the detector performance with some critical experimental parameters, namely energy resolution, pitch size and pressure. Then, we analyze the potential of the SOFT standard design (thus fixing those experimental parameters) in terms of signal detection efficiency and background rejection factors.

\subsection{Impact of the experimental setup}

In the previous chapters we have discussed the different parameters that affect the NEXT detector design: energy resolution, pixel size, operative pressure and radioactivity budget are among the most important from the point of view of the physics performance. These parameters depend on choices related to the detector, while others such as the impact of the external backgrounds, depend largely on the environment (the depth of the laboratory, for example) and the shielding.

In the reminder of this section we focus only on the ``internal backgrounds'', coming from  detector materials. The dominant sources are the readout planes and the pressurized vessel walls. The analysis of the impact of these backgrounds will be used to define the optimal parameters for NEXT.

\subsection*{Energy resolution}
\begin{figure}[ptb]
\centering
\includegraphics[width=0.7\textwidth]{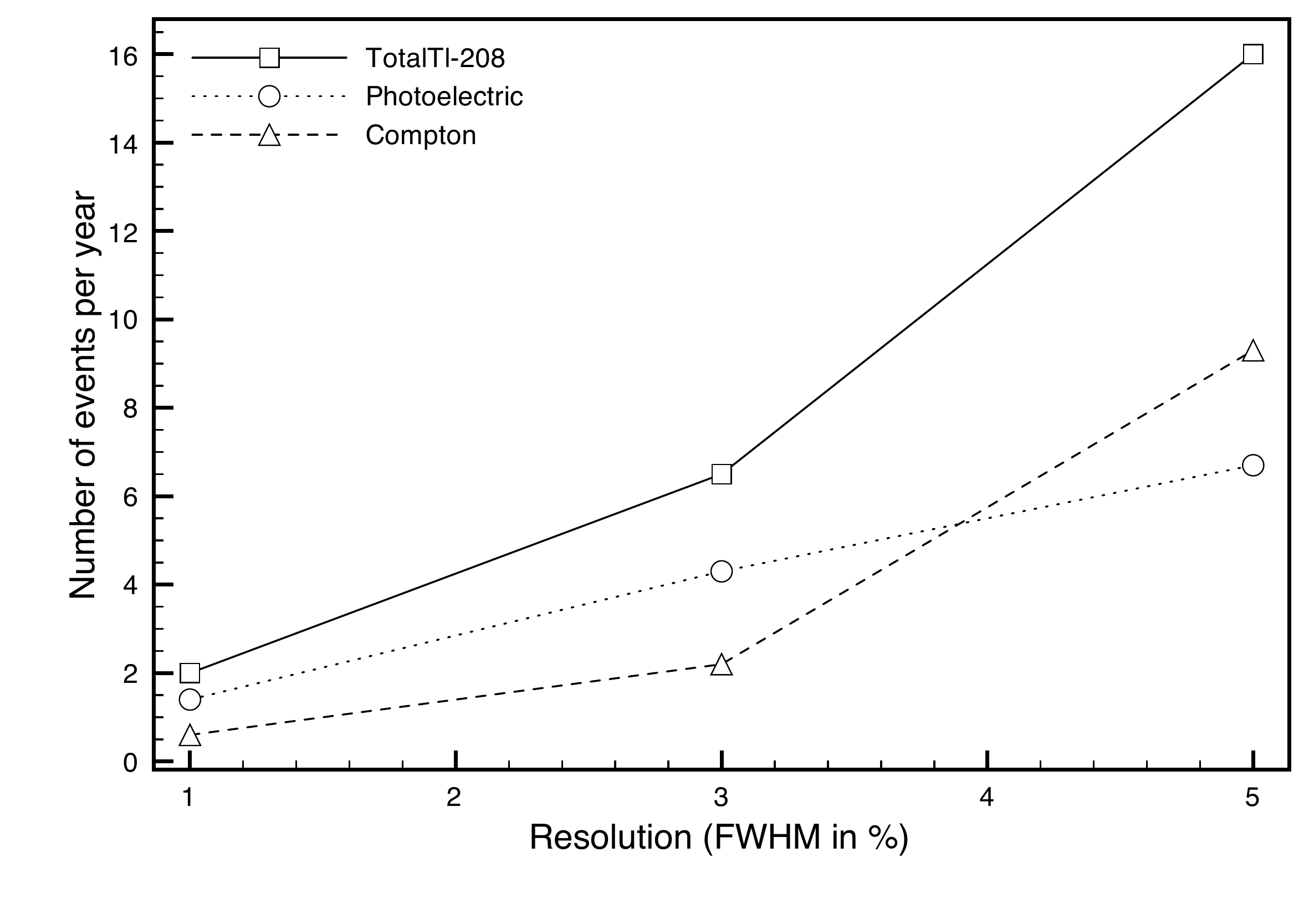}
\includegraphics[width=0.7\textwidth]{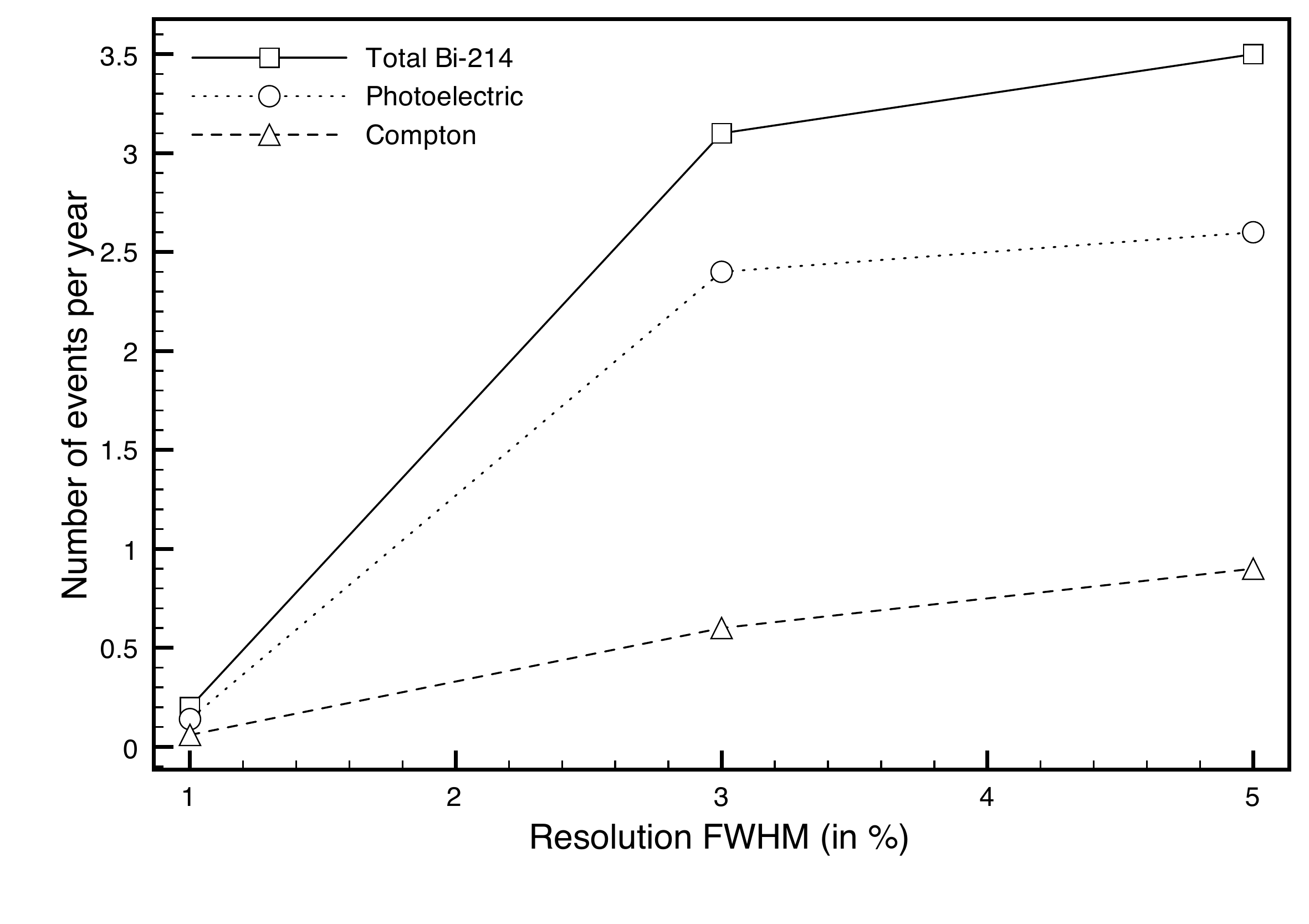}
\caption{\small Effect of resolution in the dominant source of background (events emanating from vessel).
Top: \TL\ events. Bottom: \BI\ events.}
\label{fig:eres}
\end{figure}

Along this document we have argued often that energy resolution is a must and much effort in NEXT design will be devoted to achieve better than 1\% FWHM at \qbb. To quantify the effect of resolution in the background it is enough to consider the dominant source of background (events emanating from the vessel) and study how it increases with increasing resolution, as shown in Figure \ref{fig:eres}. Notice the rapid increase of the background with degrading resolution. For \TL\ events the largest contribution comes from Compton events where the scattered photon is integrated in the track, while for \BI\ ones the largest contribution comes from the photoelectric peak, laying very near the signal peak.

\subsection*{Tracking pitch}
\begin{figure}[tb]
\centering
\includegraphics[width=0.7\textwidth]{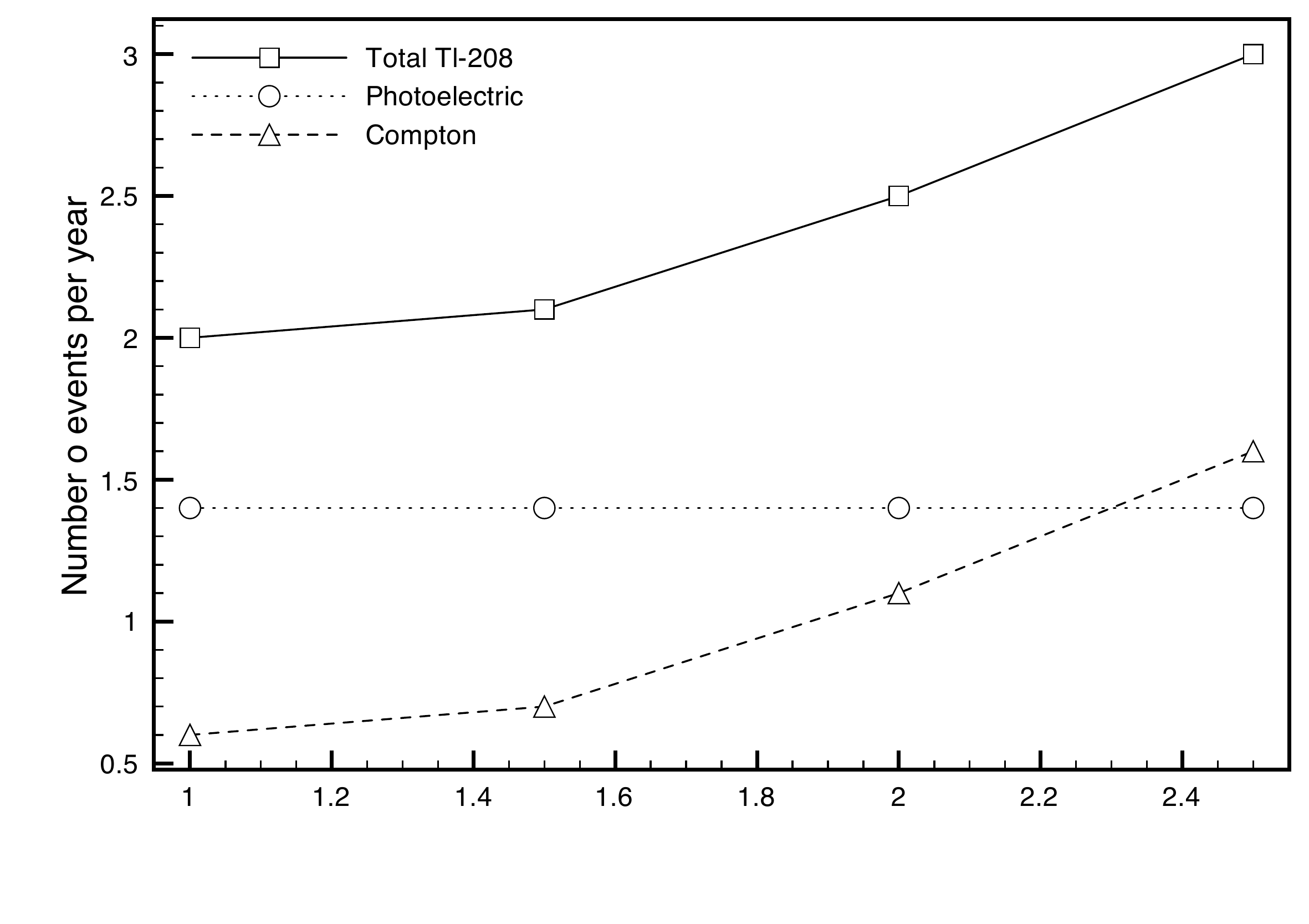}
\caption{\small Effect of pitch in the dominant source of background (\TL\ events emanating from vessel).}
\label{fig:pitch}
\end{figure}

A second fundamental parameter is the tracking pitch. A larger pitch translates into a worse two-track (or track-blob) separation and therefore one expects that backgrounds increase with increasing pitch. As before, we quantify the effect of the pitch in the background by considering the dominant source (in this case is enough to count \TL\ events emanating from the vessel) and studying how they increase with increasing pitch. As shown in Figure \ref{fig:pitch} the dependence with the pitch is much less dramatic than the dependence with energy resolution. As expected, the increase of the background is due to Compton events where the scattered photon is integrated in the track with larger pitch. A pitch of 1.5 cm is found to be a reasonable possibility, implying 10,000 channels for a ASTPC of R = 70 cm. 

\subsection*{Dependence with pressure}
A non trivial question is to choose the operative pressure of NEXT. Higher pressure results in a smaller detector (for the same mass), but at the same time requires smaller pitch if the track pixelization is to be maintained constant. On the other hand, the transverse diffusion does not decrease very much with distance, so the natural pitch is not far from 1 cm. This provides an optimal pixelization at 10 bar (with some 15--30 hits per track), while at 5 bar one could compensate with somewhat larger pitch, and at 20 bar with somewhat smaller. As pressure increases the technical problems (with electric field intensity, feedthroughs and vessel design) become more complex, and as pressure decreases the detector become bigger. All this suggests 10 bar as a good compromise.

Background levels are not expected to change dramatically with pressure. At lower pressure one has more surface to volume ratio (a minus from the background point of view) but less density and thus less interactions in the gas (a plus). Both effects keep compensating each other as pressure increases. Therefore, the choice of the operative pressure attends mostly to technical and economical considerations, rather than background ones. 



\subsection{Signal selection efficiency and background rejection power} \label{sec:rejpower}

We analyze now the performance of the standard design of the NEXT SOFT detector in terms of the signal identification and background rejection power. We assume a pressure of 10 bar, a track pitch of 1.5 cm, an energy resolution of 1 \% FWHM at $\Qbb=2480$ keV, and a vessel containing 100 kg of xenon.

In order to estimate the efficiency of the selection cuts and the capability to reject background, \bbonu\ and background data samples have been generated with the Monte Carlo tools described in Sec. \ref{sec:mc}. Signal events are simulated in the volume inside the vessel, meanwhile internal backgrounds (\TL\ and \BI\ decays) are simulated in the vessel walls and in the readout planes. Then, we apply the selection procedure described in Sec.\ \ref{sec:sel}, up to the second cut, and estimate accordingly its performance. 

Tables \ref{tab:BI}, \ref{tab:TL} and \ref{tab:eff} summarize the results. Rejection factors for \BI\ and \TL\ backgrounds are listed in Tables \ref{tab:BI} and \ref{tab:TL}, respectively, showing the contribution of each specific cut. Contribution of Compton and photoelectric events in the final sample of events passing the cuts are also presented. Total suppression factor is $\sim$4$\times 10^{-7}$ for \BI\ and $\sim$4$\times 10^{-6}$ for \TL. Signal detection efficiency due to selection cuts is about 40\%, as can be seen in Table \ref{tab:eff}.




\begin{table}[p]
\begin{center}
\begin{tabular}{ccc}
\hline \hline
Selection criteria & Events in readout & Events in vessel \\
\hline 
Initial & $10^8$ & $10^8$ \\ 
Fiducial Volume & 244296 & 195457 \\
ROI & 213 & 176 \\ 
One track & 37 & 35\\ \hline 
Photoelectric Events & 17 & 22 \\ 
Compton Events & 20 & 13 \\ \hline
Total suppression & $3.7 \times 10^{-7}$ & $3.5 \times 10^{-7}$\\
\hline \hline
\end{tabular}
\caption{\small Suppression of the \BI\ events by the selection cuts and the energy window (ROI). Suppression factors for each cut are displayed in rows 3 to 5. Rows 6 and 7 show contribution of Compton and Photoelectric events to the final sample of events passing the cuts, respectively.}
\end{center}
\label{tab:BI}
\end{table}%

\begin{table}[p]
\begin{center}
\begin{tabular}{ccc}
\hline \hline
Selection criteria & Events in readout & Events in vessel \\
\hline
Initial & $10^7$ & $10^7$ \\
Fiducial Volume & 3125234 & 2610556 \\ 
ROI & 2123 & 1648 \\
One track & 40 & 35\\ \hline 
Photoelectric events & 30 & 24 \\
Compton events & 10 & 11 \\ \hline
Total suppression & $4 \times 10^{-6}$ & $3.5 \times 10^{-6}$\\
\hline \hline
\end{tabular}
\caption{\small Suppression of the \TL\ events by the selection cuts and the energy window (ROI). Suppression factors for each cut are displayed in rows 3 to 5. Rows 6 and 7 show contribution of Compton and Photoelectric events to the final sample of events passing the cuts, respectively.}
\end{center}
\label{tab:TL}
\end{table}%


\begin{table}[p]
\begin{center}
\begin{tabular}{ccc}
\hline \hline
Selection criteria & Signal events  & Selection efficiency \\
\hline
Initial & $10^6$ & 1 \\ 
Fiducial Volume & 741249 & 0.74 \\
ROI & 589405 & 0.59 \\ 
One track & 409279 & 0.41\\ 
\hline \hline
\end{tabular}
\caption{\small Selection efficiency for the signal, according to the energy window (ROI) and the event selection cuts.}
\end{center}
\label{tab:eff}
\end{table}%

Notice the third cut in Section \ref{sec:sel} has not been applied in this study. On top of the above selection one applies now the condition that the track ends in two blobs. However, this selection criterium implies a pattern recognition algorithm which identifies the blobs and match them to the track with a certain goodness. The NEXT Collaboration is working on the optimization of such an algorithm, and preliminary results yield a rejection factor of about 1/50. The rejection power of this cut measured by the St. Gotthard TPC was roughly a factor 1/30, hence consistent with our preliminary estimations. Although the aim is to improve the value achieved by Gotthard, along this document we conservatively assume a suppression of 1/30 due to signal topology.


\section{Background estimates}

Taking into account background rejection numbers shown in the previous Section, we present now a realistic estimation of the backgrounds expected in the NEXT experiment. The contamination levels in the detector materials quoted in this Section are conservative since values are taken from  available measurements. The NEXT collaboration plans to reduce these levels by searching for more radiopure materials.

\subsection{Internal backgrounds}

We examine the impact of the \TL\ and \BI\ backgrounds in the SOFT ASTPC. This detector uses
250 PMTs in the cathode, each one with a radioactivity budget of 0.3 mBq \TL\ and 0.4 mBq \BI. The anode is composed of some 10,000 MPCCs, whose total radioactivity budget is estimated in 30 mBq \TL\ and 40 mBq \Bi. The vessel is assumed to be made of low radioactivity steel, 1 cm thick, with a mass of 740 kg and a radioactivity budget of 1 mBq/kg for \TL\ and the same quantity for \BI. The total budget is summarized in Table \ref{tab:bu}.

\begin{table}[p]
\begin{center}
\begin{tabular}{ccc}
\hline \hline
Radioactivity source & \TL\ (mBq) & \BI\  (mBq) \\ \hline
Cathode & 75 & 100 \\ 
Anode & 30 & 40 \\ 
Vessel & 740 & 740 \\ \hline \hline
\end{tabular}
\caption{\small Radioactivity budget for the SOFT ASTPC. \label{tab:bu}}
\end{center}
\end{table}

According to previous section, the final rejection factor for \TL\ events emanating from the readout planes is $1.3 \times 10^{-7}$ (recall an extra factor of 1/30 coming from the signal topology). For the \TL\ events emanating from the vessel the rejection factor is slightly better due to the vessel self-shielding, yielding $1.2 \times 10^{-7}$. Notice the contribution of \BI\ is one order of magnitude smaller and therefore irrelevant for the
present discussion. 

To compute the number of background events in the ROI per year, we multiply the radioactivity 
budget by the relevant time period (1 year = $3.15 \times 10^7$~ seconds) and by the 
suppression factor. As an example, the number of \TL\ events emanating from the cathode that 
pass all the cuts is: 
\begin{equation}
75 \times 10^{-3} \cdot 3.15 \times 10^7 \cdot 1.3 \times 10^{-7} = 307 \times 10^{-3} \sim 0.3
\end{equation}
Thus one gets about 1.2$\times 10^{-4}$~ \TL\ events/keV/kg/year. 

The contribution of the vessel is much larger:
\begin{equation}
740 \times 10^{-3} \cdot 3.15 \times 10^7 \cdot 1.3 \times 10^{-7} = 3030 \times 10^{-3} \sim 3
\end{equation}
or 1.2$\times 10^{-3}$~ \TL\ events/keV/kg/year. 
  
Tables  \ref{tab:tb} and \ref{tab:table22} summarize the internal backgrounds in NEXT. Table \ref{tab:tb} shows the contribution of \TL\ and \BI\ contamination in both readout planes and vessel, meanwhile  Table \ref{tab:table22} shows the effect of each individual selection cut over the background. Notice
the last table presents the initial number of events as the number of events found in the ROI, taking into account different energy resolutions.

\begin{table}[p]
\begin{center}
\begin{tabular}{ccc}
\hline \hline
Radioactivity source & \TL\  & \BI\  \\ \hline
Cathode (events/year) & 0.3 & - \\ 
Anodes  (events/year) & 0.2 & - \\ 
Vessel (events/year) & 3 & 0.3 \\ 
Total (events/keV/kg/year) & $1.5 \times 10^{-3}$ & - \\
\hline \hline
\end{tabular}
\caption{\small Internal background events in NEXT, after event selection cuts. Energy resolution of 1\% FWHM at \Qbb\ is assumed. \label{tab:tb}}
\end{center}
\end{table}
 
\begin{table}[htbp]
\centering
\begin{tabular}{cccccc}
\hline\hline\multicolumn{2}{c}{Resolution}&\multicolumn{4}{c} {Events (counts/keV/kg/year)$\times 10^{-3}$}\\
 \hline $\Delta$E(keV) & FWHM  & no cuts & fiducial & one track  &all \\ 
\hline $\left[ 2467,2493\right]$& 1\%  & 1459.7$\pm$0.4 & 1029.3$\pm$0.3 & 16.21$\pm$0.04  & 1.50$\pm$0.01 \\
$\left[ 2440,2520\right]$& 3\%  &1475.0$\pm$0.2 & 1082.0$\pm$0.1 & 26.63$\pm$0.03 & 4.9$\pm$0.01 \\
$\left[ 2415,2545\right]$& 5\% & 1461.2$\pm$0.2 & 1079.9$\pm$0.1 & 29.36$\pm$0.02  &5.5$\pm$0.01\\ \hline \hline
\end{tabular}
\caption{\small Internal backgrounds in NEXT, according to different selection cuts and energy resolutions. }
\label{tab:table22}
\end{table}

\subsection{External backgrounds}
While one can minimize the internal backgrounds by choosing radiopure components, it  is also necessary to suppress the external background, which comes mainly from the laboratory walls, but also from underground muons and neutron activation. The suppression of these backgrounds is achieved by shielding the detector.  

\subsection*{$^{208}$Tl from laboratory walls \label{external}}

Photons from \TL\ decay are the most important external background, therefore it is illustrative to examine the interaction cross sections for  photons of 2614\,keV in matter. Table \ref{table00} shows the fractional probability of 2614 keV photon interactions in xenon, stainless steal and lead. About 80\% of the external photons will interact in a 5\,cm--thick steel shielding, while almost all of them will be absorbed in a 15\,cm--thick lead shielding. Conversely, once in gas most of the photons will cross the chamber without interacting. In all cases the most probable interaction is the Compton scattering. 

\begin{table}[h]
\centering
\begin{tabular}{ccccc}
\hline\hline & Ph. absorption& Compton int. & pair creation & mean path (cm) \\ \hline
Xe &3\% &  82 \% & 15 \%  & 332 \\ 
SS &0.2 \% & 92 \%  & 7 \% & 2.3 \\
Pb &7 \%  & 69 \%  & 22 \% & 1.4\\ \hline\hline
\end{tabular}
\caption{\small Fractional probability of any interaction of 2614 keV photons in xenon, stainless steel and lead. Mean path in media is also presented.}\label{table00}
\end{table}

The 2.614 MeV $^{208}$Tl photon flux measured in the LSC is 0.13 $\gamma$/cm$^2$/s, distributed homogeneously and isotropically in a sphere of 1.2\,m surrounding the detector and shielding setup.
As expected from the above discussion, for a light shielding of 5 cm lead the number of background events per year is intolerably high, of the order of several thousand. A thicker lead shielding is needed to reduce this background down to a negligible level compared to the internal backgrounds. This is achieved with a shield of 15 cm, leading to 0.5 counts per year due to the \TL\ external background. 

\subsection*{The problem of shielding}

The thick lead shield of 15 cm, needed to reduce the background emanating from the laboratory walls below one count, implies about 16 tons of lead. Even assuming low-activity lead (1 mBq/kg) and taking into account the high degree of self-shielding, the resulting background, summarized in Table \ref{table2} is much larger than the internal NEXT background. 


The conclusions are not surprising, but sobering. Shielding is a major issue for an experiment like NEXT, and the collaboration is exploring different possibilities to suppress external background.

\begin{table}[h]
\centering
\begin{tabular}{ccccc}
\hline\hline
\multicolumn{1}{c}{Source and}&\multicolumn{4}{c} {Events (counts/keV/kg/year)$\times 10^{-3}$}\\
  origin & no cuts & fiducial & one track  &all \\ \hline
$^{208}$Tl (shielding)&  1840.4 $\pm$0.4& 1438.4$\pm$0.3 & 41.06 $\pm$0.05 & 7.17 $\pm$ 0.02 \\
$^{214}$Bi (shielding) & 16.00$\pm$0.01& 13.92$\pm$0.00 &3.22$\pm$0.00& 0.63$\pm$0.00 \\ \hline
     Total shielding & 1856.4$\pm$0.4& 1452.3$\pm$0.3 & 44.28 $\pm$0.05 & 7.8 $\pm$ 0.02 \\\hline\hline
\end{tabular}
\caption{\small Events in the detector due to $^{208}$Tl and $^{214}$Bi (1 mBq/kg)  in lead shielding, according to the different cuts.}\label{table2}
\end{table}

\subsection*{Underground muons}
Underground muons are able to produce high energy photons. Cosmic muons passing through matter lose energy mainly by
ionization, which is transferred to electrons in the traversed
medium. In addition, there are radiative losses which result from
muon bremsstrahlung, direct pair production by muons and
muon-nuclear interactions.

At the LSC underground muons present a mean energy of 290 GeV 
and a flux intensity of around $5\times10^{-7}$ (cm$^{-2}$s$^{-1}$). The differential 
spectrum is constant for small energies  but steepens to reflect the surface muon spectrum 
for energies higher than 500\,GeV with an angular dependence for the muon intensity of
$I(\theta)=I(0)\cos^2\theta$, $\theta$ being the vertical angle.

Our studies show this background to be largely negligible, and it can be further reduced with the use of active vetoes.  

\section*{Neutron activation}

Sea level neutrons are able to activate radioactive isotope in the detector gas and materials. We have studied this production in copper and xenon. For a time $t_E$ of exposure and a decay time $t_D$ underground, the number of radioactive nuclei is calculated as
\begin{equation}\label{eq1}
    N(t_E+t_D)=\frac{A}{\lambda}\big(1-e^{-\lambda t_E}\big)e^{-\lambda t_D},
\end{equation}
where $\lambda$ is the decay constant of the radioactive isotope and $A$ the activation computed with a MC code based on semi-empirical formulas using the sea level neutron spectrum.
The study shows that very few of the produced radioactive isotopes will decay emitting an energy above 2\,MeV and almost none of them in the form of an isolated photon or electron. Therefore, even a cosmoisotope such as $^{60}$Co, very disturbing for some of the $\beta\beta$ experiments, can be neglected since it decays $\beta^-$ emitting also two photons. The only worrisome cosmoisotope is $^{56}$Co produced in the reaction $^{63}$Cu(n, 4n+$\alpha$)$^{56}$Co with a half life of 77 days, which produce energetic photons. However, due to its short half life, the computed contribution is the order of 0.3\,mBq during the first year stored underground and around 10\,$\mu$Bq during the second year.

Once materials and gases are stored underground, the radioactive capture of medium energy neutrons can produce radioactive isotopes. The number of radioactive nuclei produced after a time $t$ is 
\begin{equation}\label{eq2}
    N(t)=\frac{A}{\lambda}\big(1-e^{-\lambda t}\big),
\end{equation}

Assuming a Watt evaporation spectrum for both fission and $\alpha$-n neutrons, normalized to LSC neutron flux, $3.8\times10^{-6}$\,cm$^{-2}$s$^{-1}$, we have computed the interaction production rate for $^{137}$Xe, the only concerning isotope produced in $^{136}$Xe activation, to be 0.6 nuclei per year and kg. This number is equal to the activity due to the short half life of $^{137}$Xe, 3.8\,min. So, in case of 100\,kg of $^{136}$Xe, we could have around 60 decays per year. However, not all them are dangerous: only those able to create single track with an energy in the resolution window.
The $^{137}$Xe isotope presents two main decays: a beta emission with $Q_{\rm max}$=4173\,keV (67\% of the times), and another beta emission with $Q_{\rm max}$=3717.5\,keV (30\% of the times) plus a photon of 455.5\,keV with a mean life of 3.8\,min. In case of our 100\,kg experiment, simple calculations reduce to 0.5 events per year the concerning decays for a resolution of FWHM=1\% (to 1.6 events per year  for FWHM=3\%). The use of 20\,cm (40\,cm) of neutron shielding would decrease by 1 (3) orders of magnitude these numbers.

\section{Physics case of the NEXT experiment}
Taking into account background estimates, signal detection efficiency and background rejection factors from previous sections, we estimate in this section the sensitivity of the NEXT experiment to the \bbonu\ process, assuming the exchange of a light Majorana neutrino. This physics potential is expressed in terms of sensitivity or exclusion plots, which quote a sensitivity to the effective parameter \mbb\ (defined in Chapter 1) under some experimental conditions (exposure, energy resolution, background levels,...). Results are given for a 90\% confidence level (C.L.). The Nuclear Matrix Element used to compute \mbb\ is taken from Reference \cite{Rodin:2007}.

\begin{figure}[p]
\centering
\includegraphics[width=0.5\textwidth]{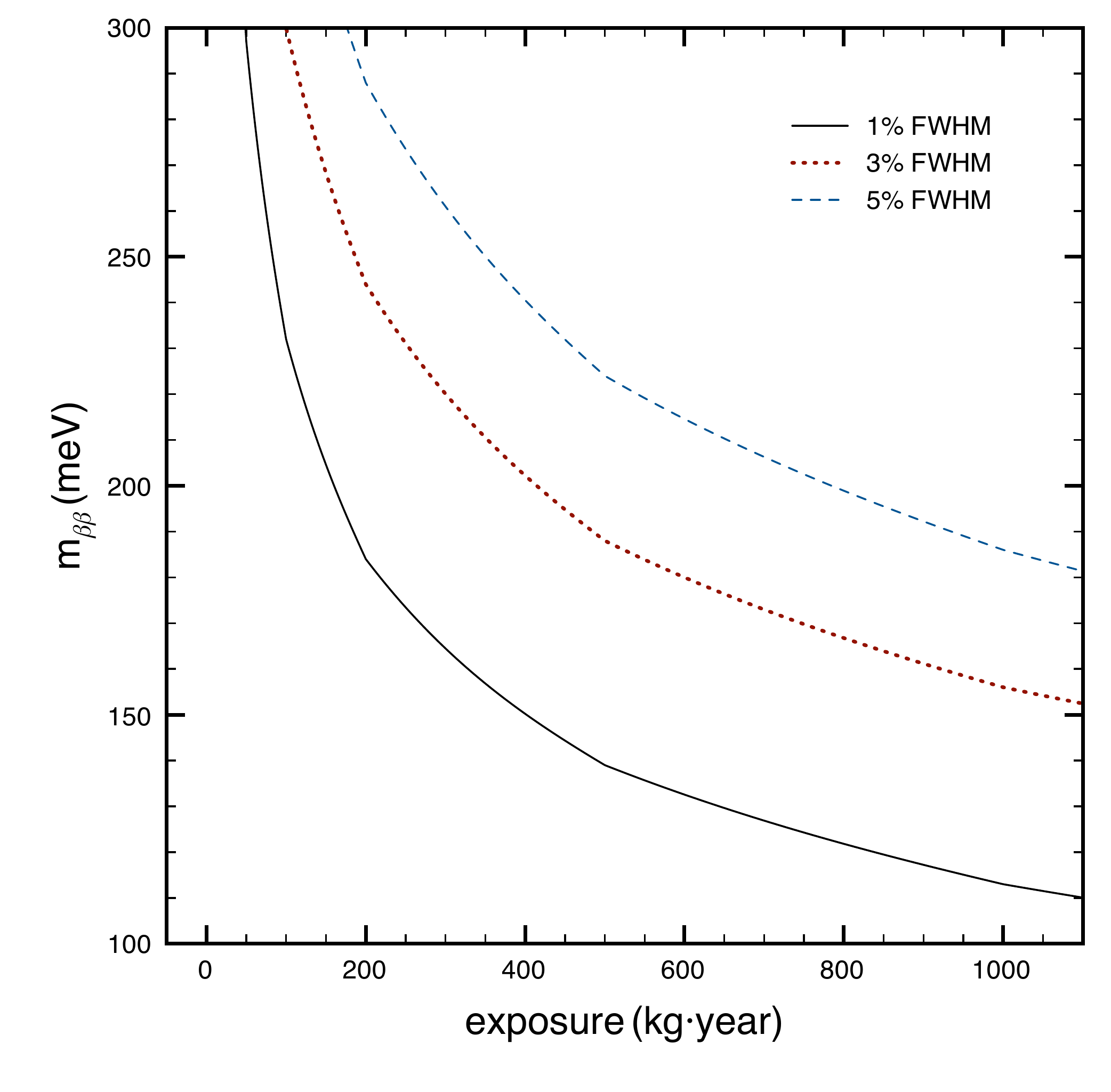}
\caption{\small Sensitivity to m$_{\beta\beta}$ at 90\% C.L. of the SOFT ASTPC as a function of the exposure. Solid, dashed and dotted lines show results for energy resolutions of 1\%, 3\% and 5\% FWHM at \Qbb, respectively.}
\label{fig:sensi-eres}
\end{figure}

The best case for the NEXT experiment is that one in which all external background have been suppressed down to a negligible level by means of shielding. For this particular scenario, the sensitivity achieved by the asymmetric SOFT TPC can be seen in Table \ref{tab:sens} and Figure \ref{fig:sensi-eres}, where the effect of the energy resolution is also shown. 

\begin{table}[bt]
\begin{center}
\begin{tabular}{ccc}
\hline\hline
Exposure (kg$\cdot$year) & $T_{1/2}^{0\nu}$ (10$^{25}$ y) & \mbb\ (meV) \\ \hline
100   & 2.11 & 232 \\
200   & 3.36 & 184\\
500   & 5.89 & 139\\
1000 & 8.91 & 113\\
\hline\hline
\end{tabular}
\caption{\small Sensitivity to \bbonu\ half-life and \mbb\ (according to central value of Nuclear Matrix Element quoted in \cite{Rodin:2007}) of the NEXT SOFT TPC. \label{tab:sens}}
\end{center}
\end{table}

\subsection{Conventional TPC versus SOFT TPC}

\begin{table}[bt]
\begin{center}
\begin{tabular}{ccc}
\hline\hline
Radioactivity source & \TL\ (mBq) & \BI\  (mBq) \\ \hline
Cathode & 375 & 500 \\ 
Anode & 375 & 500 \\ 
Vessel & 740 & 740 \\ \hline \hline
\end{tabular}
\caption{\small Radioactivity budget for the conventional STPC. \label{tab:sys}}
\end{center}
\end{table}%

\begin{figure}[p]
\centering
\includegraphics[width=0.45\textwidth]{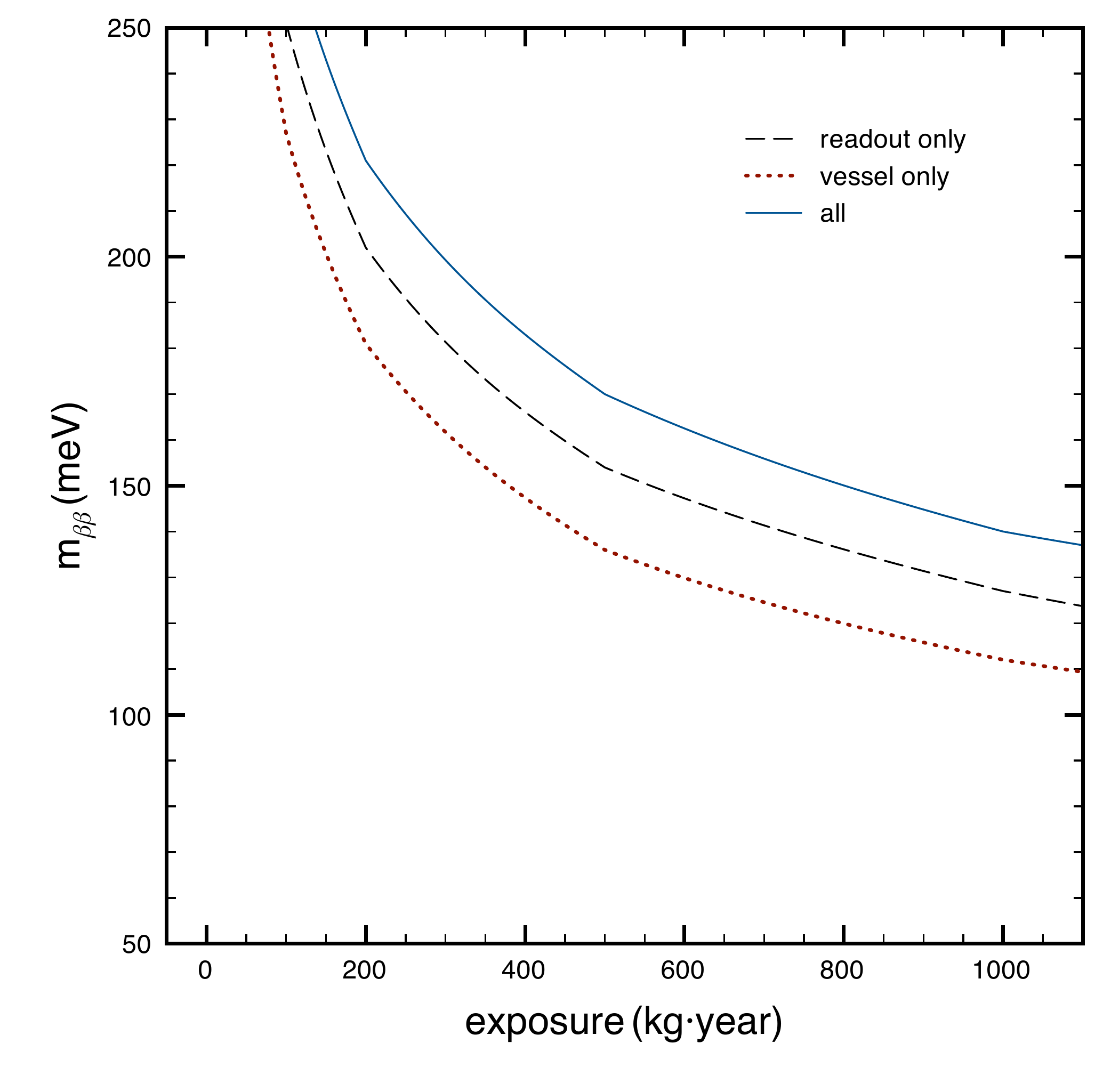}
\includegraphics[width=0.45\textwidth]{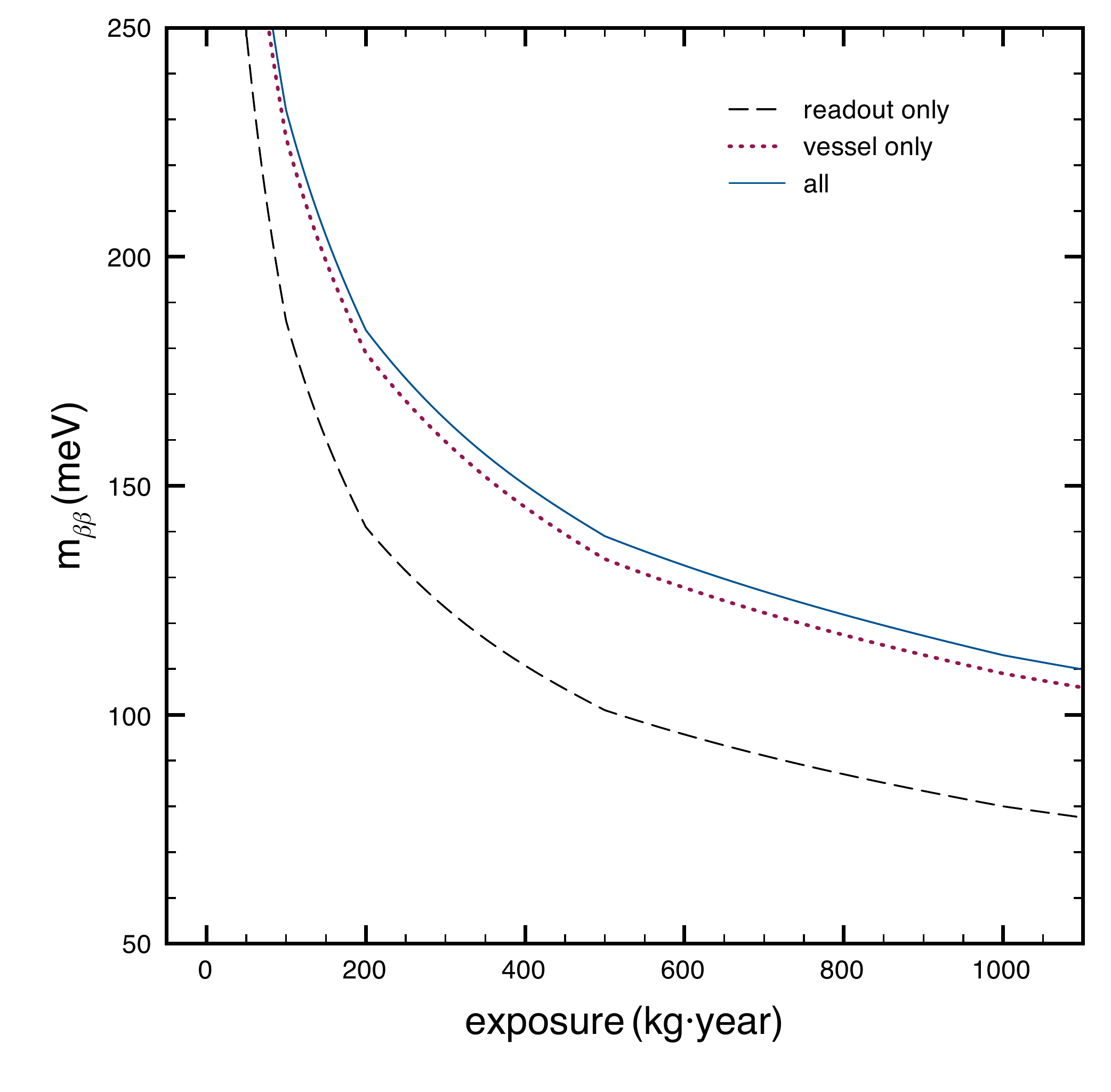}
\caption{\small Sensitivity to m$_{\beta\beta}$ at 90\% C.L. as a function of the exposure, assuming energy resolution of 1\% at \Qbb. Left: sensitivity achieved with the conventional STPC. Right: sensitivity achieved with the SOFT ASTPC. Solid line shows results when all internal backgrounds are considered, meanwhile dashed and dotted lines show results when only contamination from vessel and readout, respectively, are accounted for.}
\label{fig:sensi-sym}
\end{figure}

In the conventional, symmetric TPC described in Chapter 3, tracking and energy measurement are performed with dense arrays of 1'' R8520. About 2,500 of such PMTs are needed, resulting in the radioactivity budget shown in Table \ref{tab:sys}. Notice that the PMTs contribute now as much as the vessel, while in the case of the SOFT ASTPC the dominant source is the vessel. The left panel of Figure \ref{fig:sensi-sym} shows the sensitivity that can be achieved with 90\% C.L. as a function of the exposure. As inferred from this figure, the sensitivity is similarly limited by the backgrounds from the PMTs and the vessel. 

In the SOFT, asymmetric TPC described in Chapter 3, tracking is performed by MPPCs (or as a backup option by Micromegas) and energy measurement is performed with sparse arrays of 1'' R8520. About 250 PMTs are needed, resulting in the radioactivity budget shown in Table \ref{tab:tb}. Notice that now the PMT background contributes much less than the vessel background. The right panel of Figure \ref{fig:sensi-sym} shows the sensitivity (90\% C.L.) corresponding to this experimental setup. As expected, this approach yields much better results than the conventional TPC. 

\begin{figure}[tbhp]
\centering
\includegraphics[width=0.5\textwidth]{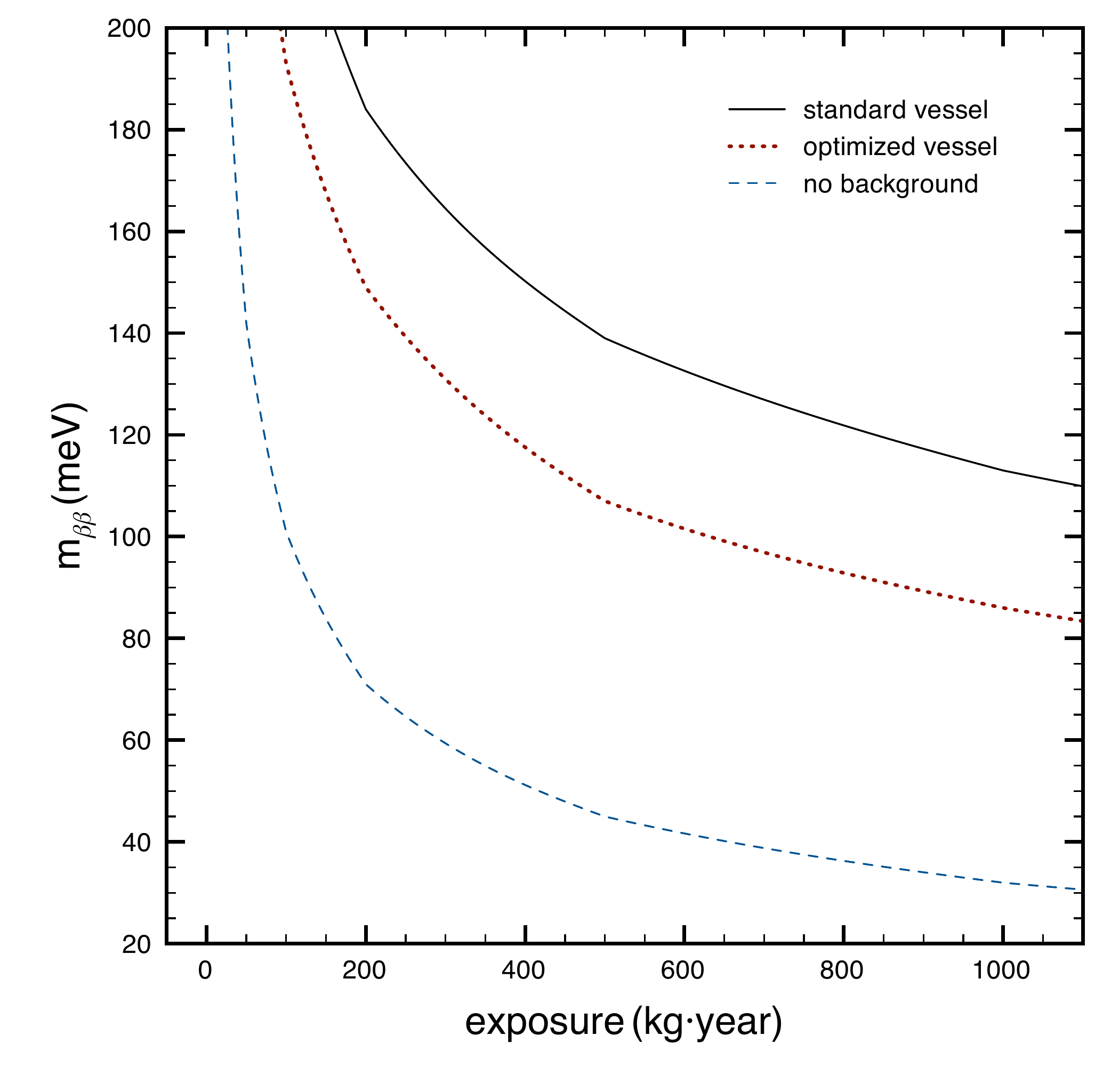}
\caption{\small Sensitivity to m$_{\beta\beta}$ at 90\% C.L. of the SOFT ASTPC as a function of 
the exposure, assuming energy resolution of 1\% at \Qbb. The solid line shows results with
the standard, 1-cm thick steel vessel. The dotted line shows results for an optimized (introducing
ten times less background) vessel. The dashed line represents the sensitivity when all backgrounds
are negligible. }
\label{fig:sensi-final}
\end{figure}

An important conclusion is that no further improvement on the radioactivity budget of the photosensors
will have a significant impact until one manages to reduced the budget coming from the vessel.
This could be achieved by procuring more radiopure steel and/or with alternative designs for the vessel
(steel with an internal lining of ultra-radiopure copper, use of lighter materials, etc).
Figure \ref{fig:sensi-final} compares the sensitivity (90\% CL) of the SOFT ASTPC with
an \emph{optimized} vessel --- ten times less background --- with that with the standard one. Assuming such an optimized vessel, sensitivity of $\mbb\sim$ 110 meV is achieved for an exposure of 500 kg$\cdot$year. For comparison purposes, sensitivity of the NEXT experiment assuming negligible backgrounds (thus depending only on exposure) is shown in same Figure.  

\section{Summary}
We have described the physics case of the NEXT experiment. As it has been proven, the NEXT experiment is capable of exploring the degenerate hierarchy of the neutrino mass, and thus confirming or refuting results from the Heidelberg-Moscow experiment. Moreover, sensitivity is expected to be improved in several ways:

\begin{itemize}
\item By improving the energy resolution (this, although theoretically possible, may be difficult).
\item By fully exploiting the power of the topological signature. We expect to be able to do much better than the conservative 1/30 rejection factor used in this analysis. Preliminary studies show that 1/50 and even 1/100 could be reachable. 
\item By procuring more radiopure steel for the vessel. A reduction of two in \TL\ and at least two in \BI\ should be possible.
\item By adding an inner lining with ultra-radiopure copper (about 1-2 cm of 100 $\mu$Bq copper). 
\item By procuring more radiopure lead for the shield or exploring the possibility of active shields. 
\end{itemize}

The full exploitation of the topological signature, together with careful selection of materials, is expected to reduce the internal background below one count per year. Achieving this level for the external backgrounds will require active shielding.

\chapter{The NEXT Project} \label{R-D}
The goal of the NEXT project is to build and operate the NEXT-100 detector at the LSC. This is an ambitious enterprise, both scientifically and technically, and it requires a strong collaboration, clear objectives and well defined deadlines. In this chapter we describe our current definition of the project, that will, however, keep evolving and improving, as the Collaboration expands and gains experience.



Finding new international partners with expertise in the field is one of the outstanding challenges of the Collaboration. 

\section{Stages of the NEXT project}
We are proposing to build a radiopure, gas high-pressure Xenon TPC based on electroluminescence, with excellent energy resolution and tracking capabilities. One needs to approach to this extremely challenging (and no less rewarding) goal in well planned stages. 
\begin{enumerate}
\item The first one involves acquisition of needed know-how and the equipment of laboratories. We call this stage NEXT-0, and its main deliverables are a set of small prototypes and a number of relevant measurements, including a measurement of the energy resolution for our PMTs, and its dependence with pressure, as well as the demonstration of the concept of tracking with photosensors. NEXT-0 also involves several studies associated with the performance of Micromegas in pure xenon and at high pressure. These setups are very valuable in the future as lab installations to test specific solutions and as characterization setups. Some of the forseen functions include testing of degassings for several materials, details studies of sensor performance and development of calibration techniques.

\item The second stage of the project (NEXT-1), has a clearly defined goal: building a (non-radiopure) TPC, large enough to contain and track electrons in the energy range relevant for \bbonu\ searches (for example those produced by Compton interactions of the 660 keV gamma from $^{137}$Cs). NEXT-1 will demonstrate the key ideas behind the SOFT concept: energy resolution, capability of tracking, topological signature, etc... It will also be essential to address some of the most challenging technical issues: electron attachment, UV light attenuation and wall reflectivity, drift field definition and gas purity. We will also test some of the choices concerning mechanics and electronics and will confirm the baseline option.

\item The third stage of the project is the construction of a radiopure TPC that we call NEXT-10, intended as a 1/10 scale version of NEXT-100. This detector is necessary to test at reasonable large scale all the final solutions for NEXT, gain experience with radiopurity, measure all expected backgrounds for NEXT (running with natural Xenon), and, if accessible, measure the \bbtnu\ mode. NEXT-10 will be operated at Canfranc and it will require a gas, veto and calibration system close to the final one. 

\item The fourth stage of the project is called NEXT-100 and implies the construction and commissioning of the NEXT-100 detector. NEXT-100 implies a larger volume and probably tighter radiopurity requirements than NEXT-10. However, there are few items that can be easyly scaled from the NEXT-10 prototype. These includes the pressurize vessel, the gas system, the calbration, veto system, electronics, and monitoring of gas purity. 

\item The fifth and last stage of the project is to conduct the physics program of NEXT. This is expected to start the first quarter of 2014.  

\end{enumerate}

\begin{figure}[tbhp]
\begin{center}
\includegraphics[width=0.99\textwidth]{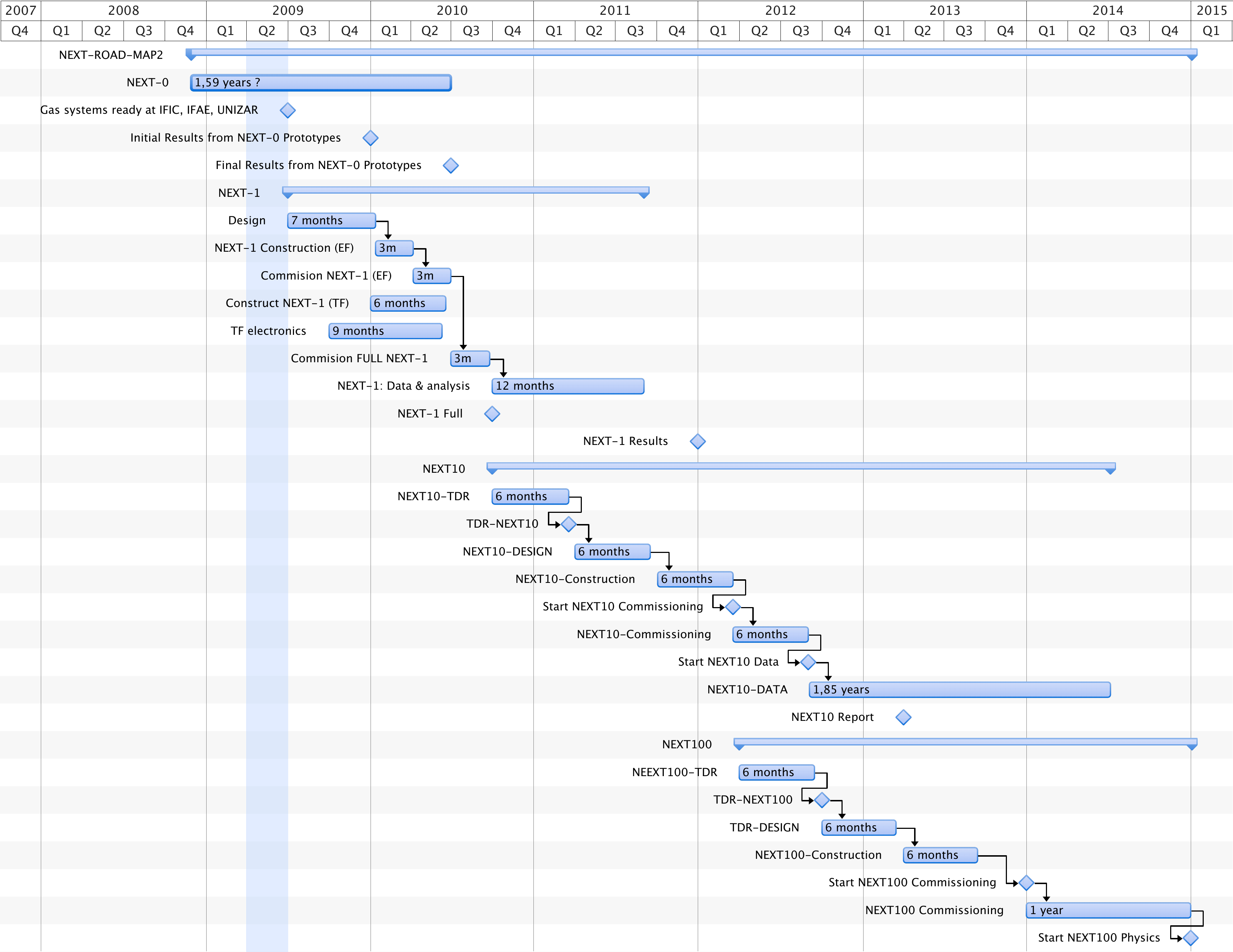} 
\end{center}
\vspace{-0.5cm}
\caption{\small Roadmap of the NEXT project.}
\label{fig:rmap} 
\end{figure}

The schedule  for the different phases of the project is shown in Figure \ref{fig:rmap}. The different phases are overlapped, taken into account that most of the phases play a role even after the next phase is started. Some obvious overlaps are comming from the development of common infrastructure in Canfranc like the gas system.

In the next sections, we present a brief summary of the different stages of the project. 

\section{NEXT-0}

The first stage of the project focuses around studying several key questions for NEXT using small prototypes built at IFIC, IFAE and UNIZAR.

\subsection{Energy resolution measurements}
\begin{figure}[tbhp]
\begin{center}
\begin{tabular}{c}
\includegraphics[width=0.8\textwidth]{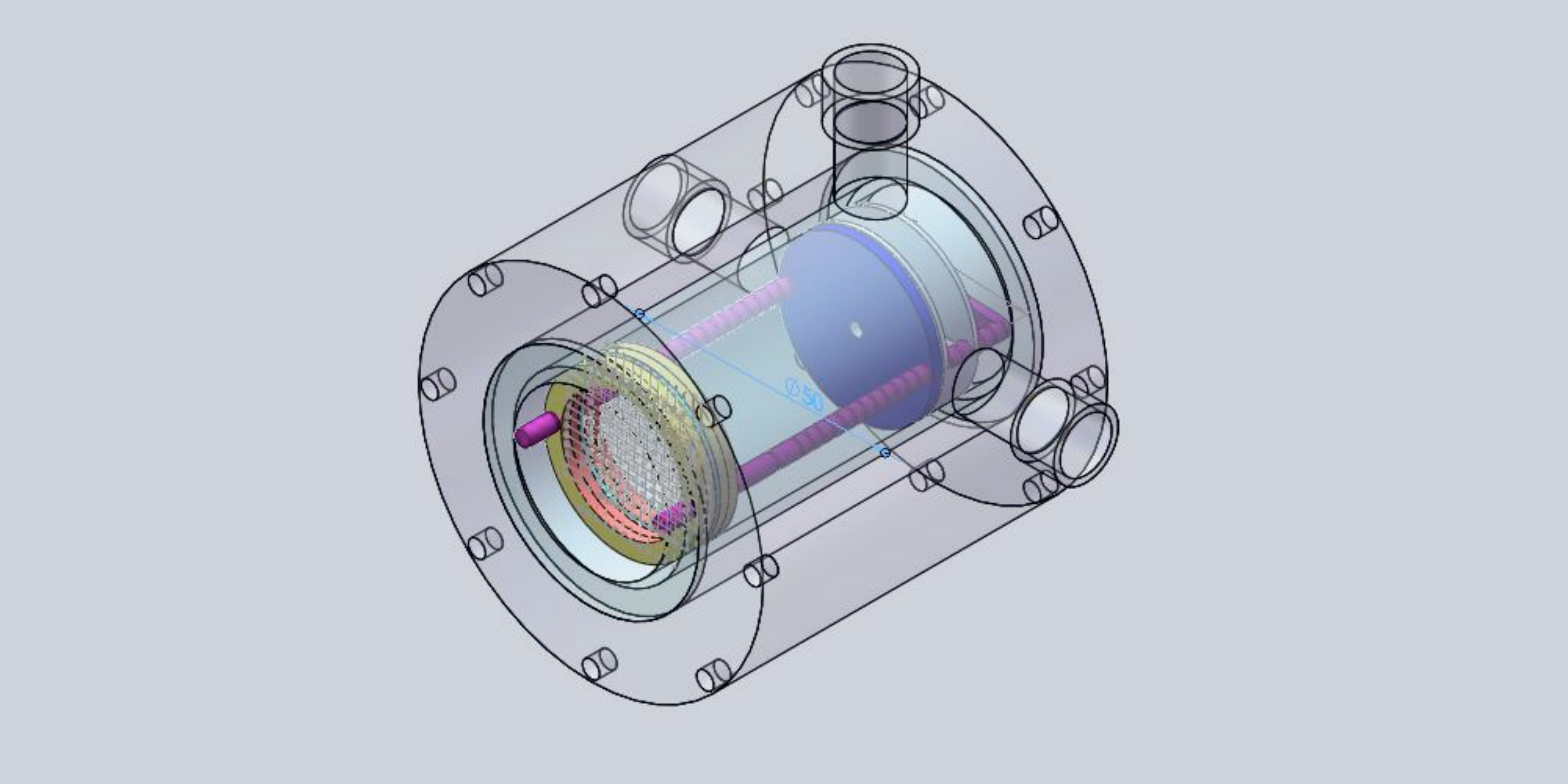} \\
\includegraphics[width=0.8\textwidth]{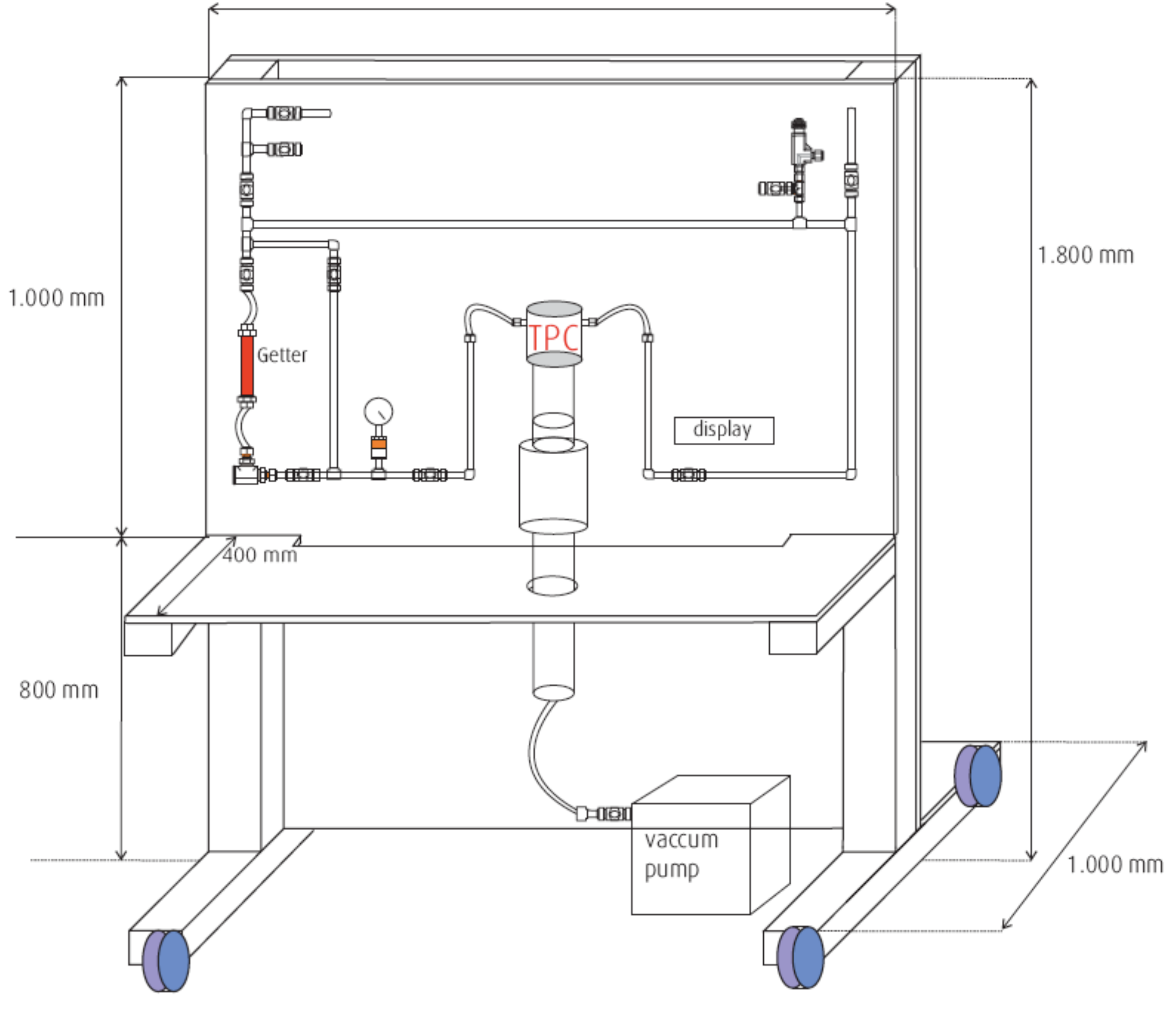} 
\end{tabular}
\end{center}
\vspace{-0.5cm}
\caption{\small Top: a scheme of the IFIC-0 Chamber. Bottom: the chamber together with the gas system.}
\label{fig:ific0} 
\end{figure}

IFIC is concentrated in understanding the measurement of the energy in NEXT with resolution of at least 1\% FWHM at \qbb. The first step in this program is to measure the resolution as a function of the energy (using different gamma sources) and as a function of the pressure, using a small 1-PMT chamber, called IFIC-0
(Figure \ref{fig:ific0}, top). The PMT (a R8520 from Hamamatsu Photonis) is coupled to the chamber using an optical window capable to withstand pressures of up to 15 bar. The setup includes a close gas system (Figure \ref{fig:ific0}, bottom) and is capable of reaching a vacuum in the range of $10^{-7}$~torr, essential to 
guarantee initial good gas purity.

The system has been designed by IFIC, UPV \& CIEMAT and is under construction (part of the system is built by Vacuum systems, a Valencia-based company and another part is built at CIEMAT). The gas system has been developed in collaboration between IFIC and Abelló-Linde. We expect to start commissioning the chamber and the gas system before the summer 2009. Many operational parameters will be tested: operative EL voltage, gas purity, energy resolution as a function of pressure. A first study of the reflectivity of TTX will also be made.

This studies will also be conducted at Coimbra, which will use a different approach (the PMT window will be placed inside the chamber). The combined results should lead to at least one publication before the end of the year.

The measurement of energy with 1\% resolution requires a refined calibration of the PMTs. IFIC has already acquired 18 R8520 (needed for NEXT-1). These will be calibrated during the next few months at IFIC. The readout electronics for the PMTs will be developed by the ITACA group (UPV), and the on-line software and DAQ by CIEMAT. CIEMAT will also study how to reinforce the R8520 to make it capable of standing at least 10 bar. 

\subsection{Tracking function with photosensors}

\begin{figure}[tbhp]
\begin{center}
\begin{tabular}{c}
\includegraphics[width=0.6\textwidth]{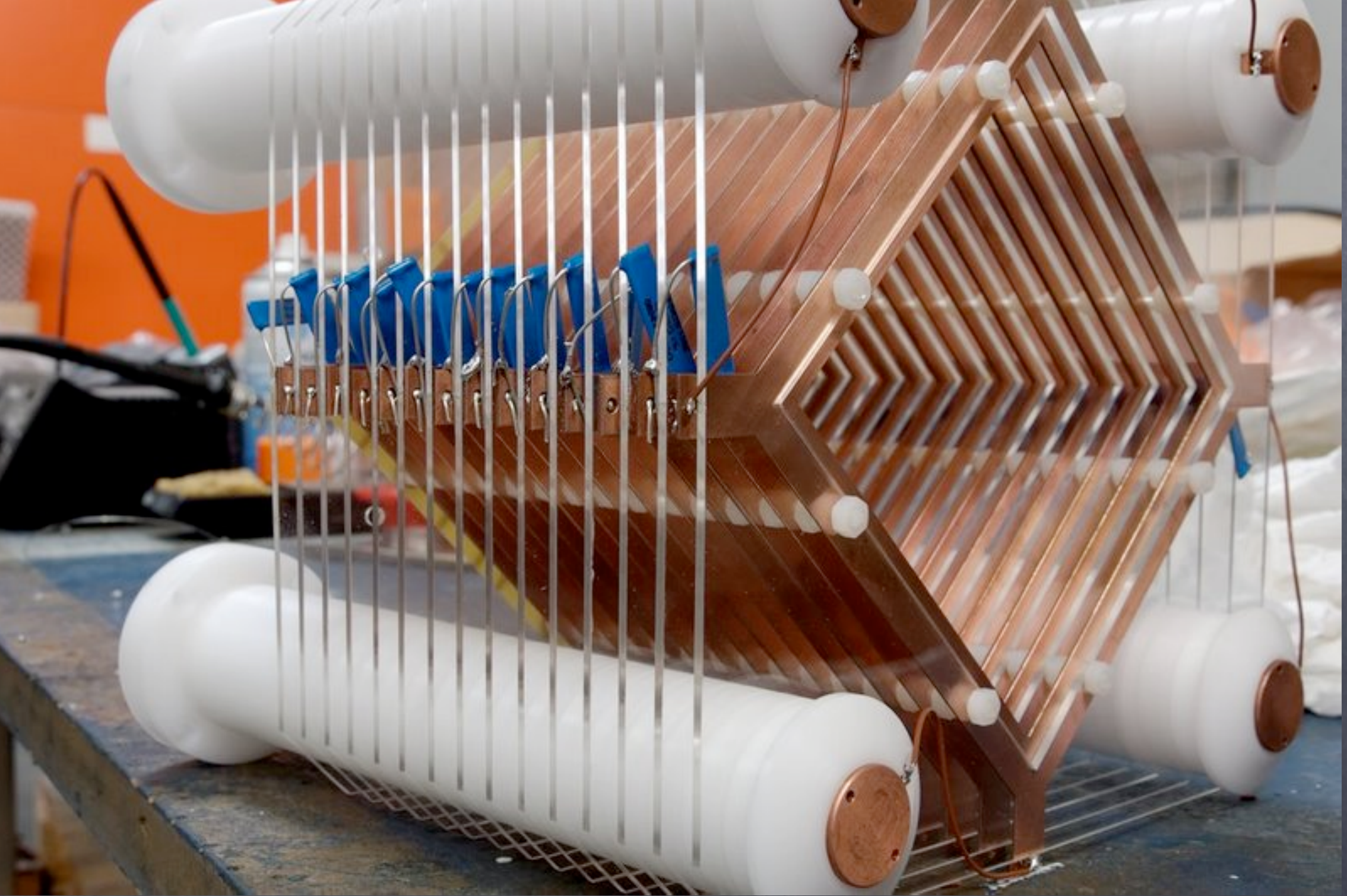} \\
\includegraphics[width=0.6\textwidth]{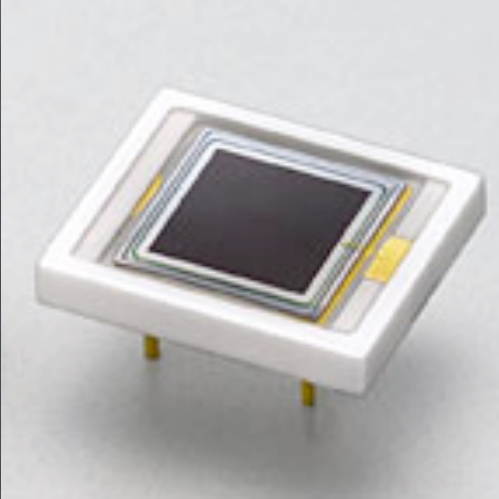} 
\end{tabular}
\end{center}
\vspace{-0.5cm}
\caption{\small Top: a detail of the field-cage of the IFAE chamber. Bottom: Hamamatsu APD.}
\label{fig:apd} 
\end{figure}

Building a tracking system based on EL and photosensors is a new concept. Before building NEXT-1, we want to test as many aspects as possible of this concept.

To this extend, few MPPCs from Hamamatsu will be characterized at IFIC, Coimbra and CIEMAT, addressing also the issue of shifting the light from UV to Blue (coating with TPB or WLS crystals are two interesting possibilities). 

At the same time, we will use the prototype chamber at IFAE (IFAE-0) to study a prototype tracking pixel plane that will use first LAAPDS and later MPPCs.
The prototype IFAE-0 has been already contructed and it is being commissioned. The chamber, see Fig.~\ref{fig:apd}, is able to sustain 10 atmospheres and has an inner active volume of 10$\times$10$\times$15 cm$^3$ that is enough to identify electron tracks of low energy.  IFAE is also contructing a small chamber to be able to study degassing properties of materials and check the performance of single photosensors. The IFAE-0 setup also includes a close gas system developed together with the IFIC group and based on the experience of a similar setup at the Coimbra University.  Large Area Avalance Photo Diodes (LAAPDS) are an interesting option as tracking photosensors due to their large area, large Q.E. to the UV (no WLS needed) and pressure resistance. There exists considerable experience in the use of this sensors both at IFAE and Coimbra, and we have around 16 units available from Hamamatsu. It is then possible to build a small tracking plane in a short time, avoiding some of the complications (such as WLS) that must be carefully understood with MPPCs. Our goal is to understand tracking (sensors intercalibration, position resolution, transverse dispersion of the EL light, ...) with proven technology (APD) using the prototype chamber at IFAE(Fig.\ref{fig:apd}). In parallell, we will develop the techniques for detecting DUV light with MPPCs.

\subsection{Studies of performance of Micromegas}

\begin{figure}[tbhp]
\begin{center}
\includegraphics[width=0.6\textwidth]{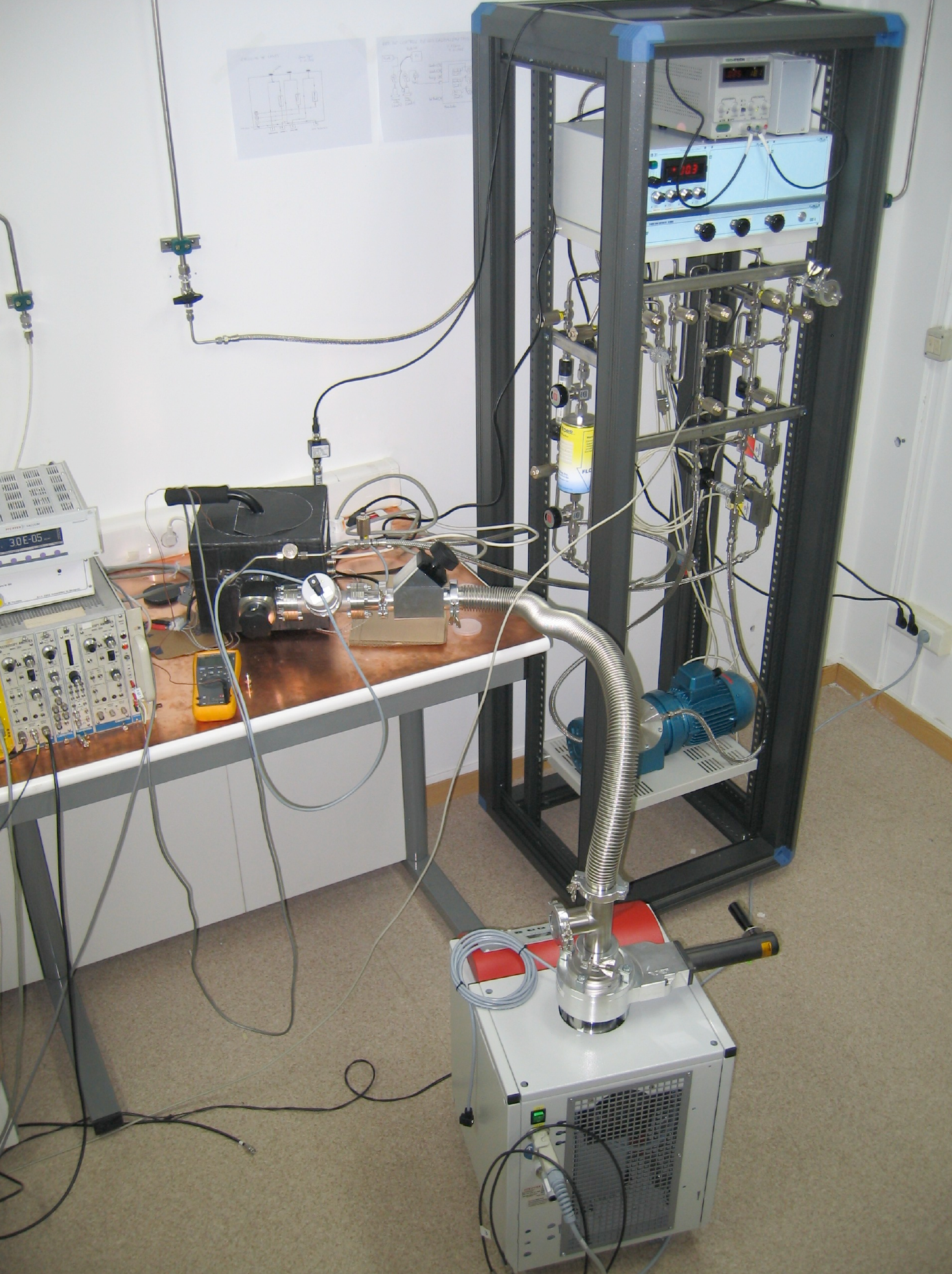} 
\end{center}
\vspace{-0.5cm}
\caption{\small The setup for NEXT-0 studies at UNIZAR.}
\label{fig:setup} 
\end{figure}

The UNIZAR group has initiated an R\&D program whose main deliverable during the NEXT-0 phase is the demonstration that Micromegas can be operated in an stable way in pure xenon and at 10 bar pressure. Other goals include the investigation of the possible use of a quencher (for example a small amount of CF$_4$), which does not degrade resolution, does not quench the UV light and improves operation, including the possibility of a smaller transverse diffusion. These studies involve a detailed understanding of gas purity and the control of impurities that can cause electron attachment and are, therefore, of uppermost importance, not only for the Micromegas development but for the general understanding of the NEXT TPC.

\subsection{Gas system}

\begin{figure}[tbhp]
\begin{center}
\begin{tabular}{c}
\includegraphics[width=0.8\textwidth, angle=-90]{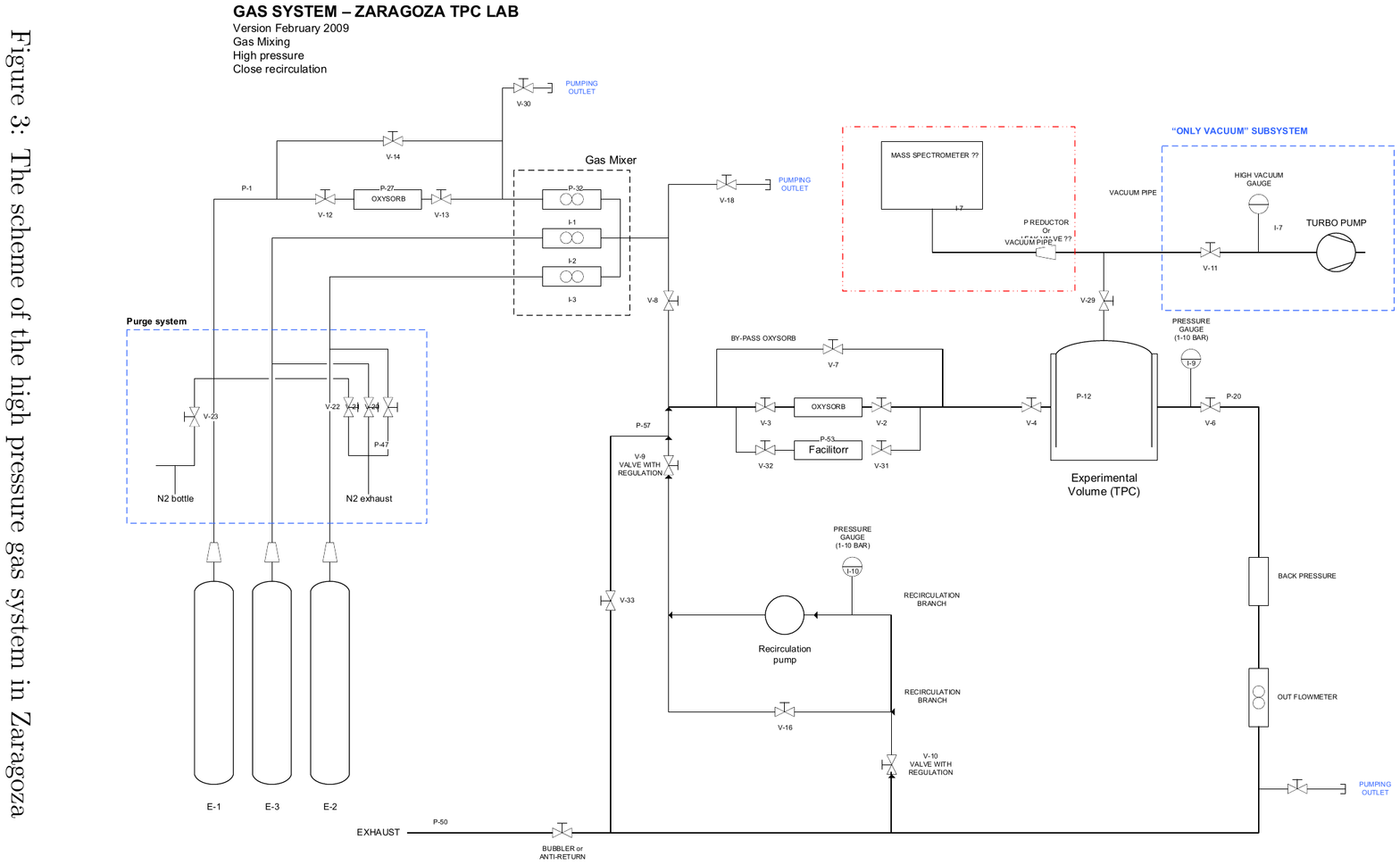} \\
\includegraphics[width=0.8\textwidth]{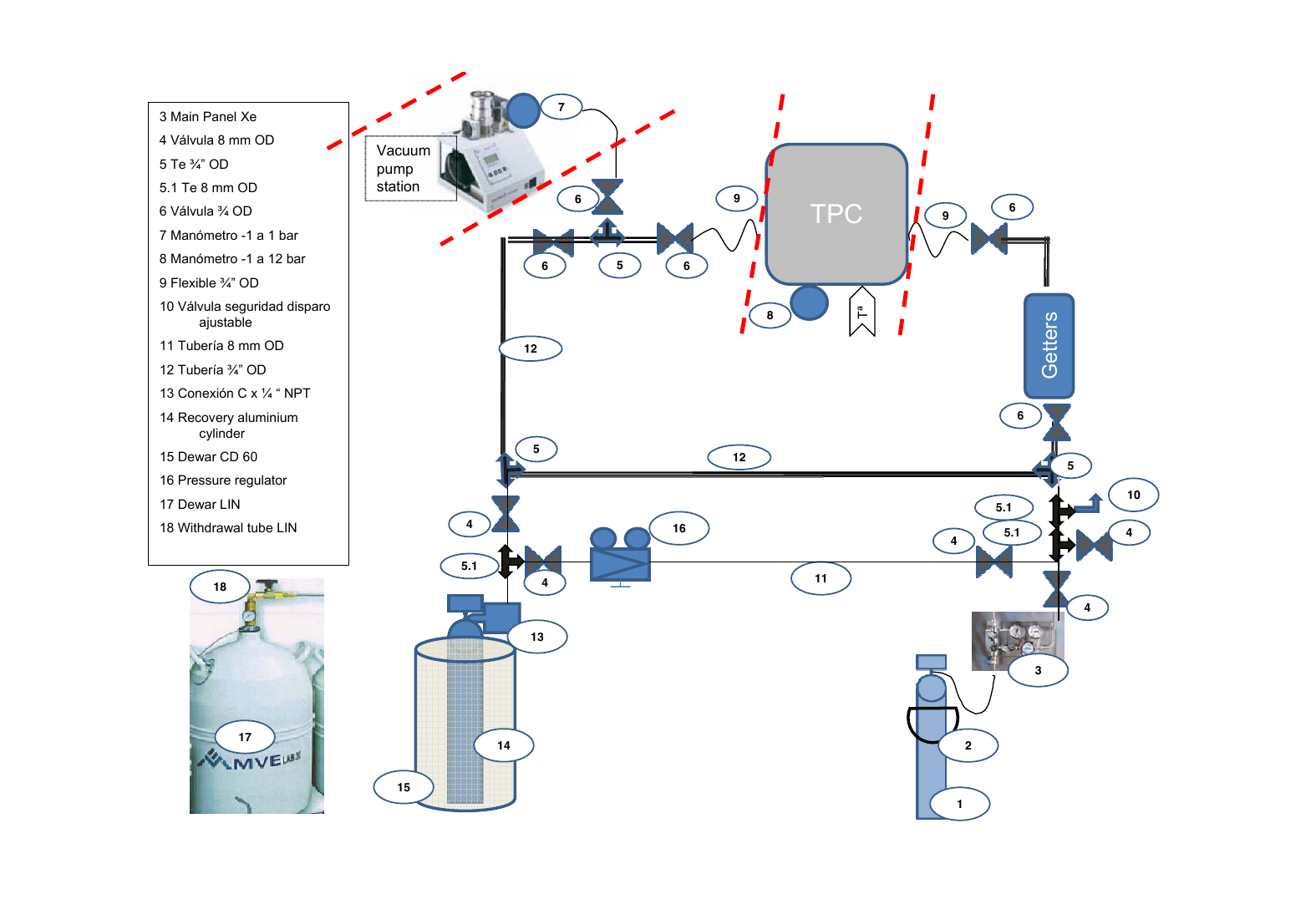} 
\end{tabular}
\end{center}
\vspace{-0.5cm}
\caption{\small Top: a scheme of the gas system at UNIZAR. Bottom: a scheme of the gas system
at IFIC/IFAE.}
\label{fig:gas} 
\end{figure}

The gas system is one of the key elements that must be controlled by the experiment. Figure \ref{fig:gas} shows
the schemes of the gas systems at Zaragoza and IFAE/IFIC. Both systems try slightly different approaches that will be throughly tested, so that the gas system to be installed at the LSC (for NEXT-10) incorporates all the gained know-how. The gas system in Zaragoza uses a pump to recirculate and pressurize the gas circuit, while IFIC and IFAE uses the convection of the gas from a hot point in the circuit that was succesfully used at the Coimbra University for years. The other difference appears during the xenon recovery procedure. IFAE uses schemme closer to that of Coimbra while IFIC uses a commercial product. The three cases have a similar cleaning procedure based on hot getters. 

\subsection{Software and simulation}

Simulations are fundamental at this stage of the project to help during the design of the different prototypes. Issues like the proper pixelization, the impact of the walls reflectivity, the dynamic range of the photosensors, shapping and sampling time of the electronics, etc. The main deliverables of the Software project in NEXT during the NEXT-0 phase are:
\begin{enumerate}
\item To define and implement the event data model for NEXT.
\item To develop a Monte Carlo simulation capable of generating the basic physics, detector response and digitisation.
\item To develop a prototype framework for analysis.
\end{enumerate}

Most of this deliverables are well under way. For example, a detailed Monte Carlo simulation developed at IFIC and UNIZAR has already produced the large data used in the physics studies presented in Chapter 4.

Simulations will also allow us to understand the response of the various prototypes used to study the energy resolution and tracking function in NEXT.

\subsection{Radiopurity}

The following milestones have been fixed for NEXT-0 phase:
\begin{enumerate}
\item  To search for available information on radiopurity of relevant materials in existing databases and
compilations of data.
\item To create a database to manage the requests of new measurements and to store the results.
\item To assess the need and possibility of performance of other techniques than gamma spectroscopy with
HPGe detectors underground for measuring radiopurity of components.
\item To determine the availability of existing detectors owned by the group of the University of Zaragoza and operated at the Canfranc Underground laboratory (several Ge and NaI detectors) for screening materials related to NEXT and to perform some measurements if possible.
\end{enumerate}

The last point is of particular importance. We need to start building the NEXT detector no later than 2011, and the LSC main installations will not be available until 2010.  We believe that is mandatory for NEXT to find a solution to operate in the old LSC installations, using existing detectors, in order to gain experience and assess the radiopurity o some of the key components of the detector (PMTs, MPPCs, Micromegas, among others).  

\section{NEXT-1}

The goal of the NEXT-1 phase of the project is to demonstrate that NEXT can achieve both the energy function with the target resolution of at least 1\% and the tracking function. A non-radiopure prototype will be built and commissioned in 2010, and operation will extend until 2011. NEXT-1 will remain a test facility so we can continue some of the developments on the scope of NEXT-100 while building the NEXT-10 in case the nominal performance is not achieved. 

\subsection{NEXT-1 prototype dimensions and equipment}

\begin{figure}[tbhp]
\begin{center}
\includegraphics[width=0.7\textwidth]{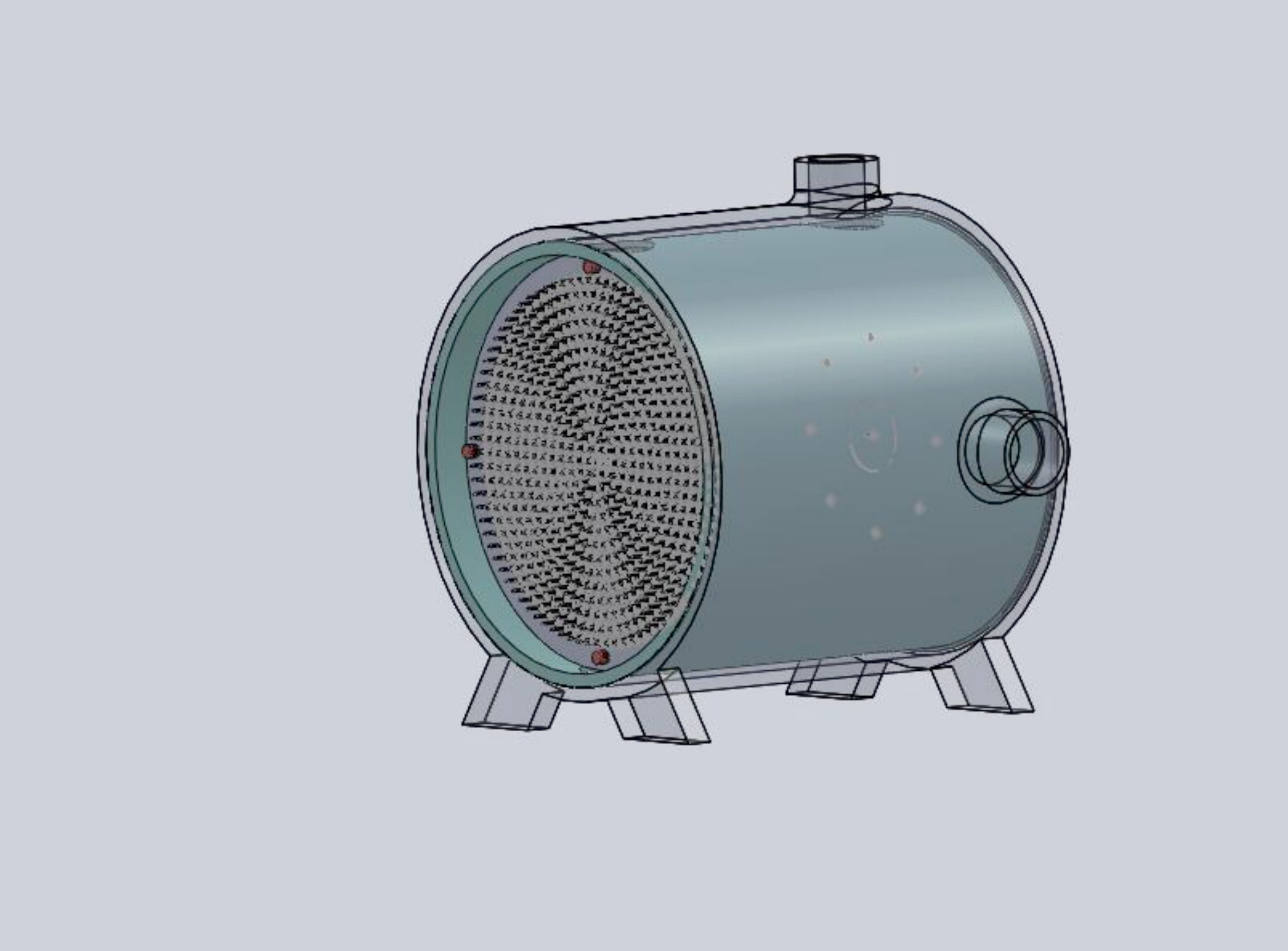} 
\end{center}
\vspace{-0.5cm}
\caption{\small A sketch of NEXT-1.}
\label{fig:next1} 
\end{figure}

NEXT-1 will be a SOFT ASTPC capable to withstand an operative pressure of at least 10 bar. The active surface will be a cylinder of R=20 cm and L=30 cm, for a total mass of natural Xenon of about 2 kg. A sketch is shown in 
Figure \ref{fig:next1}.
\begin{figure}[tbhp]
\centering
\includegraphics[width=0.7\textwidth]{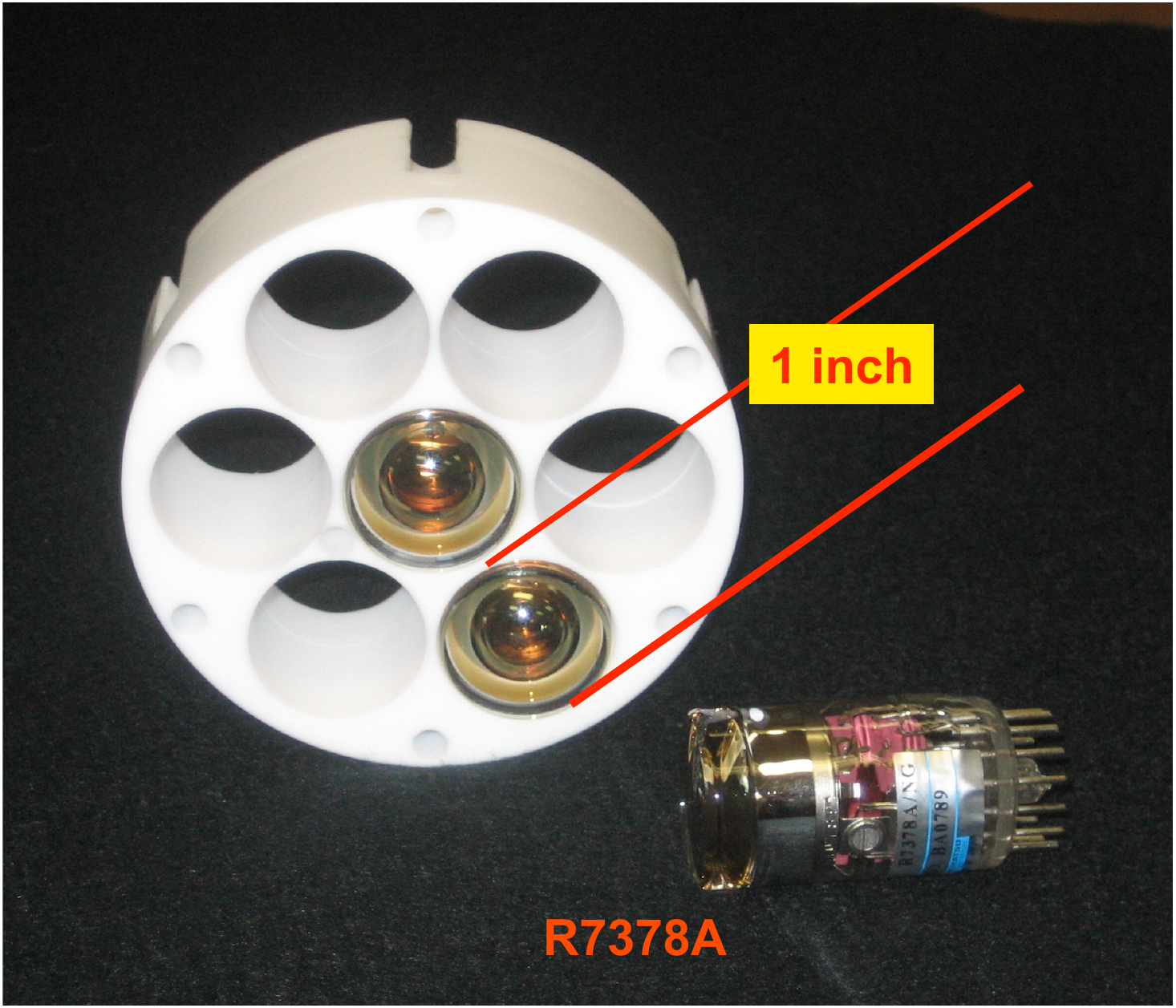}
\caption{\small Energy plane of a prototype TPC (J. White, TAMU).}
\label{fig:white1}
\end{figure}
\begin{figure}[tbhp]
\centering
\includegraphics[width=0.8\textwidth]{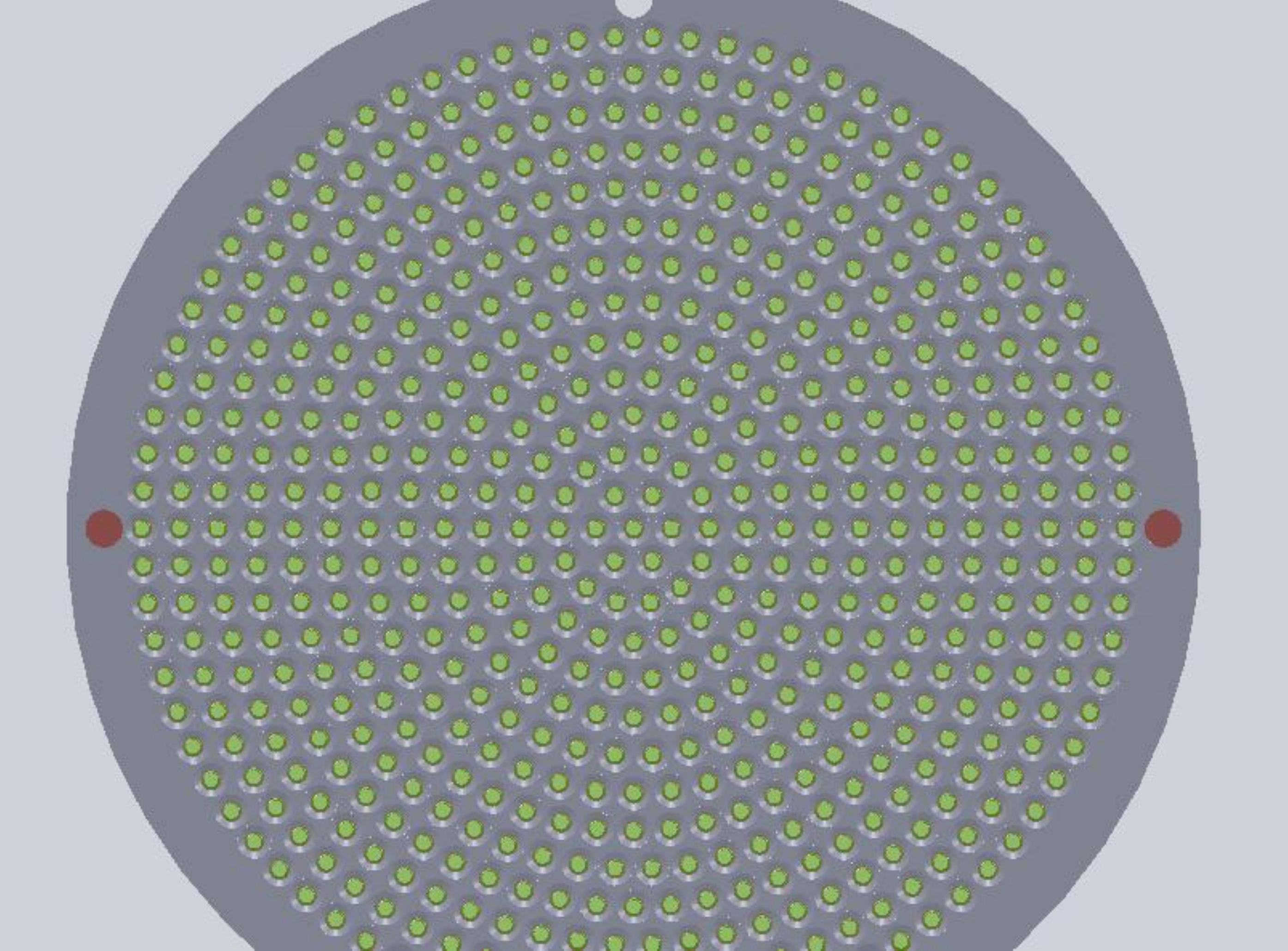}
\caption{\small Conceptual drawing of the SiPMT plane. Notice the small sensors separated by a pitch of 1 cm.}
\label{fig:sipmt}
\end{figure}

The dimensions of NEXT-1 are sufficient to fully contain Compton interactions of the 660 keV gamma from $^{137}$Cs. The maximum drift length of 30 cm is also sufficiently long to measure electron attachment and UV light attenuation. The inner walls of the field cage will be covered with a high-reflectivity material, probably TTX. A plane of 18 PMTs will be located behind the transparent cathode to read the scintillation and EL light (both direct and reflected). See Figure \ref{fig:white1} for a similar setup. The EL grids will be separated by 5 mm and will operate at an $E/P$ of 4 kV$\cdot$bar$^{-1}\cdot$ m$^{-1}$. A tracking plane of MPPCs, containing about 400 sensors at a pitch of 1~cm will be installed behind the anode (Figure \ref{fig:sipmt}). If the results of the NEXT-0 phase of the project are encouraging concerning the operation of Micromegas in pure Xenon at high pressure, we will also test a plane of Micromegas with a similar pixelization.

NEXT-1 will demonstrate convincingly the capability of NEXT-1 to achieve at least 1\% energy resolution, convoluting not only the intrinsic Fano factor (one would extrapolate a value as low as 0.4\% FWHM at \Qbb\ from this fluctuation source) but also several other effects: light attenuation (due to the presence of impurities, such as water), fluctuations due to variations on reflectivity, and fluctuations associated to the calibration of the PMTs and readout electronics.

\subsection{Technical issues}

During NEXT-1 phase, we should also demonstrate that:

\begin{enumerate}
\item One can drift the ionization charge in pure Xenon for at least 1 m without degrading the performance due to attachment.
\item Photosensors and electronics short-term stability. 
\item Water and other contaminants are present in so small proportions that UV light is not attenuated in the TPC.
\item Reflectivity of the TPC walls is large anough for our purposes. 
\item Signals can be taken out via feedthroughs or preprocessed with front-end electronics near the readout planes.
\item Gas operation.
\item Long-term stability like degradation of reflective walls or photosensor response degradation due to DUV radiation.
\item Uniformity and stability of the drift electric field.
\end{enumerate}

\subsection{Gas system}

 The gas system will be very close to those used in the NEXT-0 prototypes. The larger inner volume might require large recirculation flux but no big step is expected at this phase. The main activities during this phase of the project will concentrate in the design of the NEXT-10 gas system. This includes most probably tools to calibrate gas properties and monitor their stability.

It is also relevant at this stage of the project to devote some effort to the definition of the sequrity protocols for the operation of vacuum and pressurize systems inside the laboratory. 

Coimbra University has been contributing to the gas system for the XENON collaboration and they are in a good position to play a leading role in this project.

\subsection{Calibration}

 NEXT-1 will be the first prototype to exercise calibration concepts. We need to perform four important tasks: 

\begin{enumerate}

\item The intercalibration of the energy measurement photosensors. 

\item The absolute energy calibration. 

\item The intercalibration of the tracking photosensors. 

\item The characterization of the gas properties: photon absorption, electron dispersion, velocity and attenuation.

\end{enumerate}

 As part of the calibration we need to understand also the thresholds obtained from the different technical solutions that have been adopted. NEXT-1 is also the ideal place to test the different calibration solutions that we will adopt for the NEXT-10 prototype. The design of the ancillary around the calibration (electronics, monitor chambers, radioactive source handling, etc\ldots) will take place in this phase previous to the contruction for NEXT-10 prototype. 

Some safety issues like the hadling of radioactive sources has to be developed with the LSC at this stage. 

\subsection{Photosensors}

 During this phase the photosensor mechanical integration and calibration will be developed. It is critical to define the protocols to build large amount of them into large areas under extrict conditions of cleanliness and radiopurity. NEXT-1 can be  used as a prototype to test some mechanical solutions if needed.

\subsection{Electronics}

 The final design of the electronics will be fixed during this phase. Prototypes will be used at the NEXT-1 phase so we can correct design or implementation problems. The next level, NEXT-10, will require a large amount of electronics and we have to consider delays in the construction and testing procedures. This is even more relevant if we finally take the decision to build custom electronics. 

\subsection{Radiopurity campaign}

In 2010 an intense program of radiopurity measurements must be conducted at the LSC with the aim of screening the materials that will be used for the first radiopure prototype (NEXT-10). At this stage we will have already a conceptual design of the NEXT-10 including materials and many of the detector components will be screened in Canfranc: TPC inner volume material, photosensors, electronics, cables, etc\ldots

\section{NEXT-10}

The goal of the NEXT-10 phase of the project is to demonstrate that NEXT can achieve both the energy function with the target resolution of at least 1\% and the tracking function while being a radiopure detector. In addition, NEXT-10 will allow us to measure the backgrounds of NEXT-100. NEXT-10 is expected to have a total fiducial mass of about 10~kg of depleted or enriched Xenon. The inner size of the prototype will be of the order of 0.2 m$^3$.

 Unless we find that the \bbtnu\ mode is accessible the detector will run with natural xenon. NEXT-10 will incorporate solutions that we consider final or close to final for the energy and tracking planes and will be constructed from radiopure components. 

The detector will be used to collect ample statistics to understand our backgrounds (this is easy to achieve by simply not shielding it from the gammas emanating from the laboratory wall), as well as the effect of passive and perhaps active shielding.

NEXT-10 will be built in 2011 and commissioned and operated in 2012 and further on. 

\subsection{ Gas system}

 The gas system is an evolution of the ones installed for the NEXT-1. The location inside the Canfranc laboratory adds complexity to the problem. The main diferences with previous gas systems are: 

\begin{itemize}

\item Large recirculation flux
\item Large recovery containers. Probably the recovery techinque should be changed from liquid Nitrogen to pressure pump. 
\item Technique for gas purification might differ. 
\item Radon purifier might be neeeded. 
\item Enriched Xenon gas will be used so all fast recovery techniques will be on place.

\end{itemize}

The NEXT-10 gas system is almost a one to one scale of the final one. NEXT-10 will be operated in parallel with NEXT-100 for some time and we need separated circulation gas systems.

One of the critical issues to be developed in coordination with the mechanics working group are the safety and the fast recovery system in the account of a large gas leak. Both the early detection and the protocol to minimize loses will be included in the early stages of the mechanics and gas system designs.

NEXT-10 will also be the final test of the ability of the gas system to keep the required purity. The understanding of light absortion and electron attachment will be finally confirmed at this phase. The scale of the detector and gas systems will allow us to extrapolate to the final detector. 

\subsection{ Active and passive Veto }

 NEXT-10 will be the first full prototype requiring the external active and pasive veto. It is very important to understand the level of rejection achieved with the selected solution. Due to its smaller size we could also consider the possibility to operate the detector with and without shielding to understand the real impact of the shielding. For the long run, NEXT-10 might be a good testbench to measure and understand the laboratory background and the impact on the NEXT physics. Another issues to be considered are the mechanical supports and infrastructure that will be common to the later NEXT-100 detector. 

\subsection{ Mechanics }

  Some of the basic mechanics parameters will be tested at this stage. The pressurize vessel with all services feedthroughs can be tested with this last prototype. There are other issues like the uniformity of the drift electric field, cathode high voltage (up to 100 kV) feedthroughs, gas in and outlets,... These points will be finally fixed at this project phase and transported to the final NEXT-100 detector. 
 
The construction of the detector vessel and the assembly of photosensors will be done in a clean room environment. The usage of the LSC clean room is a good option for NEXT-10. This could serve as a good trainning for the aseembly of the NEXT-100 detector that is larger in volume and complexity.

 The detector is suficiently large to show material deformation and mechanical assembly precission  similar to the final NEXT100. Early tests of this requirement will allow to improve the design for the final NEXT-100. 

\subsection{ Calibration }

  A prototype of the final calibration system will be installed at the NEXT-10. This prototype will be very similar from the size and the performance to the NEXT-100. The goal of the calibration system will be to obtain the target energy resolution and required stability. This includes intercalibration of sensors, absolute energy calibration, electronics calibration and gas monitoring. 

\subsection{Photosensors}

  NEXT-10 represents the first attempt to operate large amount of photosensors ($\sim$900 MPPCs and $\sim$50 PMT's ). The design of the testing procedures and the machanical integration will be done during this phase. The construction of readout planes has to be done under clean conditions. The procedures were developed already in the previous phase. 
  
\subsection{ Electronics }

 Final design of the front-end and back-end electronics will be installed at the NEXT-10 prototype. The scale of the system will be already one forth of the total. Data Acquisition system, trigger (if any)  and data shipping will be installed in LSC at this stage. NEXT-10 is also an excellent bench mark to understand issues relevant to the DAQ like data reduction, event rate and data storage schemmes.

\subsection {Slow Control}

NEXT-10 will have the level of complexity and safety similar to the final detector requiring a full slowcontrol system running by this time. We do not expect big changes except on the number of monitor parameters when we operate NEXT100. That implies that most of the development has be finished by the starting of NEXT-10 such that the Slow Control will be installed at this stage. 

It is crucial the monitoring and the testing of safety aspects of the experiment like the gas leakage detection and High Voltage control. We envisage the same system to control and monitor the NEXT-10 and NEXT-100 detectors.

\subsection{ NEXT-10 at the LSC }

The main prototype will sit in an space of 2$\times$2 m$^2$ (1 m$^2$ central to accommodate a TPC vessel of a few kg of Xenon at 10 bar pressure, plus a surrounding area to install different shielding configurations of a max of 50 cm thickness). This central area must be under the action of the crane, together with some extra empty space to assist in the installation of the heavy pieces of the setup. The rest of the space is used for gas equipment, electronics, clean air flow cabinet and general working area.    

\section{NEXT-100}

The TDR for NEXT-100 will be prepared during 2012 and construction will proceed during 2013. We expect  to commission the detector by the end of 2013. NEXT-100 will be operated in Hall A. 

\subsection{ Active and Pasive veto }

  The larger NEXT-100 volume imply additional mechanical difficulties in the construction of the veto and shielding. The external dimensions of the NEXT-100 detector might be as large as 3 meters high and 2 meters long. The shield and veto has to be designed also to provide access to the readout of the detector and allow openning of the chamber for maintenance.   

\subsection{ Mechanics }

  NEXT-100 detector will be scaled up from the NEXT-10 prototype. However, the change in scale and the new location inside LSC imply changes in the detector layout. Veto and shielding, gas distribution system and electronics has to be adapted from those in NEXT-10.
 
\subsection{ Gas system }

  Gas system will be separated from the one in NEXT-10 prototype to allow performing tests with the NEXT-10 prototype. The new gas system will not be very different from he one of the previous phase but we expect to apply some changes based on the accumulated experience. The main difference will be the system size. The recovery system, and gas storage will be a factor of 10 larger. Recirculation flux and gas purification will be also larger.  

\subsection {Photosensors}

 The production of photosensor readout planes is one of the most consuming manpower  tasks in this phase. The need to perform the integration under clean and radiopure conditions. NEXT will actively search for installations at the different labs, including the clean room in Canfranc, to define an efficient assembly line.

\subsection{ NEXT-100 in LSC }

The activities regarding the installation and commissioning of NEXT100, will require an additional larger space in hall A, as well as other needs in terms of electrical power, gas delivery and others. Preliminary estimates of the required space is shown in Table\ref{tab:SpaceRequirements}. The biggest element of the whole experiment will be the final high pressure TPC, which together with its shielding could possibly need a space of 20--30 m$^2$. If we need to go to active/radiopure veto systems (liquid scintillator, water), this need for space may grow up to 100 m$^2$ or more.

\begin{table}[ht]

\begin{center}
\begin{tabular}{cc}
\hline \hline
 Support structure  & Area (m$^2$)\\\hline 
 Mobile Clean Room  & 3.3  \\
  Gas Bottle storage & 3.2 \\
\hline
 Total & 6.5\\
\hline\hline
\end{tabular}
\end{center}

\vspace{0.5cm}

\begin{center}
\begin{tabular}{cc}
\hline\hline
 Fixed structures & Area (m$^2$) \\\hline 
 Cabin for gas    & 0.7    \\ 
 Xenon recovery system & 6.5 \\
 Tank for storage and fast recovery & 2.3  \\
 Electronic racks & 3.5 \\
\hline 
  Total & 13.0 \\
\hline\hline
\end{tabular}
\end{center}

\vspace{0.5cm}

\begin{center}
\begin{tabular}{cc}
\hline\hline
Detector Estimation & Area (m$^2$)\\\hline 
NEXT-10 detector  & 2.3  \\ 
NEXT-100 detector & 21.5 \\
\hline
Total &  23.8\\
\hline\hline
\end{tabular}
\end{center}

\vspace{0.5cm}
\caption{\small Preliminary estimation of the key elements used for the space requirements of the final NEXT 100 detector.}
\label{tab:SpaceRequirements}  
\end{table}

\listoffigures
\listoftables
\bibliographystyle{amsalpha}
\bibliography{biblio}

\newcommand{\etalchar}[1]{$^{#1}$}
\providecommand{\bysame}{\leavevmode\hbox to3em{\hrulefill}\thinspace}
\providecommand{\MR}{\relax\ifhmode\unskip\space\fi MR }
\providecommand{\MRhref}[2]{%
  \href{http://www.ams.org/mathscinet-getitem?mr=#1}{#2}
}
\providecommand{\href}[2]{#2}
\begin{thebibliography}{BdSD{\etalchar{+}}99}

\bibitem[A{\etalchar{+}}99]{IGEX}
C.~E. Aalseth et~al., \emph{{Neutrinoless double-beta decay of Ge-76: First
  results from the International Germanium Experiment (IGEX) with six
  isotopically enriched detectors}}, Phys. Rev. \textbf{C59} (1999),
  2108--2113.

\bibitem[A{\etalchar{+}}02]{antiKlapdorA}
\bysame, \emph{{Comment on 'Evidence for Neutrinoless Double Beta Decay'}},
  Mod. Phys. Lett. \textbf{A17} (2002), 1475--1478.

\bibitem[A{\etalchar{+}}03]{GEANT4}
S.~Agostinelli et~al., \emph{Geant4: A simulation toolkit}, Nucl. Instrum.
  Meth. \textbf{A506} (2003), 250.

\bibitem[A{\etalchar{+}}04]{GERDA}
I.~Abt et~al., \emph{{A new Ge-76 double beta decay experiment at LNGS}},
  [arXiv:hep-ex/0404039].

\bibitem[A{\etalchar{+}}05a]{CUORE}
R.~Ardito et~al., \emph{{CUORE: A cryogenic underground observatory for rare
  events}}, [arXiv:hep-ex/0501010].

\bibitem[A{\etalchar{+}}05b]{NEMO3A}
R.~Arnold et~al., \emph{{Technical design and performance of the NEMO 3
  detector}}, Nucl. Instrum. Meth. \textbf{A536} (2005), 79--122,
  [arXiv:physics/0402115].

\bibitem[A{\etalchar{+}}08a]{PDG}
C.~Amsler et~al., \emph{{Review of particle physics}}, Phys. Lett.
  \textbf{B667} (2008), 1.

\bibitem[A{\etalchar{+}}08b]{Xenon10A}
J.~Angle et~al., \emph{{First Results from the XENON10 Dark Matter Experiment
  at the Gran Sasso National Laboratory}}, Phys. Rev. Lett. \textbf{100}
  (2008), 021303, arXiv: 0706.0039 [astro-ph].

\bibitem[A{\etalchar{+}}08c]{Aprile:2008rc}
E.~Aprile et~al., \emph{New measurement of the relative scintillation
  efficiency of {Xenon} nuclear recoils below {10 keV}}, arXiv:0810.0274
  [astro-ph].

\bibitem[ABBD06]{Aprile-book}
E.~Aprile, A.~I Bolotnikov, A.~I. Bolozdynya, and T.~Doke, \emph{Noble gas
  detectors}, Wiley-VCN, 2006.

\bibitem[ABC09]{Aprile:2009yh}
Elena Aprile, Laura Baudis, and for the~XENON100 Collaboration, \emph{{Status
  and Sensitivity Projections for the XENON100 Dark Matter Experiment}}, arXiv:
  0902.4253 [astro-ph.IM].

\bibitem[AIEE08]{Avignone:2008}
Frank~T. Avignone~III, Steven~R. Elliott, and Jonathan Engel, \emph{{Double
  Beta Decay, Majorana Neutrinos, and Neutrino Mass}}, Rev. Mod. Phys.
  \textbf{80} (2008), 481--516, arXiv:0708.1033 [nucl-ex].

\bibitem[AKZ05]{Avignone:2005cs}
F.~T. Avignone, G.~S. King, and Yu.~G. Zdesenko, \emph{{Next generation
  double-beta decay experiments: Metrics for their evaluation}}, New J. Phys.
  \textbf{7} (2005), 6.

\bibitem[B{\etalchar{+}}97]{Bolozdynya:96}
A.~Bolozdynya et~al., \emph{{A high pressure xenon self-triggered scintillation
  drift chamber with 3D sensitivity in the range of 20-140 keV deposited
  energy}}, Nuc. Inst. Meth. \textbf{A385} (1997), 225.

\bibitem[B{\etalchar{+}}01]{EXO}
M.~Breidenbach et~al., \emph{{EXO, an advanced Enriched Xenon double-beta decay
  Observatory (Letter of Intent)}}, 2001,
  \url{http://www-project.slac.stanford.edu/exo/}.

\bibitem[Bau07]{Baudis:2007dq}
L.~Baudis, \emph{Direct detection of cold dark matter}, arXiv:0711.3788
  [astro-ph].

\bibitem[BdSD{\etalchar{+}}99]{Borges:99}
F.~I. G.~M. Borges, J.~M.~F. dos Santos, T.~H. V.~T. Dias, F.~P. Santos, P.~J.
  B.~M. Rachinhas, and C.~A.~N. Conde, \emph{Operation of gas proportional
  scintillation counters in a low charge multiplication regime}, Nucl. Instr.
  Meth. \textbf{A422} (1999), 321--325.

\bibitem[Bia99]{Magboltz}
S.~F. Biagi, \emph{{Monte Carlo simulation of electron drift and diffusion in
  counting gases under the influence of electric and magnetic fields}}, Nucl.
  Instrum. Meth. \textbf{421} (1999), 234--240.

\bibitem[BR97]{Bolotnikov:97}
A.~Bolotnikov and B.~Ramsey, \emph{The spectroscopic properties of
  high-pressure xenon}, Nucl. Inst. Meth. \textbf{A396} (1997), 360--370.

\bibitem[C{\etalchar{+}}03]{Conti:2003av}
E.~Conti et~al., \emph{{Correlated fluctuations between luminescence and
  ionization in liquid xenon}}, Phys. Rev. \textbf{B68} (2003), 054201,
  [arXiv:hep-ex/0303008].

\bibitem[C{\etalchar{+}}07]{Coelho:07}
L.~C.~C. Coelho et~al., \emph{{Xenon GPSC high-pressure operation with
  large-area avalanche photodiode readout}}, Nuc. Inst. Meth. \textbf{A575}
  (2007), 444.

\bibitem[Che08]{SNO+}
Mark~C. Chen, \emph{{The SNO+ Experiment}}, arXiv: 0810.3694 [hep-ex].

\bibitem[CMS75]{Charpak:75}
Georges Charpak, S.~Majewski, and F.~Sauli, \emph{{The scintillating drift
  chamber: A new tool for high accuracy, very high rate particle
  localization}}, Nucl. Instr. Meth. \textbf{126} (1975), 381.

\bibitem[Con04]{Conde:04}
C.~A.~N. Conde, \emph{Gas proportional scintillation counters for x-ray
  spectrometry}, John Wiley \& sons, 2004, Chapter 4.2 in {\em X-ray
  Spectrometry: Recent Technical Advances}, K. Tsuji, J. Injuk and R. van
  Greiken, eds.

\bibitem[CP67]{Conde:67}
C.~A.~N. Conde and A.~J. P.~L. Policarpo, Nucl. Instr. Meth. \textbf{53}
  (1967), 7--12.

\bibitem[CP81]{Charpak:81}
H.~N. Charpak, G.~Ngoc and J.~P.~L. Policarpo, \emph{Neutral radiation
  detection and localization}, 1981, US Patent 4,286,158, August 25, 1981.

\bibitem[D{\etalchar{+}}05]{Akimov:2005mq}
Akimov D. et~al., \emph{{EXO: An advanced Enriched Xenon double-beta decay
  Observatory}}, Nucl.\ Phys.\ Proc.\ Suppl. \textbf{138} (2005), 224.

\bibitem[EV02]{Elliott:2002}
Steven~R. Elliott and Petr Vogel, \emph{{Double beta decay}}, Ann. Rev. Nucl.
  Part. Sci. \textbf{52} (2002), 115--151.

\bibitem[Fan47]{FANO}
U.~Fano, \emph{{Ionization Yield of Radiations. 2. The Fluctuations of the
  Number of Ions}}, Phys. Rev. \textbf{72} (1947), 26--29.

\bibitem[G{\etalchar{+}}03]{MAJORANA}
Richard Gaitskell et~al., \emph{{White paper on the Majorana zero-neutrino
  double-beta decay experiment}}, [arXiv:nucl-ex/0311013].

\bibitem[GGM08]{Gonzalez-Garcia:2007}
M.~C. Gonzalez-Garcia and M.~Maltoni, \emph{Phenomenology with massive
  neutrinos}, Physics Reports \textbf{460} (2008), 1, arXiv:0704.1800 [hep-ph].

\bibitem[Gio98]{Giomataris:1998}
Y.~Giomataris, \emph{{Development and prospects of the new gaseous detector
  'Micromegas'}}, Nucl. Instr. Meth. \textbf{A419} (1998), 239--250.

\bibitem[Giu07]{Giunti:2006}
Carlo Giunti, \emph{{Theory and phenomenology of neutrino mixing}}, Nucl. Phys.
  Proc. Suppl. \textbf{169} (2007), 309--320, [arXiv:hep-ph/0611125].

\bibitem[GM35]{GoeppertMayer:1935qp}
M.~Goeppert-Mayer, \emph{{Double beta-disintegration}}, Phys. Rev. \textbf{48}
  (1935), 512--516.

\bibitem[GRRC96]{Giomataris:1995}
Y.~Giomataris, P.~Rebourgeard, J.~P. Robert, and Georges Charpak,
  \emph{{MICROMEGAS: A high-granularity position-sensitive gaseous detector for
  high particle-flux environments}}, Nucl. Instrum. Meth. \textbf{A376} (1996),
  29--35.

\bibitem[K{\etalchar{+}}04]{Kowina:2004}
P.~Kowina et~al., \emph{{}}, Eur. Phys. J. A \textbf{22} (2004), 293.

\bibitem[KK{\etalchar{+}}01]{Heidelberg-Moscow}
H.~V. Klapdor-Kleingrothaus et~al., \emph{{Latest results from the
  Heidelberg-Moscow double-beta-decay experiment}}, Eur.\ Phys.\ J.\ A
  \textbf{12} (2001), 147, [arXiv:hep-ph/0103062].

\bibitem[KK{\etalchar{+}}04]{Klapdor03}
\bysame, \emph{{Search for neutrinoless double beta decay with enriched Ge-76
  in Gran Sasso 1990-2003}}, Phys.\ Lett. \textbf{B586} (2004), 198,
  [arXiv:hep-ph/0404088].

\bibitem[KKK06]{Klapdor06}
H.~V. Klapdor-Kleingrothaus and I.~V. Krivosheina, \emph{The evidence for the
  observation of $0\nu\beta\beta$ decay: The identification of $0\nu\beta\beta$
  events from the full spectra}, Mod.\ Phys.\ Lett. \textbf{A21} (2006), 1547.

\bibitem[KR07]{Kaufmann:2007zz}
L.~Kaufmann and A.~Rubbia, \emph{{The ArDM project: A direct detection
  experiment, based on liquid Argon, for the search of Dark Matter}}, Nucl.\
  Phys.\ Proc.\ Suppl. \textbf{173} (2007), 141.

\bibitem[L{\etalchar{+}}98]{Gotthard98}
R.~Luscher et~al., \emph{Search for beta beta decay in xe-136: New results from
  the gotthard experiment}, Phys. Lett. \textbf{B434} (1998), 407.

\bibitem[M{\etalchar{+}}07]{Monteiro:07}
C.~M.~B. Monteiro et~al., \emph{{Secondary scintillation yield in pure xenon}},
  JINST \textbf{2} (2007), 5001.

\bibitem[Maj37]{Majorana:1937vz}
Ettore Majorana, \emph{{Theory of the Symmetry of Electrons and Positrons}},
  Nuovo Cim. \textbf{14} (1937), 171--184.

\bibitem[Men05]{Mena:2005}
Olga Mena, \emph{{Unveiling Neutrino Mixing and Leptonic CP Violation}}, Mod.
  Phys. Lett. \textbf{A20} (2005), 1--17, [arXiv:hep-ph/0503097].

\bibitem[Nyg07]{Nygren:2007zz}
D.~Nygren, \emph{{Optimal detectors for WIMP and $0\nu\beta\beta$ searches:
  Identical high-pressure xenon gas TPCs?}}, Nucl. Instrum. Meth. \textbf{A581}
  (2007), 632--642.

\bibitem[Nyg09]{Nygren:2008}
\bysame, \emph{{High pressure Xenon Gas Electroluminescent TPC for
  $0\nu\beta\beta$ Decay Search}}, Nucl. Instrum. Meth. \textbf{A} (2009).

\bibitem[Pah08]{SUPERNEMO}
R.~Benton Pahlka, \emph{{The SuperNEMO Experiment}}, arXiv: 0810.3169 [hep-ex].

\bibitem[PTZ00]{DECAY4}
O.~A. Ponkratenko, V.~I. Tretyak, and Yu.~G. Zdesenko, \emph{{The Event
  generator DECAY4 for simulation of double beta processes and decay of
  radioactive nuclei}}, Phys. Atom. Nucl. \textbf{63} (2000), 1282--1287,
  [arXiv:nucl-ex/0104018].

\bibitem[Rac37]{Racah:1937qq}
G.~Racah, \emph{{On the symmetry of particle and antiparticle}}, Nuovo Cim.
  \textbf{14} (1937), 322, (also in Klapdor-Kleingrothaus, H.~V.: {\it Sixty
  years of double beta decay}, 110-116).

\bibitem[RFSV06]{Rodin:2007}
V.~A. Rodin, A.~Faessler, F.~Simkovic, and P.~Vogel, \emph{{Assessment of
  uncertainties in QRPA 0nu beta beta-decay nuclear matrix elements}}, Nucl.
  Phys. \textbf{A766} (2006), 107--131, arXiv: 0706.4304 [nucl-th].

\bibitem[SK79]{Suzuki:1979km}
M.~Suzuki and S.~Kubota, \emph{{Mechanism of proportional scintillation in
  argon, krypton and xenon}}, Nucl. Instrum. Meth. \textbf{164} (1979),
  197--199.

\bibitem[SV82]{Schechter:1982}
J.~Schechter and J.~W.~F. Valle, \emph{{Neutrinoless double beta decay in
  SU(2)$\times$U(1) theories}}, Physical Review \textbf{D25} (1982), 2951.

\bibitem[V{\etalchar{+}}93]{Gotthard93}
J.~C. Vuilleumier et~al., \emph{{Search for neutrinoless double beta decay in
  Xe-136 with a time projection chamber}}, Phys. Rev. \textbf{D48} (1993),
  1009--1020.

\bibitem[ZDT02]{antiKlapdorB}
Yu.~G. Zdesenko, F.~A. Danevich, and V.~I. Tretyak, \emph{{Has neutrinoless
  double beta decay of Ge-76 been really observed?}}, Phys. Lett. \textbf{B546}
  (2002), 206--215.

\end{thebibliography}

\end{document}